\documentclass[english,aip,letter,superscriptaddress,floatfix,jcp,twocolumn]{revtex4}
\usepackage[T1]{fontenc}
\usepackage[latin9]{inputenc}
\usepackage{verbatim}
\usepackage{amsmath}
\usepackage{amssymb}
\usepackage{graphicx}
\usepackage{esint}

\makeatletter

\providecommand{\tabularnewline}{\\}

\@ifundefined{textcolor}{}
{%
 \definecolor{BLACK}{gray}{0}
 \definecolor{WHITE}{gray}{1}
 \definecolor{RED}{rgb}{1,0,0}
 \definecolor{GREEN}{rgb}{0,1,0}
 \definecolor{BLUE}{rgb}{0,0,1}
 \definecolor{CYAN}{cmyk}{1,0,0,0}
 \definecolor{MAGENTA}{cmyk}{0,1,0,0}
 \definecolor{YELLOW}{cmyk}{0,0,1,0}
}

\usepackage{ae,aecompl}

\makeatother

\usepackage{babel}
\begin{document}

\title{Brownian Dynamics without Green's Functions}

\author{Steven Delong}

\affiliation{Courant Institute of Mathematical Sciences, New York University,
New York, NY 10012}

\author{Florencio Balboa Usabiaga}

\affiliation{Departamento de Física Teórica de la Materia Condensada and Condensed
Matter Physics Center (IFIMAC), Univeridad Autónoma de Madrid, Madrid
28049, Spain}

\author{Rafael Delgado-Buscalioni}

\affiliation{Departamento de Física Teórica de la Materia Condensada and Condensed
Matter Physics Center (IFIMAC), Univeridad Autónoma de Madrid, Madrid
28049, Spain}

\author{Boyce E. Griffith}

\affiliation{Leon H. Charney Division of Cardiology, Department of Medicine, New
York University School of Medicine, New York, NY 10016}

\affiliation{Courant Institute of Mathematical Sciences, New York University,
New York, NY 10012}

\author{Aleksandar Donev}

\email{donev@courant.nyu.edu}

\affiliation{Courant Institute of Mathematical Sciences, New York University,
New York, NY 10012}
\begin{abstract}
We develop a Fluctuating Immersed Boundary (FIB) method for performing
Brownian dynamics simulations of confined particle suspensions. Unlike
traditional methods which employ analytical Green's functions for
Stokes flow in the confined geometry, the FIB method uses a fluctuating
finite-volume Stokes solver to generate the action of the response
functions ``on the fly''. Importantly, we demonstrate that both
the deterministic terms necessary to capture the hydrodynamic interactions
among the suspended particles, as well as the stochastic terms necessary
to generate the hydrodynamically-correlated Brownian motion, can be
generated by solving the steady Stokes equations numerically only
once per time step. This is accomplished by including a stochastic
contribution to the stress tensor in the fluid equations consistent
with fluctuating hydrodynamics. We develop novel temporal integrators
that account for the multiplicative nature of the noise in the equations
of Brownian dynamics and the strong dependence of the mobility on
the configuration for confined systems. Notably, we propose a random
finite difference approach to approximating the stochastic drift proportional
to the divergence of the configuration-dependent mobility matrix.
Through comparisons with analytical and existing computational results,
we numerically demonstrate the ability of the FIB method to accurately
capture both the static (equilibrium) and dynamic properties of interacting
particles in flow.
\end{abstract}
\maketitle
\global\long\def\V#1{\boldsymbol{#1}}
\global\long\def\M#1{\boldsymbol{#1}}
\global\long\def\Set#1{\mathbb{#1}}

\global\long\def\D#1{\Delta#1}
\global\long\def\d#1{\delta#1}

\global\long\def\norm#1{\left\Vert #1\right\Vert }
\global\long\def\abs#1{\left|#1\right|}

\global\long\def\grad{\M{\nabla}}
\global\long\def\avv#1{\langle#1\rangle}
\global\long\def\av#1{\left\langle #1\right\rangle }

\global\long\def\P{\mathcal{P}}

\global\long\def\ki{k}
\global\long\def\wi{\omega}

\global\long\def\bu{\V u}
 \global\long\def\bv{\V v}
 \global\long\def\br{\V r}

\global\long\def\sM#1{\M{\mathcal{#1}}}
\global\long\def\Mob{\sM M}
\global\long\def\J{\sM J}
\global\long\def\S{\sM S}
\global\long\def\L{\sM L}

\section{\label{sec:Intro}Introduction}

Stochastic fluctuations in fluids arise from the fact that fluids
are composed of molecules whose positions and velocities are random.
One can capture thermal fluctuations using direct particle level calculations.
But even coarse-grained particle methods \cite{ParticleMesoscaleHydrodynamics,SHSD_PRL,DiffusionRenormalization_PRL}
are computationally expensive because the dynamics of individual particles
is much faster than hydrodynamic time scales. Alternatively, thermal
fluctuations can be included in the Navier-Stokes equations through
stochastic forcing terms, as proposed by Landau and Lifshitz \cite{Landau:Fluid}.
The basic idea of fluctuating hydrodynamics \cite{FluctHydroNonEq_Book}
is to add a stochastic stress tensor to the usual viscous stress tensor
\cite{OttingerBook}. This has been shown to be a very good model
of fluids down to essentially molecular scales \cite{DiffusionRenormalization,DiffusionRenormalization_PRL,StagerredFluct_Inhomogeneous,StagerredFluctHydro,StaggeredFluct_Energy,LowMachExplicit}.

The presence of suspended particles is a common feature of complex
fluids. At small scales, the motion of immersed particles is driven
by thermal fluctuations, giving rise to Brownian motion strongly affected
by hydrodynamic effects. Fluctuating hydrodynamics has been shown
to be a useful tool in modeling the dynamics of colloidal particles
and polymer chains suspended in a fluid \cite{FluctuatingHydro_FluidOnly,LatticeBoltzmann_Polymers,BD_LB_Comparison,BD_LB_Ladd,LB_SoftMatter_Review,SELM,ImmersedFEM_Patankar,ISIBM,ForceCoupling_Fluctuations,CompressibleBlobs,BrownianParticle_IncompressibleSmoothed,FIMAT_Patankar}.
By coupling a fluctuating fluid solver with immersed particles one
can model the Brownian dynamics from the short time scales, at which
sound waves play a role \cite{CompressibleBlobs}, to longer times,
at which the velocity correlations decay in a power-law manner due
to viscous dissipation. At the same time, the dynamics of interest
in many problems is the diffusive (Brownian) dynamics of the immersed
structures, which happens at much longer times due to the very small
Reynolds numbers, or more precisely, the very large Schmidt numbers
present in typical applications.

In the limit of zero Reynolds number, or more precisely, infinite
Schmidt number, the methods of Brownian \cite{BD_Fixman,BD_Hinch,BrownianDynamics_DNA,BrownianDynamics_DNA2,BrownianDynamics_OrderN,BD_IBM_Graham,BrownianDynamics_FMM}
and Stokesian dynamics \cite{BrownianDynamics_OrderNlogN,StokesianDynamics_Wall}
have dominated in chemical engineering, and related techniques have
been used in biochemical engineering \cite{HYDROLIB,SphereConglomerate,HYDROPRO,HYDROPRO_Globular}.
In this work we focus on Brownian dynamics, which can be seen as a
simplified version of Stokesian dynamics that does not include second-order
multipole terms (rotlets and stresslets) or lubrication effects in
the hydrodynamic interactions among the immersed particles. A key
common feature of this class of methods is that they simulate the
overdamped (diffusive) dynamics of the particles by using Green's
functions for steady Stokes flow to capture the effect of the fluid.
While this sort of \emph{implicit solvent} approach works very well
in many situations, it has several notable technical difficulties:
achieving near linear scaling for many-particle systems is technically
challenging \cite{BrownianDynamics_OrderN,BrownianDynamics_OrderNlogN,BrownianDynamics_FMM},
handling non-trivial boundary conditions (bounded systems) is complicated
\cite{BD_IBM_Graham} and has to be done on a case-by-case basis \cite{BrownianDynamics_DNA2,StokesianDynamics_Wall,StokesianDynamics_Slit,StokesianDynamics_Confined,BD_LB_Comparison},
and including Brownian motion requires additional specialized treatment
\cite{BD_Fixman,BD_Hinch}. Notably, combining all components together
and performing Brownian or Stokesian dynamics in complex geometry
with accurate hydrodynamics, thermal fluctuations, and near-linear
scaling requires a rather sophisticated set of tools. This is evidenced
by the fact that existing Stokesian dynamics simulations of Brownian
suspensions in even the simplest confined geometry, a slit channel,
have relied on several uncontrolled approximations \cite{SD_TwoWalls},
even though all of the expressions and tools have, in principle, been
developed \cite{StokesianDynamics_Brownian,StokesianDynamics_Slit}.

At first sight, it may appear that there is a conceptual gap between
methods based on fluctuating hydrodynamics and those based on Green's
functions. The fluid inertia, or, more precisely, the momentum diffusion
is inherently part of the fluctuating hydrodynamics formulation of
Brownian motion \cite{VACF_FluctHydro,Faxen_FluctuatingHydro,VACF_Langevin,LangevinDynamics_Theory},
while it does not appear in the equations of Brownian or Stokesian
dynamics. For example, particles suspended in a fluctuating fluid
with inertial memory exhibit a well-known power-law decay of the velocity
auto-correlation function (VACF) \cite{VACF_Langevin}, which is not
present in Brownian dynamics (BD) because BD is meant to describe
longer time scales, at which the VACF looks like a Dirac delta function.
In order to access the diffusive scaling, methods based on fluctuating
hydrodynamics, such as Lattice-Boltzmann (LB) techniques \cite{LB_SoftMatter_Review},
must ensure that the Schmidt number $\text{Sc}$ is sufficiently large
\cite{StokesEinstein}, though in practice $\text{Sc}$ is always
limited by computational efficiency considerations. Extensive testing
has confirmed that with proper care a match can be achieved between
results obtained using LB and BD methods \cite{BD_LB_Comparison,BD_LB_Ladd,LBM_vs_BD_Burkhard}.

Nevertheless, there remains a gap in the range of accessible Reynolds/Schmidt
numbers between the two classes of methods. We close this gap in this
work by designing a Fluctuating Immersed Boundary (FIB) method that
solves the overdamped (inertia-less) equations of Brownian dynamics
using an \emph{explicit solvent} representation of the fluid hydrodynamics.
Importantly, the FIB method includes confinement in nontrivial geometries
and Brownian motion consistently and with a controlled accuracy, and
has linear complexity in the number of immersed particles. The key
observation underlying the FIB method is that analytical Green's functions
can be replaced by a steady Stokes solver with a stochastic stress
tensor, as dictated by fluctuating hydrodynamics. Specifically, the
action of the required response functions (on both deterministic and
stochastic terms) is computed ``on the fly'' rather than pre-computed
analytically. The fluid solver can be used to handle nontrivial boundary
conditions, including cases where the concentration of chemical reactants
affects the fluid flow via osmo-phoretic effects \cite{Hematites_PRL,Hematites_Science}.
The stochastic increments required to simulate the Brownian motion
are generated by the fluctuating Stokes solver with no additional
effort, in arbitrary domains with a combination of standard periodic,
no-slip or slip boundaries \cite{NonProjection_Griffith}. Because
in confined systems the mobility strongly depends on the positions
of the particles relative to the boundaries, we pay special attention
to correctly capturing the well-known stochastic drift term proportional
to the divergence of the configuration-dependent mobility matrix.
In particular, we develop a random finite difference approach that
is related, but distinct from, the traditional Fixman midpoint method.

Rather closely related to our proposal is the work on the Stochastic
Immersed Boundary Method (SIBM) and its generalization the Stochastic
Eulerian Lagrangian Method (SELM) developed by Atzberger and collaborators
\cite{StochasticImmersedBoundary,SELM}, as well as the work of Maxey
and collaborators on the Force Coupling Method (FCM) \cite{ForceCoupling_Stokes,ForceCoupling_Monopole,ForceCoupling_Ellisoids}.
In work independent from ours, Keaveny has recently included thermal
fluctuations in the fluctuating FCM method \cite{ForceCoupling_Fluctuations},
and also accounted for stresslet and rotlet terms (which are not included
in our FIB method). While inertia can be included easily in both SELM
and FCM, as it can be in the Inertial Coupling Method (ICM) \cite{ISIBM}
very closely-related to the FIB method, both methods can also be used
in the steady Stokes limit \cite{ForceCoupling_Fluctuations}. At
the level of the mathematical (continuum) formulation the SELM, fluctuating
FCM and FIB methods are very similar, though the numerical techniques
used to discretize and solve the equations of motion are rather distinct,
leading to several crucial differences between the work presented
here and existing work. Specifically, we develop novel temporal integrators
that efficiently account for the dependence of the mobility on configuration,
which is crucial in confined geometries. Crucially, we do not assume
specific forms of the boundary conditions when solving the fluid (steady
or unsteady) Stokes equations, and, in particular, we do not rely
on periodic boundary conditions and using a Fourier basis (and the
associated FFTs) to diagonalize the Stokes operator \cite{StochasticImmersedBoundary,ForceCoupling_Fluctuations}.
Furthermore, we do not use Gaussian kernels as in the FCM, rather,
we employ the compact-support kernels Peskin specifically constructed
for immersed-boundary discretizations that employ a finite-difference-type
discretization of the fluid equations \cite{IBM_PeskinReview}. Note
also that we handle domain boundaries (for both deterministic \emph{and}
stochastic terms) directly in the finite-volume fluctuating Stokes
solver, unlike recent extensions to BD \cite{BD_IBM_Graham} that
handle complex boundaries by discretizing the boundary using immersed-boundary
techniques. Independently of our work, an extension to SELM to nonperiodic
domains, but using a finite-element rather than a finite-volume Stokes
solver, has recently been developed \cite{SELM_FEM}. We will defer
a more detailed comparison with this related but distinct work until
the concluding section, after we present the technical details of
the FIB method.

This paper is organized as follows. In the remainder of this section
we summarize the well-known and widely-used method of Brownian dynamics,
to the extent necessary for subsequent comparison with our FIB method.
In Section \ref{sec:FluidPartModel} we discuss the equations of motion
solved in the FIB method at the continuum level, and explain the relation
to the equations of Brownian dynamics. Then, we explain how we discretize
those equations in both space (Section \ref{sec:Discretization})
and time (Section \ref{sec:TemporalIntegrator}). In Section \ref{sec:NumericalTests}
we perform a series of validation tests confirming the accuracy and
robustness of the FIB method on a variety of tests of increasing complexity.
Several technical derivations are detailed in the Appendix.

\subsection{Brownian Dynamics}

The equations of Brownian Dynamics (BD) model the diffusive dynamics
of the positions $\V q\left(t\right)=\left\{ \V q_{1}\left(t\right),\dots,\V q_{N}\left(t\right)\right\} $
of a collection of $N$ particles via the Ito system of stochastic
differential equations,
\begin{equation}
\frac{d\V q}{dt}=\Mob\V F+\sqrt{2k_{B}T}\Mob^{\frac{1}{2}}\,\widetilde{\M{\mathcal{W}}}(t)+k_{B}T\,\left(\partial_{\V q}\cdot\Mob\right),\label{eq:BrownianDynamics}
\end{equation}
where $\Mob(\V q)\succeq\M 0$ is the symmetric positive semidefinite
(SPD) mobility matrix, relating the applied forces, $\V F(\V q)=-\partial U(\V q)/\partial\V q$
with $U(\V q)$ a conservative potential, to the resulting (deterministic)
velocity. For notational brevity we will often omit the explicit dependence
on the configuration $\V q$ or time $t$. The stochastic forcing
$\widetilde{\M{\mathcal{W}}}(t)$ denotes a vector of independent
white noise process, formally time derivatives of independent Wiener
processes. The ``square root'' of the mobility $\Mob^{\frac{1}{2}}$
is a matrix (not necessarily square) which satisfies the fluctuation
dissipation balance condition 
\begin{equation}
\Mob^{\frac{1}{2}}\left(\Mob^{\frac{1}{2}}\right)^{\star}=\Mob.\label{eq:BDFluctDissBalance}
\end{equation}
We use a superscript star throughout to denote the adjoint of a linear
operator for a suitably-weighted inner product (conjugate transpose
for matrices for the standard inner product). Throughout this paper
we will rewrite the equations of motion (\ref{eq:BrownianDynamics})
to eliminate the final ``thermal'', ``stochastic'' or ``spurious''
drift term $k_{B}T\,\left(\partial_{\V q}\cdot\Mob\right)$ by using
the \textit{kinetic }interpretation of the stochastic integral \cite{KineticStochasticIntegral_Ottinger},
denoted in this paper by the stochastic product symbol $\diamond$,
\begin{equation}
\frac{d\V q\left(t\right)}{dt}=\Mob(\V q)\V F(\V q)+\sqrt{2k_{B}T}\Mob^{\frac{1}{2}}(\V q)\diamond\widetilde{\M{\mathcal{W}}}(t).\label{eq:BDKinetic}
\end{equation}

Condition (\ref{eq:BDFluctDissBalance}) insures that the dynamics
(\ref{eq:BDKinetic}) is time-reversible with respect to the Gibbs-Boltzmann
distribution 
\begin{equation}
P_{eq}(\V q)=Z^{-1}\exp\left(-U(\V q)/k_{B}T\right),\label{eq:gibbsboltzmann}
\end{equation}
where $Z$ is a normalization constant. This may be seen by examining
the Fokker-Planck equation for the evolution of the probability distribution
for observing the state $\V q$ at time $t$ corresponding to (\ref{eq:BrownianDynamics})
or (\ref{eq:BDKinetic}), 
\begin{equation}
\frac{\partial P}{\partial t}=\frac{\partial}{\partial\V q}\cdot\left\{ \Mob\left[\frac{\partial U}{\partial\V q}P+\left(k_{B}T\right)\frac{\partial P}{\partial\V q}\right]\right\} ,\label{eq:FokkerPlanck}
\end{equation}
and noting that the term in square brackets vanishes when $P=P_{eq}.$

Developing schemes to simulate Brownian dynamics has several challenges.
One such challenge is evaluating, or more precisely, applying the
mobility matrix, which contains all of the information about hydrodynamic
interactions between the particles. This can be non-trivial to achieve
analytically even in relatively simple geometries, and the mobility
is generally approximated via a multipole expansion or infinite series
of images. Special care must be taken to insure that the truncation
of these infinite series result in a positive-semidefinite matrix
\cite{StokesianDynamics_Wall,StokesianDynamics_Slit}. Even if an
efficient application of the action of the mobility matrix is available,
one still must also be able to generate the action of $\Mob^{\frac{1}{2}}$,
typically approximated by Chebyshev polynomials as originally proposed
by Fixman \cite{BD_Fixman_sqrtM}. Finally, the thermal drift term
$k_{B}T\,\partial_{\V q}\cdot\left(\Mob(\V q)\right)$ must be calculated
or approximated in some way. This amounts to consistently discretizing
the kinetic interpretation of the stochastic integral, which is traditionally-accomplished
by using the Fixman midpoint algorithm \cite{KineticStochasticIntegral_Ottinger}.
Note however that the Fixman method (and in general the use of the
kinetic stochastic integral) requires handling the inverse of the
mobilty matrix, which can add substantial complication and cost \cite{ForceCoupling_Fluctuations}.

\subsection{Mobility Matrix}

For two well-separated spherical particles $i$ and $j$, we can approximate
the pairwise mobility, which determines the velocity on particle $i$
resulting from a force on particle $j$, with \cite{BD_LB_Ladd,StokesianDynamics_Wall}
\begin{eqnarray}
 &  & \M{\mathcal{M}}_{ij}=\M{\mathcal{M}}_{ji}=\nonumber \\
 &  & \eta^{-1}\left(\M I+\frac{a^{2}}{6}\grad_{\V r}^{2}\right)\left(\M I+\frac{a^{2}}{6}\grad_{\V r^{\prime}}^{2}\right)\M K(\V r,\V r^{\prime})\big|_{\V r^{\prime}=\V q_{i}}^{\V r=\V q_{j}},\label{eq:MobilityFaxen}
\end{eqnarray}
where $a$ is the radius of the particles. Here $\M K$ is the Green's
function for the steady Stokes problem with unit viscosity, with the
appropriate boundary conditions such as no-slip on the boundaries
of the domain. The differential operator $\M I+\left(a^{2}/6\right)\grad^{2}$
is called the Faxen operator \cite{BD_LB_Ladd} and leads to the well-known
Faxen correction to the Stokes drag law. Note that the form of (\ref{eq:MobilityFaxen})
guarantees that an SPD mobility matrix is obtained by construction.

\subsubsection{Unconfined systems}

For a three dimensional unbounded domain with fluid at rest at infinity,
$\M K(\V r,\V r^{\prime})=\M K(\V r-\V r^{\prime})$ is isotropic
and given by the Oseen tensor,
\[
\M K(\V r)=\M O(\V r)=\frac{1}{8\pi r}\left(\M I+\frac{\V r\otimes\V r}{r^{2}}\right).
\]
For many particles in an unbounded domain, applying (\ref{eq:MobilityFaxen})
to the Oseen tensor yields the far-field expression of the Rotne-Prager-Yamakawa
(RPY) tensor \cite{RotnePrager}, commonly used in Brownian dynamics
simulations. A correction needs to be introduced when particles are
close to each other in order to produce a mobility which is positive
definite \cite{RotnePrager}, 
\begin{equation}
\Mob_{ij}=\frac{1}{6\pi\eta a}\begin{cases}
C_{1}(r_{ij})\M I+C_{2}(r_{ij})\frac{\V r_{ij}\otimes\V r_{ij}}{r_{ij}^{2}}, & \quad r_{ij}>2a\\
C_{3}(r_{ij})\M I+C_{4}(r_{ij})\frac{\V r_{ij}\otimes\V r_{ij}}{r_{ij}^{2}}, & \quad r_{ij}\leq2a
\end{cases}\label{eq:RPYTensor}
\end{equation}
where $\V r_{ij}=\V q_{i}-\V q_{j}$ is the vector connecting the
particles, and $r_{ij}$ is its length, and
\begin{eqnarray*}
C_{1}(r_{ij})=\frac{3a}{4r_{ij}}+\frac{a^{3}}{2r_{ij}^{3}} & \quad & C_{2}(r_{ij})=\frac{3a}{4r_{ij}}-\frac{3a^{3}}{2r_{ij}^{3}}\\
C_{3}(r_{ij})=1-\frac{9r_{ij}}{32a} & \quad & C_{4}(r_{ij})=\frac{3r_{ij}}{32a}.
\end{eqnarray*}
The diagonal blocks of the mobility matrix, i.e., the self-mobility
can be obtained by setting $r_{ij}=0$ (giving $C_{3}(0)=1$ and $C_{4}(0)=0$)
to obtain $\Mob_{ii}=\left(6\pi\eta a\right)^{-1}\M I$, which matches
the Stokes solution for the drag for flow around a sphere. It is important
physically that $\Mob_{ij}=\Mob_{ii}$ when $r_{ij}=0$ since two
perfectly overlapping particles must behave as if there is only a
single particle at that location. 

For a single particle in an unbounded domain it is obvious that $\Mob_{ii}$
is constant and thus has vanishing divergence. For the RPY mobility
it can be shown that $\partial_{\V q}\cdot\left(\Mob(\V q)\right)=\V 0$
even for multi-particle systems (this is in fact a rather generic
consequence of the incompressibility of the flow \cite{DiffusionJSTAT}).
Note, however, that when stresslet terms are included the mobility
becomes a complicated function of configuration (see Appendix \ref{sec:Stresslets})
and the stochastic drift term must be accounted for \cite{ForceCoupling_Fluctuations}.

\subsubsection{Confined systems}

In the presence of boundaries, the Green's function may be decomposed
as 
\begin{equation}
\M K(\V r,\V r^{\prime})=\M O(\V r-\V r^{\prime})+\M K^{w}(\V r,\V r^{\prime}),\label{eq:split_Greens}
\end{equation}
where $\M K^{w}$ is the Green's function for a disturbance velocity
field enforcing the no-slip condition at the walls. One can approximate
the pairwise far-field mobility by using (\ref{eq:MobilityFaxen})
and applying the Faxen operators to $\M O$ and $\M K^{w}$ separately
\cite{StokesianDynamics_Wall}. For the diagonal blocks, we have to
consider the self-mobility in an unbounded domain separately, and
only use (\ref{eq:MobilityFaxen}) with $\M K$ replaced by $\M K^{w}$
in order to account for the disturbance velocity from the boundary
conditions \cite{StokesianDynamics_Wall} (equivalently, to account
for the hydrodynamic interactions with the image particles),
\begin{eqnarray*}
 &  & \Mob_{ii}=\frac{1}{6\pi\eta a}\M I\\
 &  & +\:\eta^{-1}\left(\M I+\frac{a^{2}}{6}\grad_{\V r}^{2}\right)\left(\M I+\frac{a^{2}}{6}\grad_{\V r^{\prime}}^{2}\right)\M{K^{w}}(\V r,\V r^{\prime})\big|_{\V r^{\prime}=\V q_{i}}^{\V r=\V q_{i}}.
\end{eqnarray*}
Note that this approach requires knowing the Green's function for
the particular geometry in question. For a single no-slip wall $\M K^{w}$
was obtained by Blake \cite{blake1971note}, but for a slit channel
with two no-slip walls there is no manageable analytical form \cite{StokesianDynamics_Slit}. 

It is important to note that even for a single particle near a boundary
the mobility strongly depends on the position of the particle relative
to the boundary and therefore the thermal drift $k_{B}T\,\partial_{\V q}\cdot\Mob$
must be accounted for \cite{StokesianDynamics_Slit}.

\section{\label{sec:FluidPartModel}Fluctuating Immersed Boundary Method}

In this section we present the continuum formulation of the equations
of motion as employed in the FIB method. At the same time, we use
operator notation that generalizes to spatially-discretized equations,
by simply replacing the continuum integro-differential operators with
sums and differences (matrices), see the discussion of Atzbeger \cite{SELM}
for more details. This makes the majority of this section directly
transferable to the semi-discrete setting presented in Section \ref{sec:Discretization}.
The operator notation we employ also enables us to treat in a unified
way different boundary conditions without requiring a specific basis
for the solution of the Stokes equations.

We consider $n$ spherical neutrally-buoyant particles of radius $a$
in $d$ dimensions, having spatial positions $\V q=\left\{ \V q_{1},\dots,\V q_{N}\right\} $
with $\V q_{i}=(q_{i}^{(1)},\ldots,q_{n}^{(d)})$. These particles
are immersed in an incompressible fluid of constant density $\rho$,
temperature $T$, and viscosity $\eta$, and described by the fluctuating
time dependent Stokes equations for the fluid velocity $\bv(\V r,t)$,
\begin{eqnarray}
\rho\partial_{t}\bv+\grad\pi & = & \eta\grad^{2}\bv+\V f+\sqrt{2\eta k_{B}T}\;\grad\cdot\mathcal{\M{\mathcal{Z}}}\label{eq:ContFluid}\\
\grad\cdot\bv & = & 0,\nonumber 
\end{eqnarray}
along with appropriate boundary conditions. Here $\V f\left(\V r,t\right)$
is a force density applied to the fluid, $k_{B}$ is Boltzmann's constant,
and $\M{\mathcal{Z}}(\V r,t)$ is a random Gaussian tensor field whose
components are white in space and time with mean zero \cite{FluctHydroNonEq_Book},
\begin{equation}
\avv{\mathcal{Z}_{ij}(\V r,t)\mathcal{Z}_{kl}(\V r^{\prime},t')}=\left(\delta_{ik}\delta_{jl}+\delta_{il}\delta_{jk}\right)\delta(t-t^{\prime})\delta(\V r-\V r^{\prime}).\label{eq:Z_cov}
\end{equation}

The coupling between the fluid and particles employed here is used
in a large number of other methods and related prior work. In particular,
the same basic equations are employed in SIBM \cite{StochasticImmersedBoundary,SIBM_Brownian}
and SELM \cite{SELM}. In the deterministic setting, Maxey and collaborators
have developed in extensive detail the use of smooth envelope or kernel
function to represent particles in flow in the context of the Force
Coupling Method (FCM) \cite{ForceCoupling_Monopole,ForceCoupling_Stokes,ForceCoupling_Channel}.
Recently, Keaveny has included fluctuations in the description in
a manner fully consistent with our presentation \cite{ForceCoupling_Fluctuations}.
Similar representations of particles have also been used with the
Lattice Boltzmann method \cite{LB_IB_Points,LB_SoftMatter_Review}.
Both Atzberger \cite{SIBM_Brownian} and Keaveny \cite{ForceCoupling_Fluctuations}
have already noted the relation to Brownian and Stokesian dynamics.
Nevertheless, for completeness and clarity and the benefit of the
reader, here we present a unified view of these somewhat disjoint
works and point out some less-appreciated but important features.

\subsection{Fluid-Particle Interaction}

In the FIB method, the shape of the particle and its effective interaction
with the fluid is captured through a smooth kernel function $\delta_{a}\left(\V r\right)$
that integrates to unity and whose support is localized in a region
of size $a$. This kernel is used to mediate two crucial operations.
First, it is used to transfer (spread) the force exerted on the particle
to the fluid. Second, it is used to impose a minimally-resolved form
of the no-slip constraint stating that the velocity of the particle
equals the local velocity of the fluid. Following Refs. \cite{DirectForcing_Balboa,ISIBM,ReactiveBlobs,CompressibleBlobs}
we term this diffuse (rather than ``point'') particle a \emph{blob}
for lack of better terminology (in polymer modeling the term bead
is used for the same concept \cite{LB_SoftMatter_Review})\emph{.}

In order to couple the fluid velocity field to the motion of immersed
particles, we introduce composite local averaging $\J(\V q)$ and
spreading $\S(\V q)$ operators. The operator $\J(\V q)$ takes a
continuous velocity field $\bv\left(\V r\right)$ and computes its
local average at the position of each particle, while $\S(\V q)$
takes the forces $\V F=\left\{ \V F_{1},\dots,\V F_{N}\right\} $
applied on the particles and computes a smooth force density field,
\begin{eqnarray}
\left(\mathcal{\J}(\V q)\bv\left(\V r\right)\right)_{i} & = & \int\delta_{a}(\V q_{i}-\V r)\bv(\V r)d\V r\label{eq:InterpOp}\\
\left(\S(\V q)\V F\right)(\V r) & = & \sum_{i}\delta_{a}(\V q_{i}-\V r)\V F_{i}.\label{eq:SpreadOp}
\end{eqnarray}
Note that $\J$ is dimensionless, and $\S$ has units of inverse volume.
The blobs are assumed to move with the locally-averaged fluid velocity,
\begin{equation}
\frac{d\V q\left(t\right)}{dt}=\J(\V q)\bv\left(\V r,t\right),\label{eq:ParticleVelocity}
\end{equation}
which is a minimally-resolved representation of the \emph{no-slip
constraint} \cite{DirectForcing_Balboa,ISIBM}. Furthermore, the applied
forces $\V F$ affect the motion of the fluid through the addition
of a continuous force density to the fluid equation (\ref{eq:ContFluid}),
\begin{equation}
\V f=\S(\V q)\V F+\V f_{\text{th}},\label{eq:SpreadForce}
\end{equation}
where $\V f_{\text{th}}$ is a thermal or stochastic forcing that
we discuss shortly. It is crucial for energy conservation and fluctuation-dissipation
balance that the coupling operators are adjoints of one another \cite{SELM,ForceCoupling_Monopole,ISIBM},
$\J=\S^{\star}$, as follows from
\begin{equation}
\sum_{i}\left(\mathcal{\J}\V v\right)_{i}\cdot\V u_{i}=\int\V v\cdot\left(\S\V u\right)d\V r=\int\sum_{i}\delta_{a}\left(\V q_{i}-\V r\right)\left(\V v\cdot\V u_{i}\right)d\V r\label{eq:adjoint_cond}
\end{equation}

In this work we focus on suspensions of spherical particles (blobs),
for which the kernel function $\delta_{a}\left(\V r\right)\equiv\delta_{a}\left(r\right)$
should be taken to be a spherically-symmetric function of width $\sim a$.
In our computational algorithm we employ the compact-support kernels
of Peskin \cite{IBM_PeskinReview}, which are of the tensor product
form $\delta_{a}\left(\V r\right)=\prod_{\alpha=1}^{d}\delta_{a}\left(r_{\alpha}\right)$
and are specifically designed to work well in the discrete context,
as discussed further in Section \ref{sec:Discretization}. Note that
a Gaussian kernel, as used in FCM \cite{ForceCoupling_Monopole,ForceCoupling_Stokes},
has the special property that it is of the tensor product form while
also being isotropic. It should be noted, however, that much more
general forms of the local interpolation and spreading operators are
possible \cite{SELM}; this has been successfully used to generalize
FCM to non-spherical particles \cite{ForceCoupling_Ellisoids} and
can also be used to further extend our FIB method. The local averaging
and spreading operators have to be modified near physical boundaries,
specifically, when the support of the kernel $\delta_{a}$ overlaps
with a boundary. A proposal for how to do that has been developed
by Yeo and Maxey \cite{ForceCoupling_Channel}, and we have found
it to be superior to an alternative proposal developed in the context
of the immersed boundary method in Ref. \cite{IBMDelta_Boundary}.
In practice a a repulsive potential is imposed between the boundaries
and the particles, which may be sufficient to keep the kernels from
overlapping the walls.

In order to ensure that the system of equations (\ref{eq:ContFluid},\ref{eq:ParticleVelocity},\ref{eq:SpreadForce})
obeys fluctuation-dissipation (i.e., that the dynamics is time reversible
with respect to an appropriate Gibbs-Boltzmann distribution), the
thermal forcing
\begin{equation}
\V f_{\text{th}}=\left(k_{B}T\right)\partial_{\V q}\cdot\S\label{eq:f_th}
\end{equation}
should be included in the fluid equations, as derived by Atzberger
\cite{SELM} and also discussed from a different perspective in Appendix
B in Ref. \cite{ISIBM} and Ref. \cite{CompressibleBlobs}. Here we
use the convention that the contraction in the divergence of an operator
is on the second index, $f_{i}^{\text{th}}=k_{B}T\:\partial_{j}\mathcal{S}_{ij}$,
consistent with Ref. \cite{SELM} but not with Ref. \cite{ISIBM};
to avoid confusion we will write things out in indicial notation when
necessary%
\footnote{Note that the mobility is symmetric so there is no notational ambiguity
in $\partial_{\V q}\cdot\Mob$.%
}. For a translationally-invariant (e.g., periodic) system and kernel,
this term can be omitted. Namely, from the definition (\ref{eq:SpreadOp})
it follows that $\V f_{\text{th}}=-\left(k_{B}T\right)\,\grad_{\V r}\sum_{i}\delta\left(\V q_{i}-\V r\right)$,
and the solution of the incompressible velocity equations is not affected
by the addition of a gradient of a scalar. This is not strictly true
in the discrete setting (see Section \ref{sub:PeriodicPDF}) and may
not generalize to confined (i.e., not translationally-invariant) systems
for particles in the vicinity of boundaries. The term (\ref{eq:f_th})
is therefore, in general, required in order to obtain discrete fluctuation-dissipation
balance and is included in our temporal integrator.

\subsection{\label{sub:OverdampedLimit}Overdamped Limit}

Equations (\ref{eq:ContFluid}), (\ref{eq:ParticleVelocity}), and
(\ref{eq:SpreadForce}) together constitute a physically-realistic
description which obeys fluctuation-dissipation balance \cite{SELM},
including in the presence of additional particle inertia \cite{ISIBM,SELM_Reduction}.
Here we are interested in the inertia-less or overdamped limit, where
the momentum of the fluid may be eliminated as a fast variable. More
precisely, we assume that the Schmidt number is very large, $\text{Sc}=\eta/\left(\rho\chi\right)\gg1$,
where $\chi\approx k_{B}T/\left(6\pi\eta a\right)$ is a typical value
of the diffusion coefficient of the particles \cite{StokesEinstein}.

Following the notation developed in \cite{DFDB}, here we use $\mathcal{\M{\mathcal{Z}}}\left(\V r,t\right)$
to denote an infinite-dimensional standard white-noise field, use
$\M{\mathcal{W}}\left(t\right)$ to denote a finite dimensional collection
of standard white noise processes that represents a spatial discretization
of $\mathcal{\M{\mathcal{Z}}}\left(\V r,t\right)$, and use $\V W$
to denote a collection of standard (mean zero and unit variance) Gaussian
variates that appears when $\M{\mathcal{W}}\left(t\right)$ is discretized
in time. For notational clarity, and to emphasize that we also consider
spatially-discretized operators in the following calculations, we
introduce symbols for the various differential operators: $\M D$
for the divergence, $\M G=-\M D^{\star}$ for the gradient, $\M L$
for the vector Laplacian, and $\widetilde{\M D}$ for the divergence
operator acting on the stochastic tensor. In the infinite-dimensional
(continuum) setting these are differential operators, while in the
finite dimensional (discrete) setting they are matrices that approximate
the corresponding differential operators (for example, using finite
differences), taking into account the boundary conditions \cite{LLNS_Staggered}.
Note that the operator $\widetilde{\M D}$ does not have to be a consistent
representation of the tensor divergence, rather, all that matters
is that the covariance of the stochastic fluid forcing $\widetilde{\M D}\mathcal{\M{\mathcal{W}}}$
obey the fluctuation-dissipation property $\M L=-\widetilde{\M D}\av{\mathcal{\M{\mathcal{W}}}\mathcal{\M{\mathcal{W}}}^{\star}}\widetilde{\M D}^{\star}$
\cite{LLNS_Staggered,SELM_FEM}. For notational simplicity, here we
assume that the components of $\M{\mathcal{W}}$ are independent,
$\av{\mathcal{\M{\mathcal{W}}}\mathcal{\M{\mathcal{W}}}^{\star}}=\M I$,
with the understanding that some modifications of either the covariance
of $\mathcal{\M{\mathcal{W}}}$, or, equivalently, the operator $\widetilde{\M D}$,
may be necessary near boundaries to preserve fluctuation-dissipation
balance for confined systems \cite{LLNS_Staggered}.

To obtain the asymptotic dynamics in the limit $\text{Sc}\rightarrow\infty$
heuristically, we delete the inertial term $\rho\partial_{t}\bv$
in (\ref{eq:ContFluid}) to obtain the (potentially discretized) fluctuating
steady Stokes equations for the velocity $\bv$ and the pressure $\pi$,
\begin{eqnarray}
\M G\pi-\eta\M L\bv & = & \V g=\S\V F+\sqrt{2k_{B}T\eta}\;\widetilde{\M D}\mathcal{\M{\mathcal{W}}}\label{eq:stokesFluid}\\
\M D\bv & = & 0,\nonumber 
\end{eqnarray}
with appropriate boundary conditions. For periodic systems we additionally
constrain the average velocity $\av{\bv}=0$ to eliminate the non-trivial
nullspace. In the following we will denote with $\L^{-1}$ the (continuum
or discrete) Stokes solution operator for the system (\ref{eq:stokesFluid})
with unit viscosity, $\V v=\eta^{-1}\L^{-1}\V g$. Note that $\L^{-1}\succeq\M 0$
is SPD because the Stokes problem (\ref{eq:stokesFluid}) is symmetric
by virtue of the adjoint relation $\M G=-\M D^{\star}$ and the Laplacian
operator $\M L$ is symmetric negative semi-definite.

In the overdamped regime, the (fast) fluid velocity evolves instantaneously
to its steady state and may be viewed as a random function of the
particles' positions, which are the relevant (slow) variables. Heuristically,
one expects that the Brownian dynamics of the particles is described
by $d\V q/dt=\V v=\eta^{-1}\mathcal{\J}\L^{-1}\V g$. A rigorous adiabatic
mode elimination procedure \cite{AdiabaticElimination_1,AveragingHomogenization}
informs us that the correct interpretation of the noise term in this
equation is the kinetic one, leading to the overdamped Langevin equation
\begin{equation}
\frac{d\V q\left(t\right)}{dt}=\mathcal{\J}(\V q)\mathcal{\L}^{-1}\left[\frac{1}{\eta}\S(\V q)\V F(\V q)+\sqrt{\frac{2k_{B}T}{\eta}}\;\widetilde{\M D}\diamond\mathcal{\M{\mathcal{W}}}(t)\right].\label{eq:OverdampedEq}
\end{equation}
This is the rigorous asymptotic limit of (\ref{eq:ContFluid},\ref{eq:ParticleVelocity},\ref{eq:SpreadForce})
as $\text{Sc}\rightarrow\infty$ \cite{StokesLaw} and it is the equation
of motion in the FIB method.

\subsection{\label{sub:FiniteSize}Relation to Brownian Dynamics}

A key observation is that (\ref{eq:OverdampedEq}) is a specific instance
of the equation of Brownian dynamics (\ref{eq:BDKinetic}), with the
identification
\begin{equation}
\Mob=\eta^{-1}\J\L^{-1}\S\quad\mbox{and}\quad\Mob^{\frac{1}{2}}=\eta^{-\frac{1}{2}}\J\L^{-1}\widetilde{\M D}.\label{eq:Mobility_FIB}
\end{equation}
To demonstrate that this choice satisfies the fluctuation dissipation
balance condition (\ref{eq:BDFluctDissBalance}), note the adjoint
relations $\J=\S^{\star}$ and $\M L=-\widetilde{\M D}\widetilde{\M D}^{\star}$.
It is important to point out that the spatially-discretized operators
we employ obey these properties even in the presence of nontrivial
boundary conditions \cite{LLNS_Staggered}. Observe also that 
\begin{equation}
-\L^{-1}\M L\L^{-1}=\L^{-1}\label{eq:LLL}
\end{equation}
as seen from their action on an arbitrary vector $\V g$,
\begin{eqnarray*}
-\L^{-1}\M L\L^{-1}\V g & = & -\L^{-1}\eta\M L\bv=\\
\L^{-1}\left(-\M G\pi+\V g\:\right) & = & \L^{-1}\V g,
\end{eqnarray*}
where we used the fact that $\L^{-1}\M G=\V 0$ since adding a gradient
forcing to the Stokes equations does not affect the velocity. This
gives
\begin{eqnarray}
\Mob^{\frac{1}{2}}\left(\Mob^{\frac{1}{2}}\right)^{\star} & = & \eta^{-1}\J\mathcal{\L}^{-1}\left(\widetilde{\M D}\widetilde{\M D}^{\star}\right)\L^{-1}\S=\nonumber \\
-\eta^{-1}\J\left(\mathcal{\L}^{-1}\M L\L^{-1}\right)\S & = & \eta^{-1}\J\L^{-1}\S=\Mob.\label{eq:ContinumFluctDiss}
\end{eqnarray}
Also note that the mobility (\ref{eq:Mobility_FIB}) is guaranteed
to be positive-semidefinite by virtue of (\ref{eq:BDFluctDissBalance}).

More explicitly, (\ref{eq:Mobility_FIB}) gives a pairwise mobility%
\footnote{Note that this is an approximation and in practice the mobility is
not pairwise additive if higher-order multipoles such as stresslets
are accounted for, see Appendix \ref{sec:Stresslets}.%
} that only depends on the position of the pair of particles under
consideration \cite{SIBM_Brownian},
\begin{equation}
\Mob_{ij}=\eta^{-1}\int\delta_{a}(\V q_{i}-\V r)\M K(\V r,\V r^{\prime})\delta_{a}(\V q_{j}-\V r^{\prime})\ d\V rd\V r^{\prime}\label{eq:GreensMobility}
\end{equation}
where we recall that $\M K$ is the Green's function for the Stokes
problem with unit viscosity and the specified boundary conditions.
Note that in our approach the self-mobility $\Mob_{ii}$ is also given
by the same formula (\ref{eq:GreensMobility}) with $i=j$ and does
not need to be treated separately. In fact, the self-mobility of a
particle in an unbounded three-dimensional domain \emph{defines} the
effective hydrodynamic radius $a$ of a blob,
\begin{eqnarray*}
 &  & \Mob_{ii}=\Mob_{\text{self}}=\frac{1}{6\pi\eta a}\M I=\\
 &  & \eta^{-1}\int\delta_{a}(\V q_{i}-\V r)\M O(\V r-\V r^{\prime})\delta_{a}(\V q_{i}-\V r^{\prime})\ d\V rd\V r^{\prime}.
\end{eqnarray*}
The value of $a$ will therefore depend on the specific kernel used,
as discussed further in section \ref{sub:DiscreteSJ}. In two dimensions,
the self-mobility $\Mob_{\text{self}}=\mu\M I$ of a disk of radius
$a$ in a periodic domain (equivalently, a periodic array of infinite
cylinders) grows logarithmically with the length of the square periodic
cell $L$ as $\mu=\left(4\pi\eta\right)^{-1}\ln\left(L/3.708a\right)$
\cite{Mobility2D_Hasimoto}. The same scaling with the system size
holds for a blob and can be used to define an effective hydrodynamic
radius for a two-dimensional blob \cite{StokesEinstein}. Note that
in two dimensions the mobility diverges for an infinite domain, in
agreement with Stokes's paradox.

Maxey \cite{ForceCoupling_Monopole} observed that (\ref{eq:GreensMobility})
consistently includes the Faxen correction to the mobility of two
well-separated particles. Let $\M J\left(\V q_{i}\right)$ denote
the local averaging operator for a particle $i$, $\M J\left(\V q_{i}\right)\V v\equiv\left(\J\left(\V q\right)\V v\right)_{i}$.
For a smooth velocity field, we can perform a second order Taylor
expansion of the velocity field,
\begin{eqnarray*}
\M J\left(\V q_{i}\right)\V v\left(\V r\right) & = & \int\delta_{a}(\V q_{i}-\V r)\bv(\V r)d\V r\\
 & \approx & \left[\M I+\left(\int\frac{x^{2}}{2}\delta_{a}\left(x\right)dx\right)\grad^{2}\right]\V v\left(\V r\right)\big|_{\V r=\V q_{i}}\\
 & = & \left(\M I+\frac{a_{F}^{2}}{6}\grad^{2}\right)\V v\left(\V r\right)\big|_{\V r=\V q_{i}},
\end{eqnarray*}
where we assumed a spherical blob, $\delta_{a}\left(\V r\right)\equiv\delta_{a}\left(r\right)$.
This shows that we can approximate the local averaging operator by
a differential operator that is identical in form to the Faxen operator
appearing in (\ref{eq:MobilityFaxen}), if we define the ``Faxen''
radius of the blob $a_{F}\equiv\left(3\int x^{2}\delta_{a}(x)\, dx\right)^{1/2}$
through the second moment of the kernel function. In general, $a_{F}\neq a$,
but for a suitable choice of the kernel one can accomplish $a_{F}\approx a$
and thus accurately obtain the Faxen correction for a rigid sphere
(for example, for a Gaussian $a/a_{F}=\sqrt{3/\pi}$ \cite{ForceCoupling_Monopole}).
Interestingly, it has been shown that the leading-order Faxen corrections
to the linear and angular velocities of an ellipsoidal particle can
also be captured remarkably accurately (to within $5\%$) by using
a stretched and rotated Gaussian for the kernel function \cite{ForceCoupling_Ellisoids}.

The calculations above show that the mobility tensor for a pair of
blobs (\ref{eq:GreensMobility}) is a good approximation to (\ref{eq:MobilityFaxen})
for well-separated blobs and thus correctly captures the mobility
up to the Rotne-Prager level even in the presence of confinement.
This can also be seen from (\ref{eq:GreensMobility}) by noting that
when the two particles are well separated, $\M K$ is a smooth function,
and is well approximated by a Taylor series, giving
\[
\Mob_{ij}\approx\eta^{-1}\left(\M I+\frac{a_{F}^{2}}{6}\grad_{\V r}^{2}\right)\left(\M I+\frac{a_{F}^{2}}{6}\grad_{\V r^{\prime}}^{2}\right)\M K(\V r-\V r^{\prime})\big|_{\V r^{\prime}=\V q_{i}}^{\V r=\V q_{j}},
\]
which matches the expression (\ref{eq:MobilityFaxen}) for well-separated
rigid spheres. At smaller distances the mobility is mollified (regularized)
in a natural way without requiring any special handling of the case
$r_{ij}<2a$ as in the traditional RPY tensor (\ref{eq:RPYTensor}).
Furthermore, a positive definite mobility tensor is obtained by construction.
Most importantly, the same continues to hold in the presence of confinement
(nontrivial boundary conditions). The boundary conditions are taken
into account by the fluid solver when computing the action of the
Green's function (\ref{eq:split_Greens}), while the regularization
and the Faxen corrections are handled via the local averaging and
spreading operators. This inherent self-consistency of the formulation
is inherited from the underlying fluctuating hydrodynamics formulation
(\ref{eq:ContFluid},\ref{eq:ParticleVelocity},\ref{eq:SpreadForce})
\cite{ISIBM}.

\subsection{Thermal Drift}

One key difference between the inertial formulation (\ref{eq:ContFluid},\ref{eq:ParticleVelocity},\ref{eq:SpreadForce})
and the overdamped limit (\ref{eq:OverdampedEq}) is the fact that
the noise in (\ref{eq:OverdampedEq}) is multiplicative and therefore
the stochastic interpretation matters and affects the temporal discretization.
Methods for integrating (\ref{eq:BDKinetic}) have been developed
in the Brownian Dynamics literature, however, here we propose a more
efficient approach which we term Random Finite Difference (RFD). We
believe this approach will find uses in Brownian Dynamics simulations
as well as related methods for fluctuating hydrodynamics \cite{ForceCoupling_Fluctuations,SELM_FEM}.
We therefore explain it here in the more general setting of solving
(\ref{eq:BDKinetic}), of which (\ref{eq:OverdampedEq}) is a special
instance. A detailed description of predictor-corrector schemes to
solve (\ref{eq:OverdampedEq}) is given in Section \ref{sec:TemporalIntegrator}. 

Of course, one can use the Ito equation (\ref{eq:BrownianDynamics})
with integrators based on the Euler-Maruyama scheme. This, however,
requires computing the stochastic drift term $k_{B}T\,\left(\partial_{\V q}\cdot\Mob\right)$,
which is difficult in general. First, we summarize the well-known
Fixman midpoint approach to approximating $\partial_{\V q}\cdot\Mob\left(\V q\right)$,
and use it to construct an RFD approach that works better in the context
of our explicit fluid method. Below we use the superscript to denote
the time step level at which quantities are evaluated, for example,
$\Mob^{n}\equiv\Mob\left(\V q^{n}\right)$ denotes the mobilitity
evaluated at the beginning of time step $n$, while $\Mob^{n+\frac{1}{2}}\equiv\Mob\left(\V q^{n+\frac{1}{2}}\right)$
denotes a midpoint approximation of the mobility during time step
$n$.

\subsubsection{Fixman's Method}

The Fixman midpoint scheme used to capture the thermal drift \cite{BD_Fixman,BD_Hinch}
can be seen as corresponding to a direct discretization of the kinetic
stochastic integral \cite{KineticStochasticIntegral_Ottinger},
\begin{eqnarray}
\V q^{n+\frac{1}{2}} & = & \V q^{n}+\frac{\D t}{2}\Mob^{n}\V F^{\, n}\nonumber \\
 &  & +\sqrt{\frac{\D t\, k_{B}T}{2}}\left(\Mob^{n}\right)^{\frac{1}{2}}\M W^{n}\nonumber \\
\V q^{n+1} & = & \V q^{n}+\D t\Mob^{n+\frac{1}{2}}\V F^{\, n+\frac{1}{2}}\nonumber \\
 &  & +\sqrt{2\D t\, k_{B}T}\Mob^{n+\frac{1}{2}}\left(\Mob^{n}\right)^{-\frac{1}{2}}\M W^{n},\label{eq:FixmanPC}
\end{eqnarray}
where $\M W^{n}$ is a vector of i.i.d. standard Gaussian variables
and
\begin{eqnarray*}
\left(\Mob^{n}\right)^{-\frac{1}{2}}\left(\left(\Mob^{n}\right)^{-\frac{1}{2}}\right)^{\star} & = & \left(\Mob^{n}\right)^{-1}.
\end{eqnarray*}
While the Fixman method is quite elegant and has been widely used
with notable success, it requires handling the inverse of the mobility
matrix, which would add significant complication to our method \cite{ForceCoupling_Fluctuations}.

In order to show that (\ref{eq:FixmanPC}) is consistent with (\ref{eq:BDKinetic})
one has to show that the first and second moments of the increment
$\V q^{n+1}-\V q^{n}$ are $O\left(\D t\right)$ with coefficients
matching the drift and diffusion terms in the Ito equation (\ref{eq:OverdampedEq}),
and higher moments should be of higher order in $\D t$. The only
nontrivial component is the stochastic drift term $k_{B}T\,\partial_{\V q}\cdot\Mob\left(\V q\right)$.
In order to compact the notation, henceforth we will index matrices
and vectors without regard for the physical particles represented.
For example, we will write $q_{i}$ to represent the scalar that is
the $i$th entry of the length $nd$ vector of positions $\V q$,
disregarding which particle this entry describes. We will likewise
consider the mobility $\Mob$ as a matrix of scalars $\mathcal{M}_{ij}$.
This allows us to use Einstein summation notation and indicial algebra.
We can show that the Fixman algorithm (\ref{eq:FixmanPC}) generates
the correct stochastic drift term from
\begin{eqnarray}
 &  & \lim_{\D t\rightarrow0}\frac{1}{\D t}\Biggl<\mathcal{M}_{ij}\left(q_{p}^{n}+\sqrt{\frac{\D t\, k_{B}T}{2}}\left(\mathcal{M}_{pr}^{n}\right)^{\frac{1}{2}}W_{r}^{n}\right)\times\nonumber \\
 &  & \sqrt{2\D t\, k_{B}T}\left(\mathcal{M}_{jk}^{n}\right)^{-\frac{1}{2}}W_{k}^{n}\Biggr>=k_{B}T\,\partial_{j}\mathcal{M}_{ij}\left(\V q^{n}\right),\label{eq:FixmanDrift}
\end{eqnarray}
where the average is over realizations of $\V W$ and the shorthand
$\partial_{j}$ denotes a partial derivative with respect to the $j$-th
component of $\V q$.

\subsubsection{\label{sub:RFD}Random Finite Difference}

The equivalence (\ref{eq:FixmanDrift}) only relies on the covariance
structure of $\M W^{n}$, and there is no reason that we must use
an increment that is related in any way to the noise term in (\ref{eq:BDKinetic}).
More generally, we can obtain a divergence of the mobility in expectation
from the general relation, 
\begin{equation}
\lim_{\epsilon\rightarrow0}\frac{1}{\epsilon}\av{\Mob\left(\V q+\epsilon\D{\V q}\right)\D{\V p}-\Mob\left(\V q\right)\D{\V p}}=\partial_{\V q}\cdot\Mob(\V q),\label{eq:DriftWithIncrements}
\end{equation}
where $\D{\V q}$ and $\D{\V p}$ are Gaussian variates with mean
zero and covariance $\av{\D{\V q}_{i}\D{\V p}_{j}}=\delta_{ij}$.
In particular, the choice $\D{\V q}=\D{\V p}$ is much simpler to
use than the Fixman method choice $\D{\V q}\sim\left(\Mob^{n}\right)^{\frac{1}{2}}\M W^{n}$
and $\D{\V p}\sim\left(\Mob^{n}\right)^{-\frac{1}{2}}\M W^{n}$. Here
$\epsilon$ is a small discretization parameter that can be taken
to be related to $\D t$ as in the Fixman method, but this is not
necessary. One can more appropriately think of (\ref{eq:DriftWithIncrements})
as a ``random finite difference'' (RFD) with $\epsilon$ representing
the small spacing for the finite difference, to be taken as small
as possible while avoiding numerical roundoff problems. The advantage
of the ``random'' over a traditional finite difference is that only
a small number of evaluations of the mobility per time step is required.
Note that the subtraction of $\Mob\left(\V q\right)\D{\V p}$ in (\ref{eq:DriftWithIncrements})
is necessary in order to control the variance of the RFD estimate.
One can use a centered difference to improve the truncation error
and obtain the correct thermal drift via the RFD
\begin{eqnarray}
 &  & \frac{1}{\delta}\av{\left(\mathcal{M}_{ij}\left(q_{k}^{n}+\frac{\delta}{2}\widetilde{W}_{k}\right)\widetilde{W}_{j}-\mathcal{M}_{ij}\left(q_{k}^{n}-\frac{\delta}{2}\widetilde{W}_{k}\right)\widetilde{W}_{j}\right)}\nonumber \\
 &  & =\partial_{j}\mathcal{M}_{ij}\left(\V q^{n}\right)+O(\delta^{2}),\label{eq:RFDMobility}
\end{eqnarray}
where $\widetilde{\M W}$ is a vector of $dn$ i.i.d. standard Gaussian
random variables and $\delta$ is a small parameter. 

While expression (\ref{eq:RFDMobility}) could be used to approximate
the drift term and may be a useful alternative to the Fixman scheme
in related methods such as the fluctuating FCM \cite{ForceCoupling_Fluctuations},
using an RFD of the form (\ref{eq:RFDMobility}) requires at least
one more Stokes solve per time step in order to evaluate the action
of $\Mob\left(\V q+\epsilon\D{\V q}\right)$. It is, however, possible
to avoid the second Stokes solve by splitting the divergence of the
mobility into two pieces,
\[
\eta\,\partial_{\V q}\cdot\Mob=\partial_{\V q}\cdot\left(\J\L^{-1}\S\right)=\left(\partial_{\V q}\M J\right):\left(\L^{-1}\S\right)+\J\L^{-1}\left(\partial_{\V q}\cdot\S\right),
\]
where colon denotes a double contraction, see (\ref{eq:2termdrift}).
We approximate the first term involving the gradient $\partial_{\V q}\M J$
using a standard two-stage Runge-Kutta (predictor-corrector) approach,
and use an RFD to approximate $\partial_{\V q}\cdot\S$, as explained
in detail in Section \ref{sec:TemporalIntegrator}.

\section{\label{sec:Discretization}Spatial Discretization}

In this section we describe our spatial discretization of (\ref{eq:OverdampedEq}),
which is constructed from components described in extensive detail
in prior work by some of us; here we only briefly summarize the key
points. The finite-volume solver used here to solve the fluctuating
Stokes equations in confined domains is taken from Ref. \cite{LLNS_Staggered},
while the discretization of the fluid-particle interaction operators
is based on the immersed-boundary method \cite{IBM_PeskinReview}
and is described in extensive detail in Ref. \cite{ISIBM}. The key
novel component here is the use of a steady Stokes fluid solver to
generate a fluctuating velocity, as also done in Refs. \cite{FIMAT_Patankar,ForceCoupling_Fluctuations,SELM_FEM}
using different techniques.

We discretize the fluid equation (\ref{eq:ContFluid}) using a standard
staggered ``marker and cell'' (MAC) grid with uniform mesh width
$h$ in a rectangular domain with an arbitrary combination of periodic,
no-slip, or free-slip boundaries. The differential operators $\M D$,
$\M G$, and $\M L$ are discretized on the staggered grid using standard
second order centered differences. The stochastic stress tensor $\mathcal{\M{\mathcal{Z}}}\left(\V r,t\right)$
is discretized as $\D V^{-\frac{1}{2}}\mathcal{\M{\mathcal{W}}}\left(t\right)$,
where the additional factor of $\D V^{-\frac{1}{2}}$ comes from the
fact that $\mathcal{\M{\mathcal{Z}}}$ is white in space \cite{DFDB}.
Adjustments to the stochastic increments are made near boundaries
to preserve the fluctuation-dissipation relation $-\widetilde{\M D}\widetilde{\M D}^{T}=\M L$
(more precisely, to ensure that $\M L=-\widetilde{\M D}\av{\mathcal{\M{\mathcal{W}}}\mathcal{\M{\mathcal{W}}}^{T}}\widetilde{\M D}^{T}$)
\cite{LLNS_Staggered}.

\subsection{\label{sub:DiscreteSJ}Discrete Local Averaging and Spreading}

The discrete operator (matrix) $\J$ averages velocities on the staggered
mesh by discretizing the integral $\int\delta_{a}(\V q_{i}-\V r)\bv(\V r)d\V r$
using a simple quadrature 
\[
\left(\mathcal{\J}\bv\right)_{i}^{\alpha}=\sum_{k}\delta_{a}\left(\V q_{i}-\V r_{k}^{\alpha}\right)v_{k}^{\alpha}\D V,
\]
where the sum is taken over faces $k$ of the grid, and $\D V$ is
the volume of a grid cell. Here $\alpha$ indexes coordinate directions
($x,y,z$) as a superscript, $\V r_{k}^{\alpha}$ is the center of
the grid face $k$ in the direction $\alpha$, and $v_{k}^{\alpha}\equiv v^{(\alpha)}\left(\V r_{k}\right)$
is the staggered velocity field. Likewise, $\S$ spreads forces to
the staggered grid, and its expression remains identical to (\ref{eq:InterpOp}),
but is evaluated only at faces of the staggered grid normal to the
component of force being spread,
\[
\left(\S\V F\right)_{k}^{\alpha}=\sum_{i}F_{i}^{\alpha}\;\delta_{a}\left(\V q_{i}-\V r_{k}^{\alpha}\right),
\]
where now the sum is over the particles. Near no-slip physical boundaries,
the discrete delta function is modified following the image-monopole
construction proposed by Yeo and Maxey, see (2.17) in Ref. \cite{ForceCoupling_Channel};
this is found to be superior to the modification proposed in Ref.
\cite{IBMDelta_Boundary}.

For a uniform grid, the matrices representing the discrete local averaging
and spreading operators are scaled transposes of each other, $\J^{T}=\D V\,\S$.
Note that these discrete operators are adjoints like their continuum
counterparts, $\J=\S^{\star}$, but in an inner product that includes
an appropriate weighting \cite{SELM} because the integral over the
domain in (\ref{eq:adjoint_cond}) is replaced by a sum over grid
points $k$,
\begin{eqnarray}
\sum_{i}\left(\mathcal{\J}\V v\right)_{i}\cdot\V F_{i} & = & \sum_{k,\alpha}v_{k}^{\alpha}\left(\S\V F\right)_{k}^{\alpha}\D V\nonumber \\
 & = & \sum_{i,k,\alpha}\delta_{a}\left(\V q_{i}-\V r_{k}^{\alpha}\right)v_{k}^{\alpha}F_{i}^{\alpha}\D V.\label{eq:adjoint_cond_discr}
\end{eqnarray}
Note that this adjoint relation is strictly preserved even in the
presence of no-slip boundaries.

In the majority of the simulations we use the four-point kernel of
Peskin \cite{IBM_PeskinReview} to discretize the kernel $\delta_{a}$,
although in some cases we employ the three-point discrete kernel function
of Roma and Peskin \cite{StaggeredIBM,DirectForcing_Balboa}. The
effective hydrodynamic radius $a$ for a given discrete kernel function
can be obtained from the self-mobility of a blob in a periodic domain.
For large periodic domains in three dimensions we numerically estimate
the effective hydrodynamic (rigid sphere) radius to be $a=\left(0.91\pm0.01\right)h$
for the three-point kernel \cite{DirectForcing_Balboa,ISIBM}, and
$a=\left(1.255\pm0.005\right)h$ for the four-point kernel \cite{ReactiveBlobs}.
In two dimensions, the effective (rigid disk) hydrodynamic radii are
estimated to be $a=\left(0.72\pm0.01\right)h$ for the three point
and $a=\left(1.04\pm0.005\right)h$ for the four point kernel \cite{StokesEinstein}.
Note that the spatial discretization we use is not perfectly translationally
invariant and there is a small variation of $a$ (quoted above as
an error bar) as the particle moves relative to the underlying fixed
fluid grid \cite{ISIBM,ReactiveBlobs}. By using the Peskin four-point
kernel instead of the three-point discrete kernel function the translational
invariance of the spatial discretization can be improved, however,
at a potentially significant increase in computational cost, particularly
in three dimensions. 

It is important to note that, perhaps unexpectedly, these Peskin kernels
give close agreement between the hydrodynamic and the Faxen radii
of the blob. For example, in three dimensions, the three-point kernel
gives $a_{F}\approx0.93h$ (this number is again not exactly constant
due to the imperfect translational invariance), as compared to $a\approx0.91h$.
Using the four-point kernel gives an even better agreement, with $a\approx a_{F}\approx1.25h$.
In particular, it is important to choose a kernel with a nonzero second
moment in order to capture the Faxen corrections in a physically-realistic
manner; this eliminates the Peskin six-point kernel \cite{IBM_PeskinReview}
from consideration.

\subsection{Stokes Solver}

In the FIB method we obtain the fluid velocity $\bv=\eta^{-1}\L^{-1}\V g$
by numerically solving the discrete steady Stokes equation
\begin{eqnarray}
\M G\pi-\eta\M L\bv & = & \V g=\S\V F+\sqrt{\frac{2k_{B}T\eta}{\D V}}\;\widetilde{\M D}\mathcal{\M{\mathcal{W}}}\label{eq:discreteStokes}\\
\M D\bv & = & 0\nonumber 
\end{eqnarray}
using a preconditioned Krylov iterative solver \cite{NonProjection_Griffith}.
Note that we can explicitly write $\L^{-1}$ using the Schur complement
of (\ref{eq:discreteStokes}), 
\begin{equation}
-\L^{-1}=\M L^{-1}-\M L^{-1}\M G\left(\M D\M L^{-1}\M G\right)^{-1}\M D\M L^{-1}.\label{eq:ExpliclitStokesSolve}
\end{equation}
In the continuum setting, and also in the discrete setting with periodic
boundary conditions, the various operators commute and one can simplify
$\L^{-1}=-\M P\M L^{-1}$, where $\M P=\M I-\M G(\M{DG})^{-1}\M D$
is the $L_{2}$ projection operator onto the subspace of (discretely)
divergence free vector fields. In general, however, for many spatial
discretizations, including the one we use, the operators do \textit{not}
commute and one must keep the full form (\ref{eq:ExpliclitStokesSolve})
\cite{DFDB,LLNS_Staggered}. %

\subsection{\label{sub:DFDB}Discrete Fluctuation Dissipation Balance}

The spatially-discretized equation of motion for the particles has
the same form as the continuum (\ref{eq:OverdampedEq}), and is an
instance of (\ref{eq:BDKinetic}) with the identification 
\begin{eqnarray}
\Mob & = & \eta^{-1}\J\L^{-1}\S\label{eq:Mobility_FIB_discr}\\
\Mob^{\frac{1}{2}} & = & \left(\eta\D V\right)^{-\frac{1}{2}}\J\L^{-1}\widetilde{\M D}.\nonumber 
\end{eqnarray}
Note that the key relation (\ref{eq:LLL}) continues to hold, $-\L^{-1}\M L\L^{-1}=\L^{-1}$,
which follows directly from (\ref{eq:ExpliclitStokesSolve}). This
can be used to show that (\ref{eq:BDFluctDissBalance}) is satisfied
\begin{eqnarray}
 &  & \Mob^{\frac{1}{2}}\left(\Mob^{\frac{1}{2}}\right)^{T}\label{eq:DFDB_monopoles}\\
 & = & -\left(\eta\D V\right)^{-1}\left[\J\mathcal{\L}^{-1}\left(\widetilde{\M D}\widetilde{\M D}^{T}\right)\L^{-1}\left(\D V\,\S\right)\right]\nonumber \\
 & = & -\eta^{-1}\J\left(\mathcal{\L}^{-1}\M L\L^{-1}\right)\S=\eta^{-1}\J\L^{-1}\S=\Mob,\nonumber 
\end{eqnarray}
where we made use of $\J^{T}=\D V\,\S$. Note that these relations
are independent of the boundary conditions and thus (\ref{eq:DFDB_monopoles})
holds in confined systems. %

\section{\label{sec:TemporalIntegrator}Temporal Discretization}

In this section we introduce our approach for temporal integration
of the spatially-discretized equations of motion. A significant challenge
is accurately capturing the thermal drift present in the Ito interpretation,
$\partial_{\V q}\cdot\Mob\left(\V q\right)$, without which the system
would not obey fluctuation-dissipation balance. This requires consistently
discretizing the kinetic integral, which can be done in multiple dimensions
using a Fixman predictor corrector scheme \cite{KineticStochasticIntegral_Ottinger}.
The Fixman scheme, however, requires applying the action of the inverse
of the mobility (or, equivalently, the action of the square root of
the inverse of the mobility), which is a complicating and a potentially
expensive step \cite{ForceCoupling_Fluctuations}. Note that in certain
cases, notably, for translationally-invariant situations such as periodic
systems, the divergence of mobility vanishes and one can use a simple
Euler-Maruyama integrator, as done in the work of Atzberger and collaborators
\cite{SELM}. This is not applicable to confined systems, however,
and here we employ the Random Finite Difference (RFD) approach introduced
in Section \ref{sub:RFD}.

Below we use the superscript $n$ to denote the current time step
and quantities evaluated at the beginning of the current time step,
and superscript $n+1$ for the updated quantities at the end of the
time step. Quantities estimated at the midpoint of the time step are
denoted with superscript $n+\frac{1}{2}$. For example, $\Mob^{n+\frac{1}{2}}\equiv\Mob\left(\V q^{n+\frac{1}{2}}\right)$
denotes a midpoint approximation of the mobility. We develop two temporal
integrators, a first-order simple midpoint method that requires only
a single Stokes solve per time step, and an improved midpoint midpoint
scheme that achieves second-order accuracy in the additive-noise (linearized)
case at the cost of requiring two Stokes solves per time step. Which
scheme allows for better tradeoff between accuracy and efficiency
will depend on the specific problem at hand, and in particular, on
the time step limitations imposed by stability considerations.

\subsection{\label{sub:SimpleMidpoint}Simple midpoint scheme}

A direct application of the RFD approach to integrating (\ref{eq:OverdampedEq})
would require evaluating the action of the mobility at two different
configurations and thus at least two Stokes solves per time step.
In order to avoid using a separate Stokes solver just to obtain the
thermal drift term, we take an alternative approach and split the
thermal drift into two pieces,
\begin{eqnarray}
 &  & \eta\,\partial_{j}\mathcal{M}_{ij}\left(\V q\right)=\partial_{j}\left(\mathcal{J}_{ik}(\V q)\mathcal{\mathcal{L}}_{kl}^{-1}\mathcal{S}_{lj}(\V q)\right)=\nonumber \\
 &  & \left(\partial_{j}\mathcal{\mathcal{J}}_{ik}(\V q)\right)\mathcal{\mathcal{L}}_{kl}^{-1}\mathcal{S}_{lj}(\V q)+\mathcal{\mathcal{J}}_{ik}(\V q)\mathcal{\mathcal{L}}_{kl}^{-1}\left(\partial_{j}\mathcal{S}_{lj}(\V q)\right),\label{eq:2termdrift}
\end{eqnarray}
where we use the implied summation convention. The two pieces can
be handled separately, and only require the derivatives of $\J$ and
$\S$. We approximate the term $\partial_{j}\mathcal{\mathcal{J}}_{ik}(\V q)$
using a predictor-corrector approach in the spirit of Runge-Kutta
algorithms such as the Euler-Heun temporal integrator for Stratonovich
equations \cite{EulerHeun}. We use an RFD of the form (\ref{eq:DriftWithIncrements})
with $\D{\V q}\sim\D{\V p}$ to calculate the term $\partial_{j}\mathcal{S}_{lj}(\V q)$.

Our basic temporal integrator for the spatially-discretized equations
(\ref{eq:OverdampedEq}) consists of first solving the steady Stokes
equations with a random forcing,
\begin{eqnarray}
 &  & -\eta\M L\bv+\M G\pi=\S^{n}\V F^{n}+\sqrt{\frac{2\eta k_{B}T}{\Delta t\D V}}\widetilde{\M D}\M W^{n}\label{eq:TdriftPC}\\
 &  & +\frac{k_{B}T}{\delta}\left[\S\left(\V q^{n}+\frac{\delta}{2}\widetilde{\M W}^{n}\right)-\S\left(\V q^{n}-\frac{\delta}{2}\widetilde{\M W}^{n}\right)\right]\widetilde{\M W}^{n},\nonumber 
\end{eqnarray}
and then advecting the particles with the computed velocity field
using a midpoint predictor-corrector scheme,
\begin{eqnarray}
\V q^{n+\frac{1}{2}} & = & \V q^{n}+\frac{\D t}{2}\J^{n}\bv\label{eq:predictor}\\
\V q^{n+1} & = & \V q^{n}+\D t\J^{n+\frac{1}{2}}\bv.\label{eq:corrector}
\end{eqnarray}
Here $\M W^{n}$ is a random vector of i.i.d. standard Gaussian random
numbers that represent stochastic fluxes of momentum, with $\M W^{n}/\sqrt{\D t}$,
loosely speaking, being a temporal discretization of $\M{\mathcal{W}}(t)$.
The auxiliary displacement $\widetilde{\M W}^{n}$ is a vector of
$nd$ i.i.d. standard Gaussian variates. The parameter $\delta$ should
be as small as possible while still resolving to numerical roundoff
the length scale over which $\S$ varies; we use $\delta\approx10^{-6}h$,
where $h$ is the grid spacing.

The first-order midpoint temporal integrator (\ref{eq:TdriftPC})-(\ref{eq:corrector})
has the advantage that we can recreate the stochastic drift $\partial_{\V q}\cdot\Mob$
by performing only two additional spreading operations and one local
averaging operation per time step, in addition to the required Stokes
solve. We use a midpoint corrector step (\ref{eq:corrector}) because
in the absence of the RFD term it gives the correct diffusion coefficient
for freely-diffusing single particles, \emph{regardless} of the time
step size. Namely, for any choice of $\Delta t$, the second moment
of the stochastic increment of the particle positions is in agreement
with the Einstein formula for the diffusion coefficient, 
\begin{eqnarray}
\text{Var}\left(\V q^{n+1}-\V q^{n}\right) & = & 2\D{t\:}k_{B}T\eta^{-1}\left(\J^{n+\frac{1}{2}}\L^{-1}\S^{n+\frac{1}{2}}\right)\nonumber \\
 & = & 2\D t\, k_{B}T\,\Mob^{n+\frac{1}{2}},\label{eq:var_dq}
\end{eqnarray}
up to correction terms coming from the RFD term in the second line
of (\ref{eq:TdriftPC}). In section \ref{sub:Diffusion-Coefficient},
we confirm that this property continues to hold to very high accuracy
when the RFD is included, even for relatively large $\D t$. Note
that a trapezoidal scheme that replaces the term $\J^{n+\frac{1}{2}}\bv$
in (\ref{eq:corrector}) with $\left(\J^{n}+\J^{n+1}\right)\bv/2$
does not have the property (\ref{eq:var_dq}) and only gives the correct
diffusion coefficient for small $\D t$.

The predictor corrector steps (\ref{eq:predictor})-(\ref{eq:corrector})
reproduce the first term on the right hand side of (\ref{eq:2termdrift}).
The added stochastic force in the Stokes solve generates the thermal
forcing (\ref{eq:f_th}), which appears in the second term on the
right hand side of (\ref{eq:2termdrift}), in expectation to order
$\delta^{2}$, 
\begin{eqnarray}
 &  & \frac{k_{B}T}{\delta}\av{\mathcal{S}_{lj}\left(\V q^{n}+\frac{\delta}{2}\widetilde{\M W}^{n}\right)\widetilde{W}_{j}^{n}-\mathcal{S}_{lj}\left(\V q^{n}-\frac{\delta}{2}\widetilde{\M W}^{n}\right)\widetilde{W}_{j}^{n}}\nonumber \\
 &  & =k_{B}T\left(\partial_{k}\mathcal{S}_{lj}\left(\V q^{n}\right)\right)\av{\widetilde{W}_{k}^{n}\widetilde{W}_{j}^{n}}+O\left(\delta^{2}\right)\nonumber \\
 &  & =k_{B}T\:\partial_{j}\mathcal{S}_{lj}\left(\V q^{n}\right)+O\left(\delta^{2}\right).\label{eq:RFDTerm}
\end{eqnarray}
In Appendix \ref{sec:WeakTemporalAccuracy} we demonstrate that the
simple midpoint scheme (\ref{eq:TdriftPC})-(\ref{eq:corrector})
is a first-order weak integrator for the equations of Brownian dynamics
(\ref{eq:OverdampedEq}).

\subsection{Improved midpoint scheme}

It is possible to obtain second order accuracy in the additive-noise
(linearized) approximation by using an additional Stokes solve in
the corrector stage, as summarized by 
\begin{eqnarray}
 &  & -\eta\M L\bv+\M G\pi=\S^{n}\V F^{n}+\sqrt{\frac{4\eta k_{B}T}{\D t\D V}}\widetilde{\M D}\M W^{n,1}\nonumber \\
 &  & \M D\bv=0\nonumber \\
 &  & \V q^{n+\frac{1}{2}}=\V q^{n}+\frac{\D t}{2}\J^{n}\bv\quad\mbox{(\mbox{predictor})}\nonumber \\
 &  & -\eta\M L\tilde{\bv}+\M G\tilde{\pi}=\S^{n+\frac{1}{2}}\V F^{n+\frac{1}{2}}+\sqrt{\frac{\eta k_{B}T}{\D t\D V}}\widetilde{\M D}\left(\M W^{n,1}+\M W^{n,2}\right)\nonumber \\
 &  & +\frac{k_{B}T}{\delta}\left[\S\left(\V q^{n}+\frac{\delta}{2}\widetilde{\M W}^{n}\right)-\S\left(\V q^{n}-\frac{\delta}{2}\widetilde{\M W}^{n}\right)\right]\widetilde{\M W}^{n}\nonumber \\
 &  & \M D\tilde{\bv}=0\nonumber \\
 &  & \V q^{n+1}=\V q^{n}+\D t\J^{n+\frac{1}{2}}\tilde{\bv}\quad\mbox{(corrector)}.\label{eq:SecondOrderDeterministic}
\end{eqnarray}
Here the independent random variables $\M W^{n,1}$ and $\M W^{n,2}$
represent the two independent Wiener increments over each half of
the time step, as explained in more detail in Ref. \cite{DFDB}. Note
that by using a midpoint corrector step we ensure that the property
(\ref{eq:var_dq}) continues to hold. Here we only include an RFD
term in the corrector step and use the initial position of the particle
in the RFD term. One can also use $\V q^{n+\frac{1}{2}}$ instead
of $\V q^{n}$ but this gains no additional accuracy.

Note that the scheme (\ref{eq:SecondOrderDeterministic}) is still
only first order weakly accurate (see Appendix \ref{sec:WeakTemporalAccuracy})
because the noise in (\ref{eq:OverdampedEq}) is multiplicative. Achieving
second-order weak accuracy in the nonlinear case requires more sophisticated
stochastic Runge-Kutta schemes \cite{WeakSecondOrder_RK}. However,
we will demonstrate in Sec. \ref{sub:ShearFlow} that the improved
midpoint scheme can sometimes give results which are significantly
more accurate because the scheme (\ref{eq:SecondOrderDeterministic})
can be shown to be second order weakly accurate for the linearized
(additive-noise) equations of Brownian dynamics \cite{DFDB}. The
improved midpoint scheme may also give improved stability in certain
cases, as we observe numerically in Section \ref{sub:ColloidalGellation}.
Note, however, that both midpoint schemes are explicit and are thus
subject to stability limits on $\D t$, dictated by the stiffness
of the applied forces $\V F\left(\V q\right)$.

\section{\label{sec:NumericalTests}Results}

In this section we test the performance of the FIB by simulating a
number of scenarios of increasing complexity. We start by confirming
that our spatial discretization gives a mobility in agreement with
known results for a single particle in a slit channel. We then confirm
that our temporal integrators preserve the correct Gibbs-Boltzmann
distribution for both single and multiparticle systems. After also
verifying that the FIB method correctly reproduces the dynamical correlations
between particles in the presence of shear flow and hydrodynamic interactions,
we compare our method to standard Brownian Dynamics on the nonequilibrium
dynamics of a colloidal cluster. Unless otherwise mentioned, the tests
were conducted using the simple midpoint temporal integrator (\ref{eq:TdriftPC})-(\ref{eq:corrector}).

We have implemented the FIB algorithm in the open source code IBAMR
\cite{IBAMR}, a parallel implementation of the immersed boundary
method. The state-of-the-art multigrid-based iterative Stokes solvers
\cite{NonProjection_Griffith} implemented in IBAMR enable us to efficiently
solve the steady Stokes equations for any combination of periodic,
no-slip or free-slip boundaries on the side of a rectangular domain,
including in the presence of thermal fluctuations \cite{LLNS_Staggered}.
Although IBAMR supports adaptive mesh refinement (AMR) for deterministic
time-dependent problems, at present only uniform grids are supported
for steady-state flows with fluctuations. Unless otherwise specified,
the simulations reported here were performed using the IBAMR implementation
of the FIB method.

For periodic domains, no iterative solvers are necessary for uniform
grids since the discrete Fourier transform diagonalizes the discrete
Stokes equations and the Fast Fourier Transform (FFT) can be used
to solve the steady Stokes equations very efficiently. This was used
by some of us to solve the inertial fluid-particle equations efficiently
on Graphical Processing Units (GPUs), as implemented in the open-source
\emph{fluam} CUDA code \cite{ISIBM}. Implementing the FIB method
in \emph{fluam} amounted to simply changing the temporal integration
scheme (for both the fluid and the particle dynamics) to the midpoint
scheme (\ref{eq:TdriftPC})-(\ref{eq:corrector}), while reusing the
core numerical implementation. Note that we only use FFTs as a linear
solver for the discrete Stokes equations, similar to what is done
in SIBM \cite{StochasticImmersedBoundary}. This means that the IBAMR
and\emph{ fluam} codes give the same results for periodic systems
to within solver tolerances. For periodic systems at zero Reynolds
number flow a much higher (spectral) spatial accuracy can be accomplished
by using a Fourier representation of the velocity and pressure, as
done by Keaveny \cite{ForceCoupling_Fluctuations}. In fact, with
proper care in choosing the number of Fourier modes kept and the help
of the non-uniform FFT algorithm \cite{NUFFT} one can construct a
spatial discretization where the truncation error is at the level
of roundoff tolerance \cite{DiffusionJSTAT}. In the presence of simple
confinement such as a slit channel with only two walls, a Fourier
representation can be used in the directions parallel to the channel
walls, along with a different basis for the direction perpendicular
to the walls. Here we do not explore such specialized geometries and
use a finite-volume Stokes solver to handle more general combinations
of boundary conditions.

While the different tests performed have different relevant timescales,
there is an important common timescale of diffusion given by the typical
time it takes a free particle to diffuse a distance $h$, where $h$
is the grid spacing. The typical value of the diffusion coefficient
of a single spherical particle in a translationally-invariant system
can be obtained from the mobility $\mu$ via the Stokes-Einstein relation,
$\M{\chi}_{\text{self}}=k_{B}T\Mob_{\text{self}}=k_{B}T\mu\M I=\chi\M I$,
and leads to $\chi\approx k_{B}T/(6\pi\eta a)$ in three dimensions,
and $\chi\approx k_{B}T\,\left(4\pi\eta\right)^{-1}\ln\left(L/3.708a\right)$
in two dimensions \cite{StokesEinstein}, where we recall that $a$
is the effective hydrodynamic radius of a blob and $L$ is the length
of the periodic domain. In three dimensions there are well-known finite
size corrections to the mobility that are taken into account in the
calculations below \cite{Mobility2D_Hasimoto,LB_SoftMatter_Review,ISIBM,ReactiveBlobs}.
Based on the estimated diffusion coefficient we can define a dimensionless
time step size through the diffusive Courant number 
\[
\beta=\frac{2\chi}{h^{2}}\Delta t.
\]
This dimensionless number should be kept small (e.g., $\beta\lesssim0.25$)
in order to prevent a particle from jumping more than one grid cell
during a single time step. Note that this time step limitation is
much weaker than the corresponding limitation in methods that resolve
the inertial dynamics, such as the Inertial Coupling method \cite{ISIBM}.
Resolving the time scale of the momentum diffusion requires keeping
$\beta_{\nu}=2\nu\D t/h^{2}=\text{Sc}\,\beta$ small, which requires
a time step on the order of $\text{Sc}\sim10^{3}-10^{4}$ smaller
than the FIB method. Note, however, that in applications the time
step may further be limited by other factors such as the presence
of stiff inter-particle potentials, as we discuss further in Section
\ref{sub:ColloidalGellation}.

\subsection{\label{sub:DeterministicMobility}Mobility in a Slit Channel}

\begin{figure}
\centering{}\includegraphics[width=0.99\columnwidth]{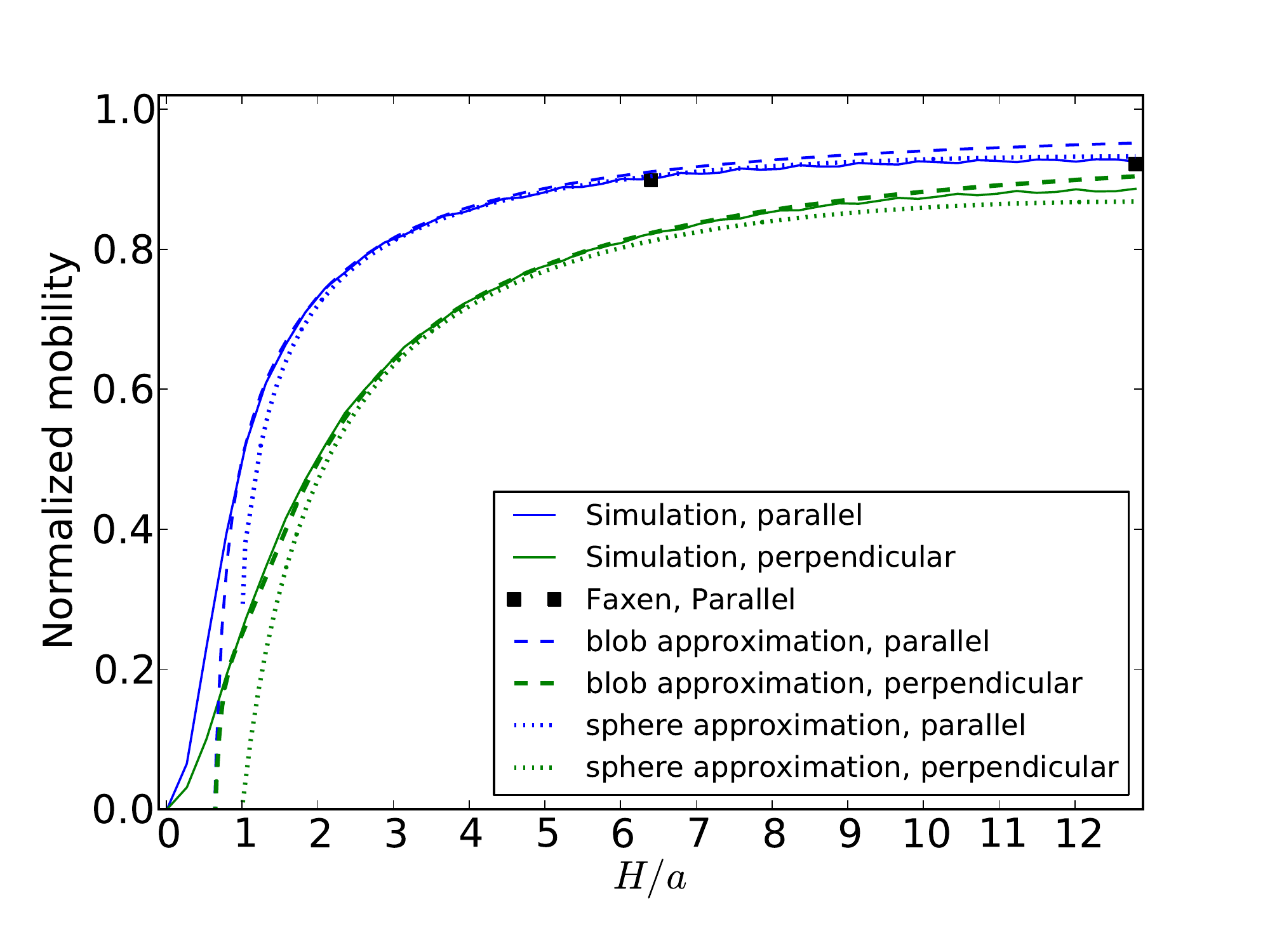}\caption{\label{fig:ChannelMobility}Mobility (relative to unbounded flow)
of a blob of hydrodynamic radius $a$ in a slit channel of thickness
$\sim25.5a$ in the directions parallel (blue lines and symbols) and
perpendicular (green lines) to the confining no-slip walls as a function
of the distance $H$ to the wall (expressed here in terms of the blob
hydrodynamic radius $a$), in three dimensions. Note that small oscillations
appear due to numerical grid artifacts. For the parallel mobility,
simulation is in excellent agreement with two exact results obtained
by Faxen (symbols), and over the range of distances where the blob
does not overlap the wall, the results are in good agreement with
the MCSA approximation for a blob in a channel (dashed lines, see
eq. (\ref{eq:single_wall})). Also shown is the MCSA approximation
for a hard sphere in a channel (dotted lines, see eq. (\ref{eq:SingleWallHardSpherePerp},
\ref{eq:SingleWallHardSphereParallel})) \cite{MobilitySlitChannel}.}
\end{figure}

The mobility of a single particle in a slit channel is affected by
the presence of the two walls. We estimate this effect by placing
a particle at multiple points across a 128$h$x128$h$x32$h$ channel
with planar no-slip walls at $z=0$ and $z=32h$, and periodic boundaries
along the $x$ and $y$ directions. For each position of the blob,
a unit force is applied either parallel and perpendicular to the wall,
the Stokes system (\ref{eq:stokesFluid}) without the stochastic momentum
flux is solved, and the resulting particle velocity is calculated,
giving the parallel $\mu_{\parallel}$ and perpendicular $\mu_{\perp}$
mobilities. The results of these calculations are reported in Fig.
\ref{fig:ChannelMobility}.

Unlike the case of a single no-slip boundary \cite{StokesianDynamics_Wall},
writing down an analytical solution for slit channels is complex and
requires numerically-evaluating the coefficients in certain series
expansions \cite{StokesianDynamics_Slit}. For the parallel component
of the mobility, Faxen has obtained exact series expansions for the
mobility at the half and quarter channel locations,
\begin{eqnarray*}
\mu_{\parallel}\left(H=\frac{L}{2}\right) & = & \frac{1}{6\pi\eta a}\Bigg[1-1.004\frac{a}{H}+0.418\frac{a^{3}}{H^{3}}\\
 &  & +0.21\frac{a^{4}}{H^{4}}-0.169\frac{a^{5}}{H^{5}}+\dots\Biggr]\\
\mu_{\parallel}\left(H=\frac{L}{4}\right) & = & \frac{1}{6\pi\eta a}\Biggl[1-0.6526\frac{a}{H}+0.1475\frac{a^{3}}{H^{3}}\\
 &  & -0.131\frac{a^{4}}{H^{4}}-0.0644\frac{a^{5}}{H^{5}}+\dots\Biggr]
\end{eqnarray*}
where $H$ denotes the distance from the blob to the nearest wall,
and $L$ is the distance between the walls. Here we neglect the corrections
coming from the use of periodic boundary conditions in the $x$ and
$y$ directions. As seen in Fig. \ref{fig:ChannelMobility}, the exact
results of Faxen are in excellent agreement with the numerical mobilities.

For other positions of the blob, we employ two different approximations.
Both of these make use the Modified Coherent Superposition Assumption
(MCSA) approximation to the unwieldy full expression for the mobility
\cite{StokesianDynamics_Slit}. This approach considers an infinite
sum of reflections of the single-wall solutions in another wall \cite{MobilitySlitChannel}.
In Fig. \ref{fig:ChannelMobility} we show two MCSA approximations
which we evaluated using (c.f. Eq. (9) in Ref. \cite{MobilitySlitChannel})
\begin{eqnarray}
\frac{\mu^{(2)}}{\mu_{0}} & = & \Biggl\{1+\sum_{n=0}^{\infty}(-1)^{n}\left[\frac{\mu_{0}}{\mu^{(1)}(nL+H)}-1\right]+\nonumber \\
 &  & +\sum_{n=1}^{\infty}(-1)^{n}\left[\frac{\mu_{0}}{\mu^{(1)}\left((n+1)L-H\right)}-1\right]\Biggr\}^{-1}\label{eq:MCSA}
\end{eqnarray}
where $\mu^{(2)}$ is either the parallel $\mu_{\parallel}^{(2)}$
or the perpendicular $\mu_{\perp}^{(2)}$ mobility in the slit channel,
$\mu_{0}$ is the mobility in an unbounded domain, and $\mu^{(1)}$
is the parallel or perpendicular single-wall mobility. Note that $\mu^{(2)}\approx\mu^{(1)}$
when the distance between the walls is very large, $L\gg H\gg a$,
as it must. Both of the the MCSA approximations are for an infinite
slit geometry, whereas we use periodic boundary conditions in the
directions parallel to the walls; we expect this has a small effect
on the value of the mobility calculated due to hydrodynamic screening,
as evidenced by the match with the exact results by Faxen.

We compare our numerical mobility to two different evaluations of
(\ref{eq:MCSA}), based on two different approximation for the single-wall
mobility $\mu^{(1)}$. The first is given by Swan and Brady \cite{StokesianDynamics_Wall},
\begin{eqnarray}
\frac{\mu_{\perp}^{(1)}(H)}{\mu_{0}} & = & 1-\frac{9a}{8H}+\frac{a^{3}}{2H^{3}}-\frac{a^{5}}{8H^{5}}\label{eq:single_wall}\\
\frac{\mu_{\parallel}^{(1)}(H)}{\mu_{0}} & = & 1-\frac{9a}{16H}+\frac{2a^{3}}{16H^{3}}-\frac{a^{5}}{16H^{5}},\nonumber 
\end{eqnarray}
as a generalization of the Rotne-Prager tensor using Blake's image
construction for a single wall \cite{blake1971note}. As such, we
expect this result to be accurate for blob particles near a single
wall, and we see good agreement in Fig. \ref{fig:ChannelMobility}.
We also use a second expression for $\mu^{(1)}$ which more closely
approximates a hard sphere. For this approximation, the perpendicular
mobility is given by a semi-empirical rational relation approximation
to an exact series of Brenner \cite{SingleWallPerpMobility},
\begin{align}
\frac{\mu_{\perp}^{(1)}(H)}{\mu_{0}}= & \frac{6\left(\frac{H}{a}\right)^{2}+2\left(\frac{H}{a}\right)}{6\left(\frac{H}{a}\right)^{2}+9\left(\frac{H}{a}\right)+2}.\label{eq:SingleWallHardSpherePerp}
\end{align}
The hard sphere approximation to the parallel single wall mobility
is given by a combination of a near-wall expression derived using
lubrication theory and a truncated expansion in powers of $a/H$ which
is more accurate further from the wall. The near-wall calculation
involves a complicated expression which we do not reproduce here (see
\cite{NearWallSphereMobility}), and it is used when $H-a\leq0.05a$.
When the blob is further from the wall, we calculate the single wall
parallel mobility from the exact power series expansion truncated
to fifth order \cite{ConfinedSphere_Sedimented}, 
\begin{align}
\frac{\mu_{\parallel}^{(1)}(H)}{\mu_{0}}= & 1-\frac{9a}{16H}+\frac{1a^{3}}{8H^{3}}-\frac{45a^{4}}{256H^{4}}-\frac{a^{5}}{16H^{5}}.\label{eq:SingleWallHardSphereParallel}
\end{align}

Both MCSA (\ref{eq:MCSA}) approximations are seen to be in very good
agreement with our numerical results away from the wall in Fig. \ref{fig:ChannelMobility}.
Near the wall, the fact that we use a minimally resolved ``blob''
model becomes relevant, and the numerical results agree more with
(\ref{eq:single_wall}) than with (\ref{eq:SingleWallHardSpherePerp},\ref{eq:SingleWallHardSphereParallel}).
The approximations (\ref{eq:single_wall}) and especially (\ref{eq:SingleWallHardSpherePerp},\ref{eq:SingleWallHardSphereParallel})
are intended to work only for $H>a$. In particular, the Swan-Brady
Rotne-Prager-Blake tensor only ensures an SPD mobility when the blobs
do not overlap the wall or each other. By contrast, our numerical
calculation does not diverge when the blob overlaps the wall, giving
instead a mobility that smoothly decays to zero as the centroid of
the blob approaches the wall, and is SPD for all blob configurations
as long as all blob centroids are inside the channel. Close to the
wall our numerical results are expected to be in close agreement with
the Rotne-Prager-Yamakawa-Blake tensor \cite{RPY_Shear_Wall}, which
is, unfortunately, not available in closed form.

\subsection{\label{sub:Diffusion-Coefficient}Diffusion Coefficient}

As explained in Section \ref{sub:SimpleMidpoint}, we chose the midpoint
form of the predictor corrector (\ref{eq:predictor},\ref{eq:corrector}),
because this gives an accurate diffusion coefficient even for large
time step size $\D t$. %
{} Here we confirm this by numerically estimating the time-dependent
diffusion coefficient of a single freely-diffusing particle in a two
dimensional periodic domain
\[
\chi(s)=\frac{1}{2ds}\av{\norm{\V q(t+s)-\V q(t)}^{2}}
\]
for a range of time step sizes. For comparison, we also try a simple
trapezoidal predictor-corrector scheme that replaces (\ref{eq:predictor},\ref{eq:corrector})
with
\begin{eqnarray}
\V q^{\star,n+1} & = & \V q^{n}+\D t\J^{n}\bv\nonumber \\
\V q^{n+1} & = & \V q^{n}+\frac{\D t}{2}\left(\J^{n}+\J^{\star,n+1}\right)\bv.\label{eq:trapPC}
\end{eqnarray}
This scheme is also a first-order weakly accurate integrator, but
does not satisfy the property (\ref{eq:var_dq}).

\begin{figure}
\centering{}\includegraphics[width=1\columnwidth]{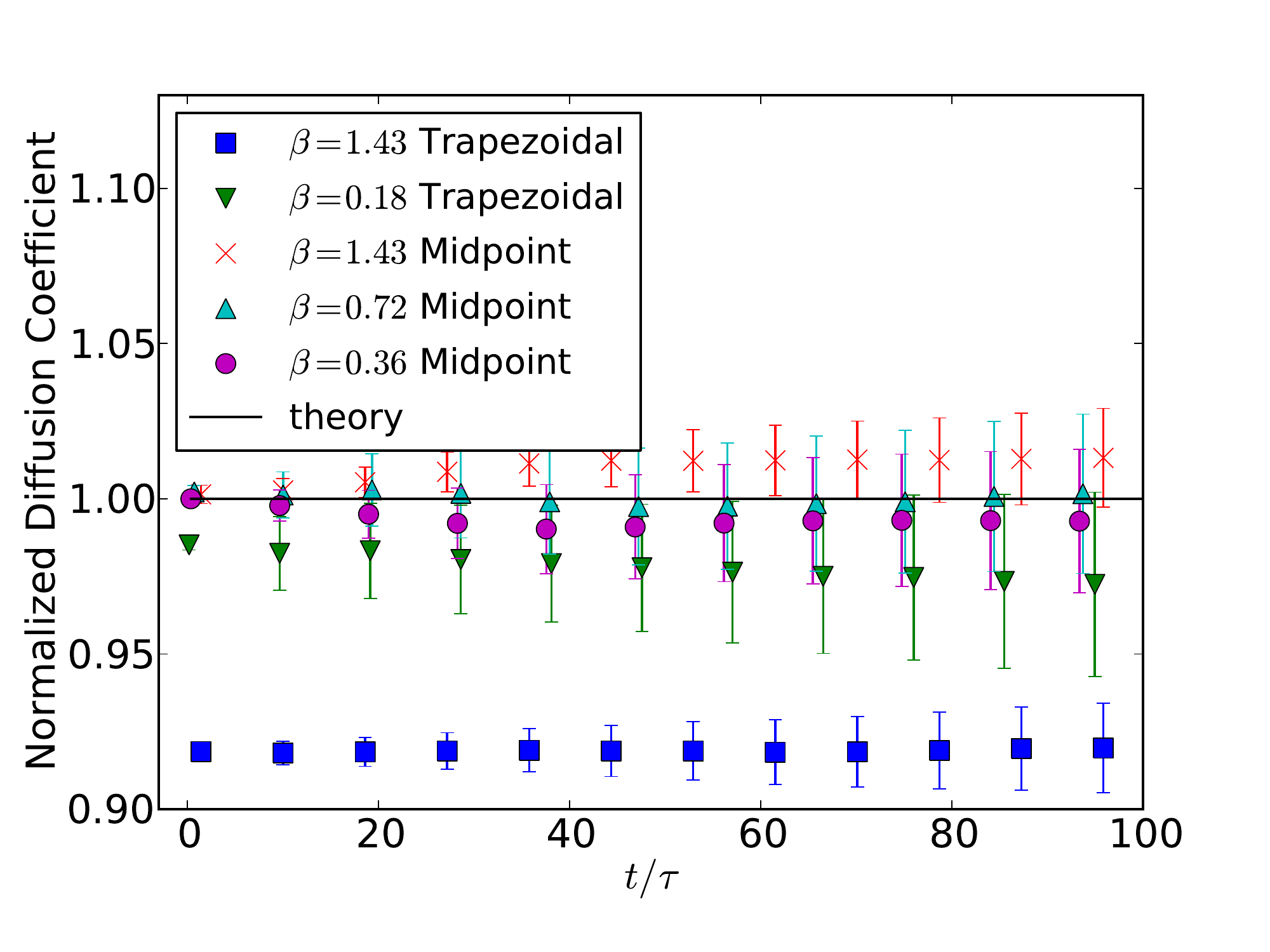}\caption{\label{fig:DiffCoeff}Normalized diffusion coefficient for multiple
time step sizes using the midpoint (\ref{eq:predictor},\ref{eq:corrector})
or the trapezoidal (\ref{eq:trapPC}) predictor-corrector schemes.}
\end{figure}

Figure \ref{fig:DiffCoeff} shows that the midpoint predictor corrector
(\ref{eq:predictor},\ref{eq:corrector}) gives a diffusion coefficient
that agrees with the theoretical result $\chi(s)=k_{B}T\mu$ independent
of $s$ to within statistical error for time step sizes as large as
$\beta=\D t/\tau=2\chi\D t/h^{2}=1.43$, where $\tau$ is the natural
diffusive time scale for this test. By contrast, the trapezoidal scheme
(\ref{eq:trapPC}) introduces a measurable truncation error already
for $\beta\gtrsim0.2$. Both schemes include an RFD term to approximate
the (small) drift term present due to the discretization showing that
the RFD term does not ruin the accuracy of the diffusion coefficient
for the midpoint scheme.

\subsection{\label{sub:ThermodynamicEquilibrium}Thermodynamic Equilibrium}

One of the most important requirements on any scheme that couples
fluctuating hydrodynamics to immersed particles is to reproduce the
Gibbs-Boltzmann distribution (\ref{eq:gibbsboltzmann}) at thermodynamic
equilibrium, independent of any dynamical parameters such as viscosity.
In prior work \cite{ISIBM}, we confirmed that when fluid and particle
inertia are consistently included in the formulation, the numerical
method reproduces the correct equilibrium distribution for both the
particle positions and the appropriate Maxwell-Boltzmann distribution
for the particle velocities. In the overdamped limit considered here
there are no velocity degrees of freedom, but the method should still
reproduce the correct Gibbs-Boltzmann distribution (\ref{eq:gibbsboltzmann})
for sufficiently small time steps. In this section we consider several
scenarios and verify that the FIB method correctly reproduces the
theoretical equilibrium distribution. As we demonstrate next, in the
case of a non-constant mobility this necessitates the proper inclusion
of the stochastic drift terms using the specialized temporal integration
techniques we developed in Section \ref{sec:TemporalIntegrator}.

\subsubsection{\label{sub:PeriodicPDF}Free Diffusion}

In the continuum setting, for a single particle in a periodic system
translational invariance implies that the mobility does not depend
on the position of the particle, and therefore $\partial_{\V q}\cdot\Mob=0$.
However, upon spatial discretization, translational invariance is
broken by the presence of a fixed Eulerian grid on which the fluid
equation is solved. Even though the Peskin kernels give excellent
translational invariance of the mobility, there is still a fraction
to a few percent (depending on the kernel) variation in the mobility
as the particle position shifts relative to the underlying grid. Here
we show that our midpoint temporal integrators correct for this and
ensure a uniform equilibrium distribution for the position of freely-diffusing
particles.

In this test, 3000 particles are allowed to diffuse freely in a periodic
two-dimensional domain of size $16h\times16h$. Because the particles
do not exert forces on each other, each of the particles is statistically
identical to an isolated particle diffusing in the same domain (even
though the particles are not independent because of the hydrodynamic
interactions \cite{StokesLaw}), and at equilibrium their positions
should be independent and uniformly distributed in the periodic domain.
A small time step size corresponding to $\beta\approx0.01$ is used
to approach the limit $\D t\rightarrow0$. The three-point Peskin
kernel is used in order to maximize the lack of translational invariance.

For testing purposes, we dropped the RFD and corrector stages in the
simple midpoint scheme (\ref{eq:TdriftPC})-(\ref{eq:corrector})
to obtain the Euler-Maruyama integrator,
\begin{eqnarray}
-\eta\M L\bv+\M G\pi & = & \S^{n}\V F^{n}+\sqrt{\frac{2k_{B}T}{\D t\D V}}\widetilde{\M D}\M W^{n}\nonumber \\
\M D\bv & = & 0\nonumber \\
\V q^{n+1} & = & \V q^{n}+\Delta t\J^{n}\bv.\label{eq:Euler-Maruyama}
\end{eqnarray}
Note that this temporal integrator is inconsistent with the kinetic
interpretation of the noise term, i.e., it is not consistent with
the Fokker-Planck equation (\ref{eq:FokkerPlanck}); it is biased
even in the limit $\D t\rightarrow0$.

\begin{figure*}
\centering{}\includegraphics[width=0.49\textwidth]{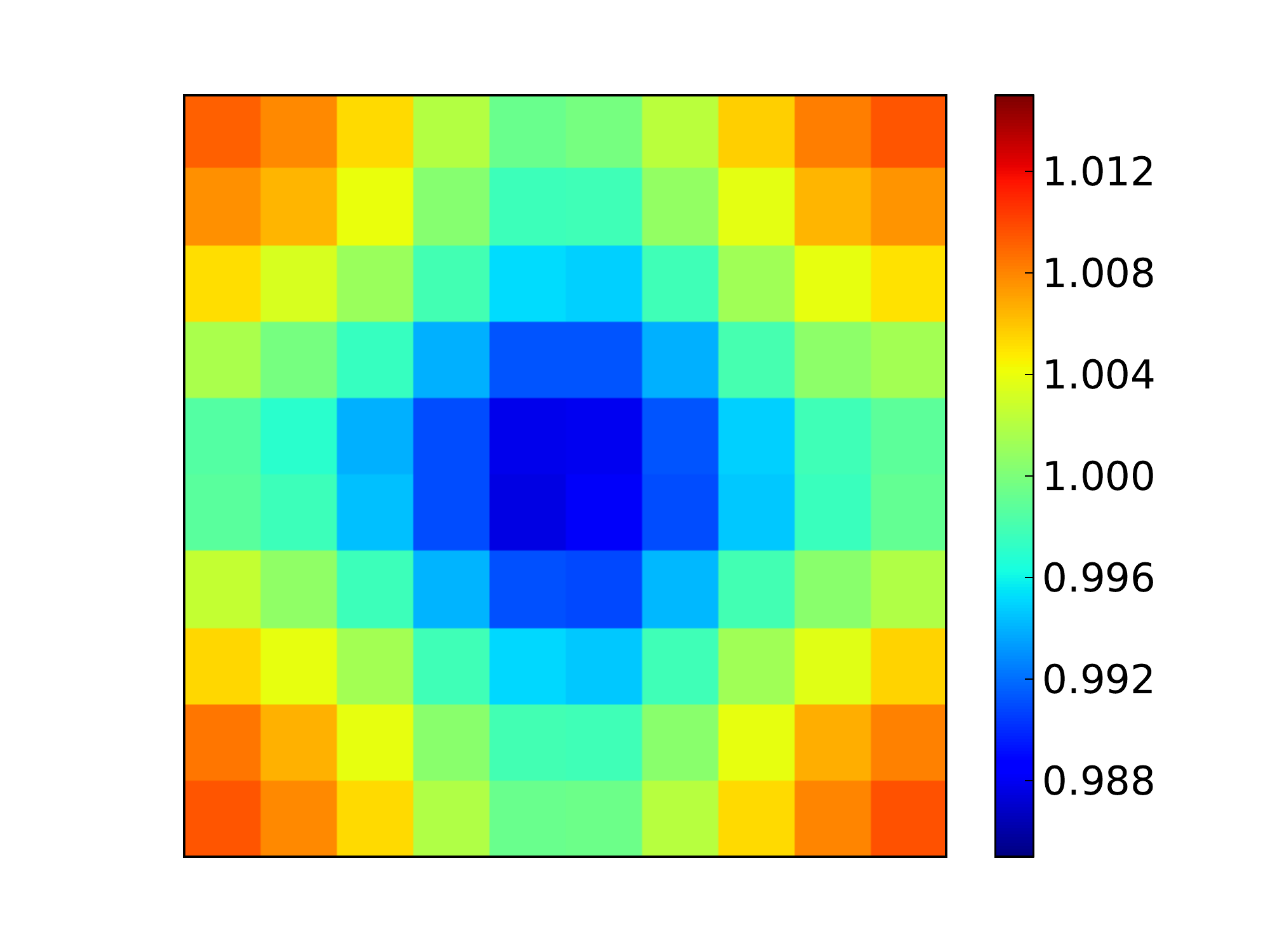}\includegraphics[width=0.49\textwidth]{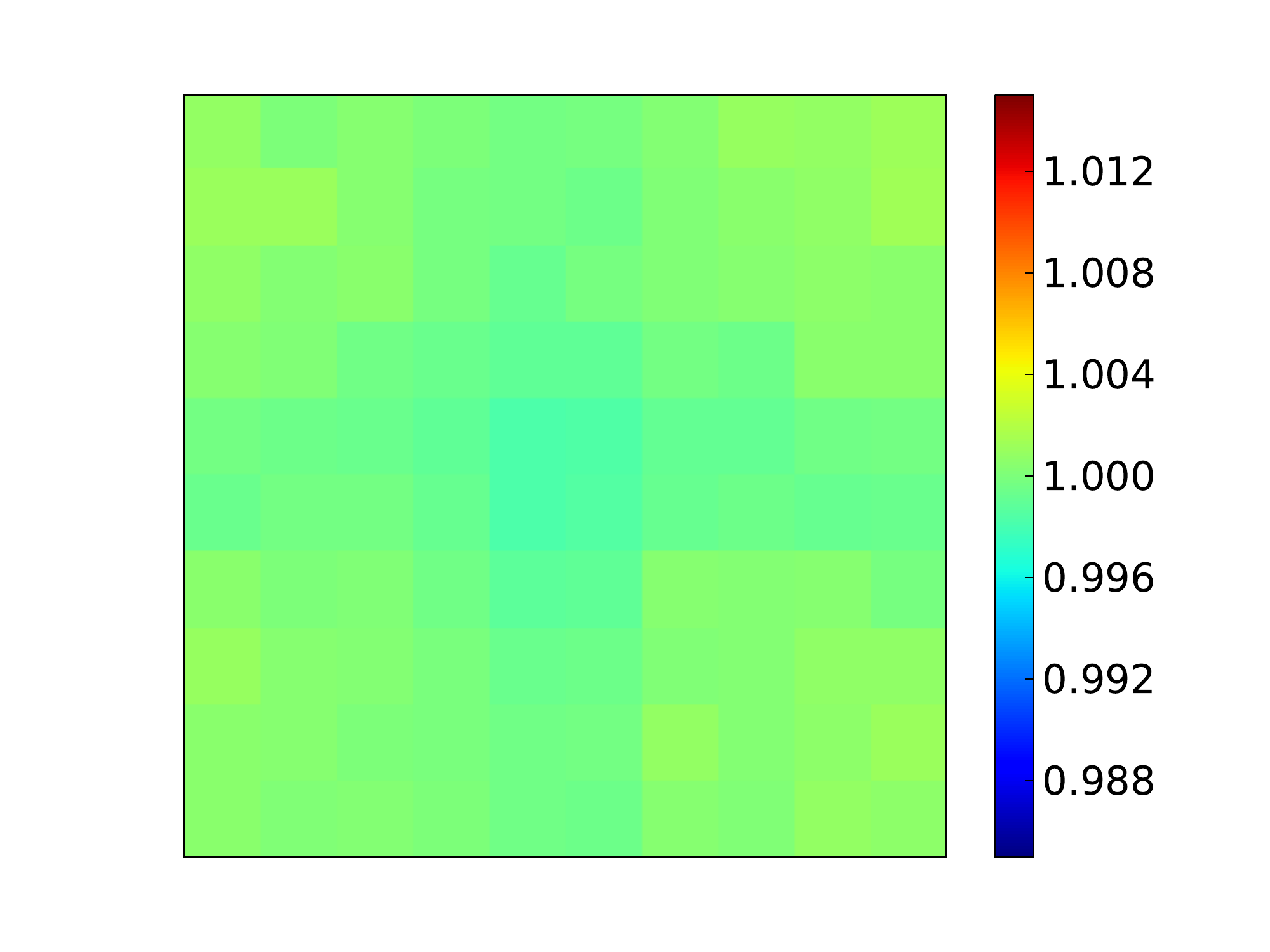}\caption{\label{fig:celldist}\textit{(Left panel)} Normalized equilibrium
probability distribution for finding a free particle at a particular
position inside a grid cell when using the Euler-Maruyama scheme (\ref{eq:Euler-Maruyama}).
A slightly nonuniform distribution is observed, in disagreement with
the correct uniform Gibbs-Boltzmann distribution. This error does
\emph{not} vanish in the limit $\D t\rightarrow0$. \textit{(Right
panel) }Using the midpoint scheme (\ref{eq:TdriftPC})-(\ref{eq:corrector})
preserves the correct distribution. The small residual artifacts disappear
in the limit $\D t\rightarrow0$. The same color scale (with variation
in the range $0.985-1.015$) is used for both panels.}
\end{figure*}

The Euler-Maruyama method was compared with our midpoint scheme (\ref{eq:TdriftPC})-(\ref{eq:corrector})
by computing an empirical histogram for the equilibrium distribution
of the position of a particle inside a cell (due to translational
invariance of the periodic grid the distribution is the same in all
grid cells). The results in Fig. \ref{fig:celldist} show small but
clear artifacts in the equilibrium distribution when using the Euler-Maruyama
(\ref{eq:Euler-Maruyama}) scheme, specifically, the particle is more
likely to be found near the corners of the grid cell instead of the
center of the grid cell. By contrast, our consistent integrator (\ref{eq:TdriftPC})-(\ref{eq:corrector})
give a uniform distribution for the position of the particle for sufficiently
small time step sizes; the same is true for the improved integrator
(\ref{eq:SecondOrderDeterministic}), not shown.

\subsubsection{Diffusion in a slit channel}

One key strength of the FIB method is the ability to handle non-periodic
boundary conditions. In this test particles are placed in a two-dimensional
channel and allowed to diffuse freely. When a particle comes within
a cutoff range $w$ from one of the two no-slip walls, it is repelled
with a harmonic potential with spring stiffness $k$,

\begin{equation}
U(H)=\frac{k}{2}(H-w)^{2}\quad\mbox{if}\quad H\leq w\quad\mbox{and zero otherwise},
\end{equation}
where $H$ is the distance of the particle from the wall. The total
potential for the equilibrium distribution is the sum of the top and
bottom wall potentials. A long equilibrium run is performed in order
to compute an empirical histogram for the marginal equilibrium distribution
$P(H)$ for finding a particle at a given distance $H$ from the nearest
wall (note that all particles are statistically identical). We perform
the simulations in two dimensions in order to maximize the statistical
accuracy. The values of the simulation parameters are given in Table
\ref{tab:ChannelEqDist}. Note that here we employ a relative large
time step size in order to test the robustness of our temporal integrators.

\begin{table}
\centering{}%
\begin{tabular}{|c|c|}
\hline 
Number of particles &
100 (midpoint) or 1 (Euler)\tabularnewline
\hline 
wall ``spring'' constant $k$ &
$6\,\left(k_{B}T/h^{2}\right)$\tabularnewline
\hline 
wall potential range $w$ &
$2h$\tabularnewline
\hline 
dimensionless time step size $\beta$ &
$0.023$\tabularnewline
\hline 
domain width $L_{x}$ &
$8h$\tabularnewline
\hline 
domain height $L_{y}$ &
$16h$\tabularnewline
\hline 
\end{tabular}\caption{\label{tab:ChannelEqDist}Parameters used for the slit channel simulation
results shown in Fig. \ref{fig:ChannelDist}.}
\end{table}

\begin{figure}
\centering{}\includegraphics[width=1\columnwidth]{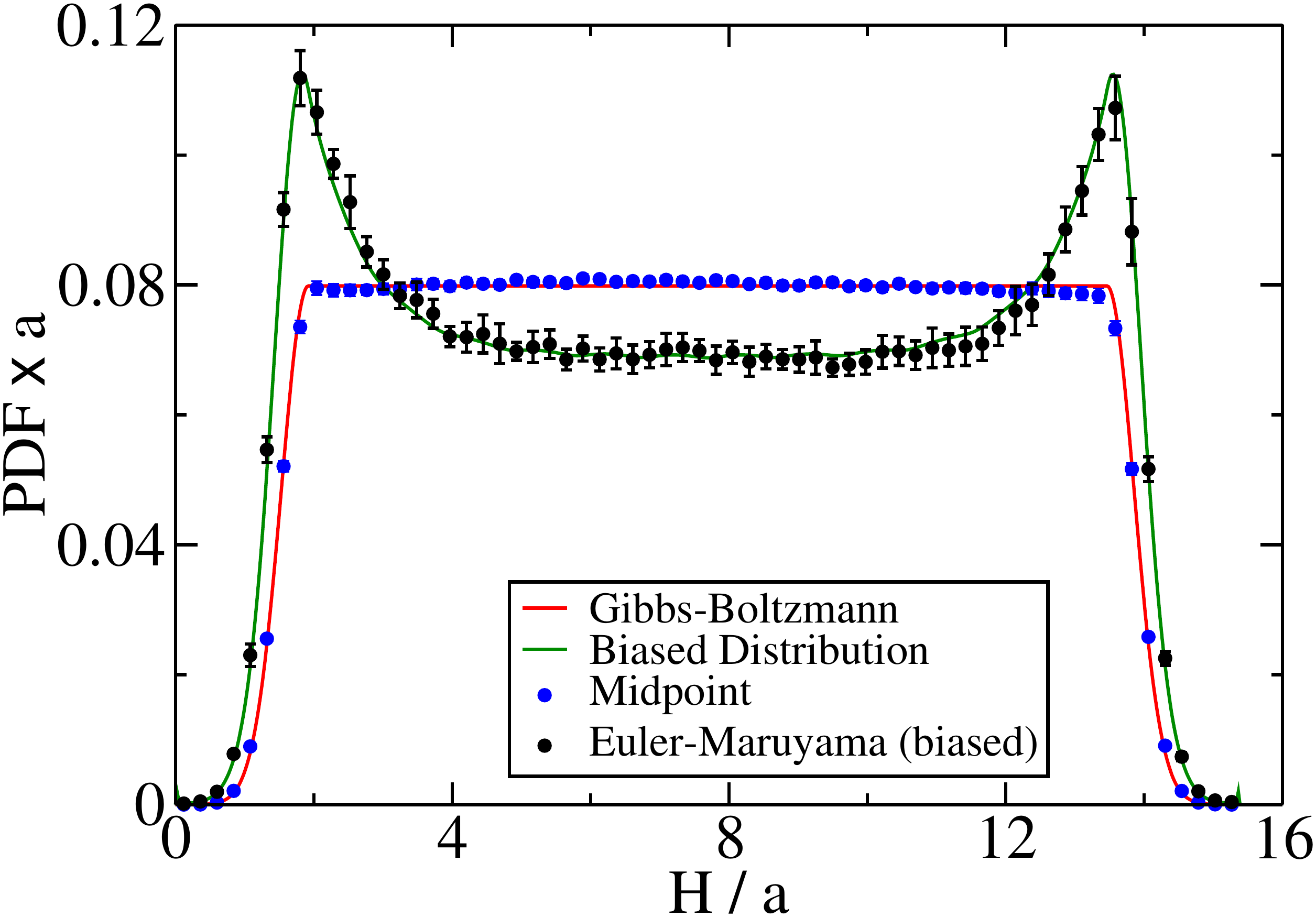}\caption{\label{fig:ChannelDist}Probability distribution of the distance $H$
to one of the walls for a freely-diffusing single blob (Euler-Maruyama
scheme (\ref{eq:Euler-Maruyama})), as well as many non-interacting
(midpoint scheme (\ref{eq:TdriftPC})) blobs, in a two dimensional
slit channel. The correct (unbiased) (\ref{eq:gibbsboltzmann}) and
the biased (\ref{eq:biasedGB}) Gibbs-Bolzmann distribution are shown
for comparison.}
\end{figure}

As illustrated in Fig. \ref{fig:ChannelDist}, the results of the
midpoint algorithm (\ref{eq:TdriftPC}) with 100 particles compares
favorably to the correct Gibbs-Boltzmann distribution $P(H)=Z^{-1}\exp\left(-U(H)/k_{B}T\right)$.
We also test the biased Euler-Maruyama scheme (\ref{eq:Euler-Maruyama})
for a single particle. This scheme does not reproduce the thermal
drift term from Eq. (\ref{eq:OverdampedEq}), and thus yields an unphysical
result where particles are more likely to be found near the boundaries
(see also discussion in Section III.C in Ref. \cite{StokesianDynamics_Slit}). 

In fact, the equilibrium distribution preserved by the biased scheme
(\ref{eq:Euler-Maruyama}) in the limit $\D t\rightarrow0$ can be
calculated analytically for a single particle. For one particle in
a slit channel, the $x$ and $y$ components of (\ref{eq:BrownianDynamics})
decouple, and the only interesting dynamics occurs in the direction
perpendicular to the channel walls. The Euler-Maruyama scheme (\ref{eq:EMPerpChannel})
is consistent with the Ito equation 
\begin{equation}
\frac{dH}{dt}=-\mu_{\perp}(H)\, U^{\prime}(H)+\sqrt{2k_{B}T\mu_{\perp}(H)}\,\mathcal{W}_{2}\left(t\right).\label{eq:EMPerpChannel}
\end{equation}
By adding and subtracting $k_{B}T\mu_{\perp}^{\prime}(H)$ we can
convert this into the kinetic stochastic interpretation,
\begin{eqnarray*}
\frac{dH}{dt} & = & -\mu_{\perp}(H)\,\widetilde{U}^{\prime}(H)+\sqrt{2k_{B}T\mu_{\perp}(H)}\,\mathcal{W}_{2}\left(t\right)\\
 &  & +k_{B}T\,\mu_{\perp}^{\prime}(H)\\
 & =- & \mu_{\perp}(H)\,\widetilde{U}^{\prime}(H)+\sqrt{2k_{B}T\mu_{\perp}(H)}\diamond\mathcal{W}_{2}\left(t\right),
\end{eqnarray*}
where the biased potential is
\[
\widetilde{U}(H)=U(H)+k_{B}T\,\ln(\mu_{\perp}(H)).
\]
This shows that the Euler-Maruyama scheme (\ref{eq:EMPerpChannel})
preserves the biased Gibbs-Boltzmann distribution corresponding to
the biased potential $\widetilde{U}(H)$,
\begin{equation}
P_{E}(H)=\widetilde{Z}^{-1}\exp\left(-\frac{\widetilde{U}(H)}{k_{B}T}\right).\label{eq:biasedGB}
\end{equation}

The biased distribution $P_{E}(H)$ is shown in Fig. \ref{fig:ChannelDist}
with $\mu_{\perp}$ calculated numerically (see Fig. \ref{fig:ChannelMobility}).
The biased distribution indeed matches the simulation results from
the Euler-Maruyama scheme, confirming that the correct equilibrium
distribution is not preserved without the RFD term and predictor-corrector
steps. At the same time, we see that the temporal integrator (\ref{eq:TdriftPC})-(\ref{eq:corrector})
preserves the correct thermodynamic equilibrium distribution even
in the presence of confinement.

\subsubsection{Colloidal suspension}

\begin{figure}
\begin{centering}
\includegraphics[width=1\columnwidth]{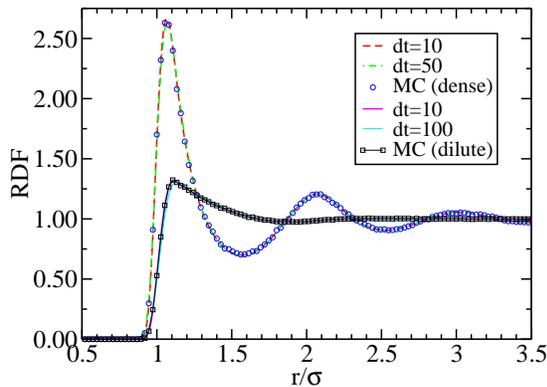}
\par\end{centering}

\caption{\label{fig:RDF}Radial pair correlation function $g(r)$ for a suspension
of particles interacting with repulsive Lennard-Jones potentials at
packing fractions $\phi\approx0.13$ (dilute) and $\phi\approx0.42$
(dense). Results from two different time step sizes are compared to
Monte Carlo (MC) simulations.}
\end{figure}

In this section we verify that our FIB algorithm gives the correct
equilibrium distribution $P\left(\V q\right)$ for a multi-particle
system by computing the radial (pair) distribution function (RDF)
$g(r)$ for a periodic collection of $N$ colloidal particles interacting
with a pairwise repulsive truncated Lennard-Jones (LJ) potential $V(r)$,
\[
U\left(\V q\right)=\sum_{i,j=1}^{N}\, V\left(\norm{\V q_{i}-\V q_{j}}\right),
\]
as described in more detail in Section 4.1 in Ref. \cite{ISIBM}.
The parameters used for these simulations are given in Table \ref{tableRDF},
and the GPU-based code \emph{fluam} with the three point kernel is
used for these simulations\emph{ }\cite{ISIBM}. In the left panel
of Fig. \ref{fig:RDF} we compare $g(r)$ between a simulation where
the particles are immersed in an incompressible viscous solvent, and
a standard computation of the equilibrium RDF using a Monte Carlo
algorithm to sample the equilibrium distribution (\ref{eq:gibbsboltzmann}).
We test the FIB algorithm at two different densities, a dilute suspension
corresponding to a packing fraction based on the LJ diameter of $\phi\approx0.13$,
and a dense suspension (close to the freezing point) at packing fraction
$\phi\approx0.42$. Note that while the minimally-resolved model here
cannot accurately model the dynamics (hydrodynamic interactions) at
high packing fractions \cite{BrownianDynamics_OrderNlogN,ForceCoupling_Stokes},
we do obtain the correct equilibrium properties because our formulation
and numerical scheme obey discrete fluctuation-dissipation balance
for any interaction potential and any viscosity.

\begin{table}
\centering{}%
\begin{tabular}{|c|c|}
\hline 
grid spacing $\D x$  &
1 \tabularnewline
\hline 
grid size &
$32^{3}$\tabularnewline
\hline 
shear viscosity $\eta$  &
1 \tabularnewline
\hline 
time step size $\Delta t$  &
Variable\tabularnewline
\hline 
temperature $k_{B}T$  &
$10^{-3}$\tabularnewline
\hline 
LJ strength $\epsilon$  &
$10^{-3}$\tabularnewline
\hline 
LJ / hydro diameter $\sigma$  &
2 \tabularnewline
\hline 
number of particles $N$ &
1000 (dilute) or 3300 (dense) \tabularnewline
\hline 
\end{tabular} \caption{\label{tableRDF}Parameters used in the colloidal suspension equilibrium
simulations shown in Fig. \ref{fig:RDF}. }
\end{table}
As seen in Fig. \ref{fig:RDF}, we obtain excellent agreement with
the Monte Carlo calculations even for time steps close to the stability
limit. The Brownian time scale here is%
\footnote{Perhaps a more relevant diffusive length scale to use is the typical
inter-particle gap, which can be substantially smaller than $a$ for
dense suspensions.%
}
\[
\tau_{B}=\frac{a^{2}}{\chi}=\frac{6\pi a^{3}\eta}{k_{B}T}\approx2\cdot10^{4},
\]
and the time step size is primarily limited (to $\D t\lesssim100$,
corresponding to $\beta=0.005$, for the dilute suspension, and $\D t\lesssim50$
for the denser suspension) by stability requirements relating to the
presence of the stiff LJ repulsion between the particles. Note that
the time step size in these simulations is substantially larger than
those required in the Inertial Coupling scheme developed by some of
us in Ref. \cite{ISIBM} (there, a time step of $\D t=1$ was used).

\subsection{\label{sub:ShearFlow}Particles in Shear Flow}

In this section we verify the the FIB method correctly models the
\emph{dynamics} of hydrodynamically-interacting Brownian particles
by computing time correlation functions of the positions of particles
in shear flow. Particles are anchored with a harmonic spring to their
initial locations and subjected to shear flow, as can be experimentally
realized by using optical tweezers to apply the potential \cite{Hydro_Tweezers}.
Brammert, Holzer, and Zimmerman have performed theoretical analysis
of this system \cite{holzer2010dynamics,bammert2010dynamics} and
provide explicit expressions used to test the accuracy of our scheme.
The numerical results presented below demonstrate the ability of our
midpoint schemes to correctly reproduce the effect of hydrodynamic
interactions between distinct immersed particles.

Note that it is not possible to have an unbounded system in a finite-volume
approach; a finite system is necessary and it is most convenient to
use a large but finite periodic system %
\footnote{In Green's function's based approaches there is no difficulty in dealing
with unbounded three dimensional systems (at rest at infinity) since
the Oseen tensor is the required response function and is easy to
compute. Note however that when simulating periodic domains (e.g.,
colloidal suspensions) one requires the Green's functons for Stokes
flow in a periodic domain, and these are \emph{not} simple to compute
\cite{RotnePrager_Periodic,BrownianDynamics_OrderN}, unlike in our
approach, which handles boundary conditions naturally.%
}. In these tests, we add a background shear flow with velocity $\V u$
to the \emph{periodic} fluctuating component $\bv$ calculated by
the steady Stokes solver; this mimics common practice in Brownian
dynamics simulations of polymer chains in flow \cite{BrownianDynamics_DNA}.
We will define $y$ as the direction of shear, and $x$ as the direction
of flow. The background flow is of the form $\V u\left(x,y\right)=(\dot{\gamma}y,0,0)$
for some constant $\dot{\gamma}$. Note that the total flow $\V u+\bv$
is a solution to the Stokes equations with the same forces that generated
$\bv$, but with boundary conditions modified to match the added background
flow. The resulting velocity of the particle is then $\J(\V q)\bv+\J(\V q)\V u$.
With a spherically symmetric kernel and the constant-shear flow $\V u$,
we have $\J(\V q)\V u=\V u(\V q)$. To implement the addition of the
background flow, we calculate $\J\bv$ without any modification for
the shear flow, and then separately add $\V u(\V q)$. The temporal
scheme is then the same as (\ref{eq:TdriftPC})-(\ref{eq:corrector})
but with $\J^{n}\bv$ replaced by $\J^{n}\bv+\V u\left(\V q^{n}\right)$
and likewise $\J^{n+\frac{1}{2}}\bv$ is replaced by $\J^{n+\frac{1}{2}}\bv+\V u\left(\V q^{n+\frac{1}{2}}\right)$.
Note that here $\V q$ is the position of the particle not on the
periodic torus but in an unbounded domain obtained by periodically
replicating the fixed unit cell. In our tests the particles are localized
to a single unit cell and do not interact with periodic image particles;
in more general situations such as sheared suspensions more complicated
approaches (reminiscent of Lees-Edwards boundary conditions commonly
employed in molecular dynamics) are necessary to account for the lack
of periodicity in shear flow \cite{Atzberger_Shear}. Alternatively,
one can use periodic flows of the form $u_{x}\sim\sin ky$ with $k$
sufficiently small (i.e., periodic box sufficiently large) to approach
the limit $k\rightarrow0$.

\begin{table}
\begin{centering}
\begin{tabular}{|c|c|}
\hline 
domain width, $L$ &
$64h$\tabularnewline
\hline 
\hline 
hydrodynamic radius, $a$ &
$1.04h$\tabularnewline
\hline 
spring constant, $k$ &
1$\left(k_{B}T/h^{2}\right)$\tabularnewline
\hline 
time step size $\frac{\D t}{\tau}$ &
0.22, 0.11, 0.02\tabularnewline
\hline 
diffusive CFL number $\beta$ &
0.45, 0.22, 0.05\tabularnewline
\hline 
Weissenberg number &
1.0\tabularnewline
\hline 
\end{tabular}
\par\end{centering}

\caption{\label{tab:ShearParameters1}Parameters for the simulation of a single
particle in shear flow.}
\end{table}

\begin{table}
\begin{centering}
\begin{tabular}{|c|c|}
\hline 
domain width, $L$ &
$32h$\tabularnewline
\hline 
\hline 
well separation, $b$ &
$5h$\tabularnewline
\hline 
hydrodynamic radius, $a$ &
$1.25h$\tabularnewline
\hline 
spring constant, $k$ &
10$\left(k_{B}T/h^{2}\right)$ \tabularnewline
\hline 
time step size $\frac{\D t}{\tau}$ &
0.3, 0.08\tabularnewline
\hline 
diffusive CFL number$\beta$ &
0.06, 0.015\tabularnewline
\hline 
Weissenberg number &
1.0\tabularnewline
\hline 
\end{tabular}
\par\end{centering}

\caption{\label{tab:ShearParameters2}Parameters for the simulation of two
particles in shear flow. }
\end{table}

\begin{figure*}[!t]
\centering{}\includegraphics[width=0.49\textwidth]{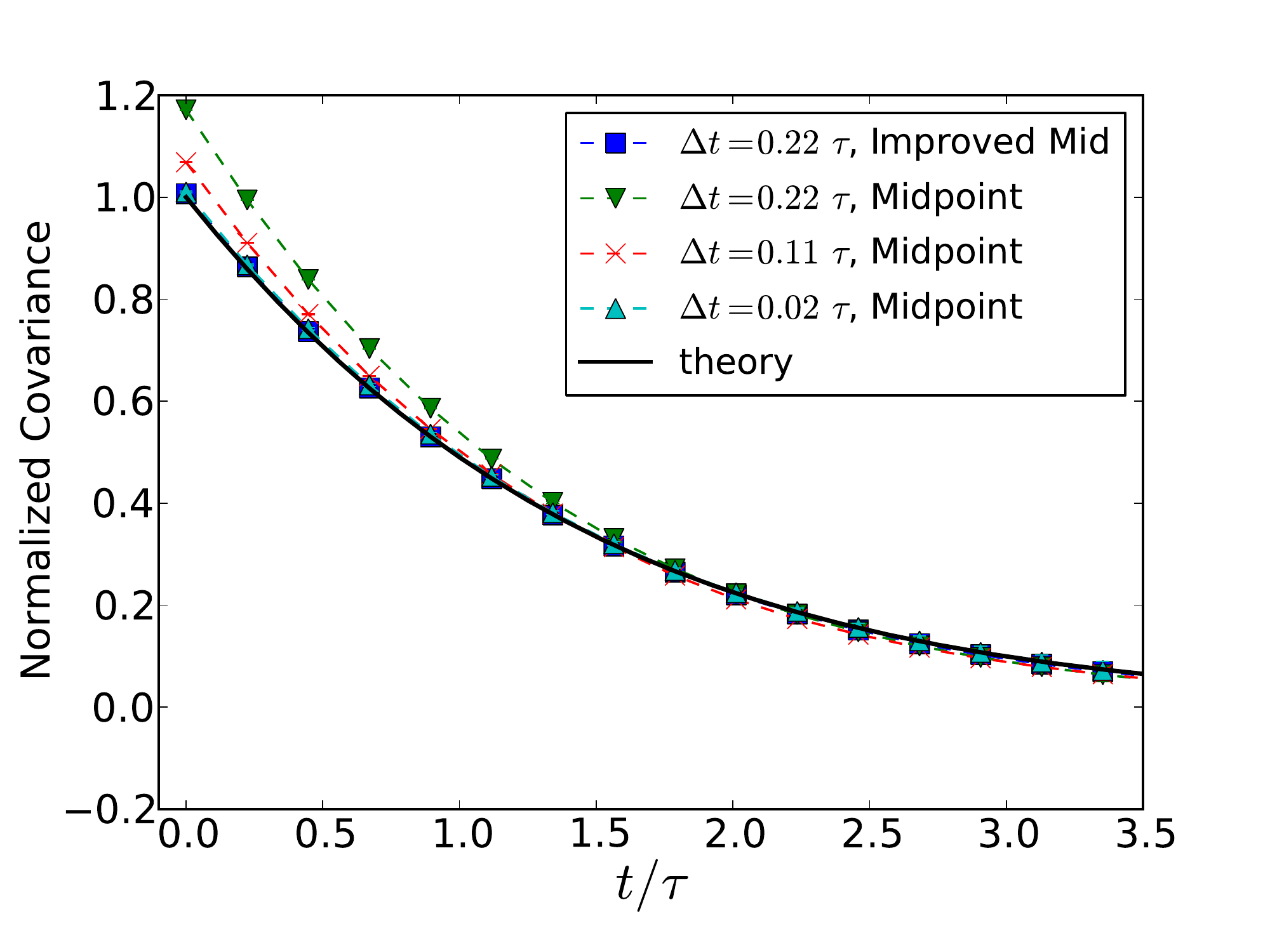}\includegraphics[width=0.49\textwidth]{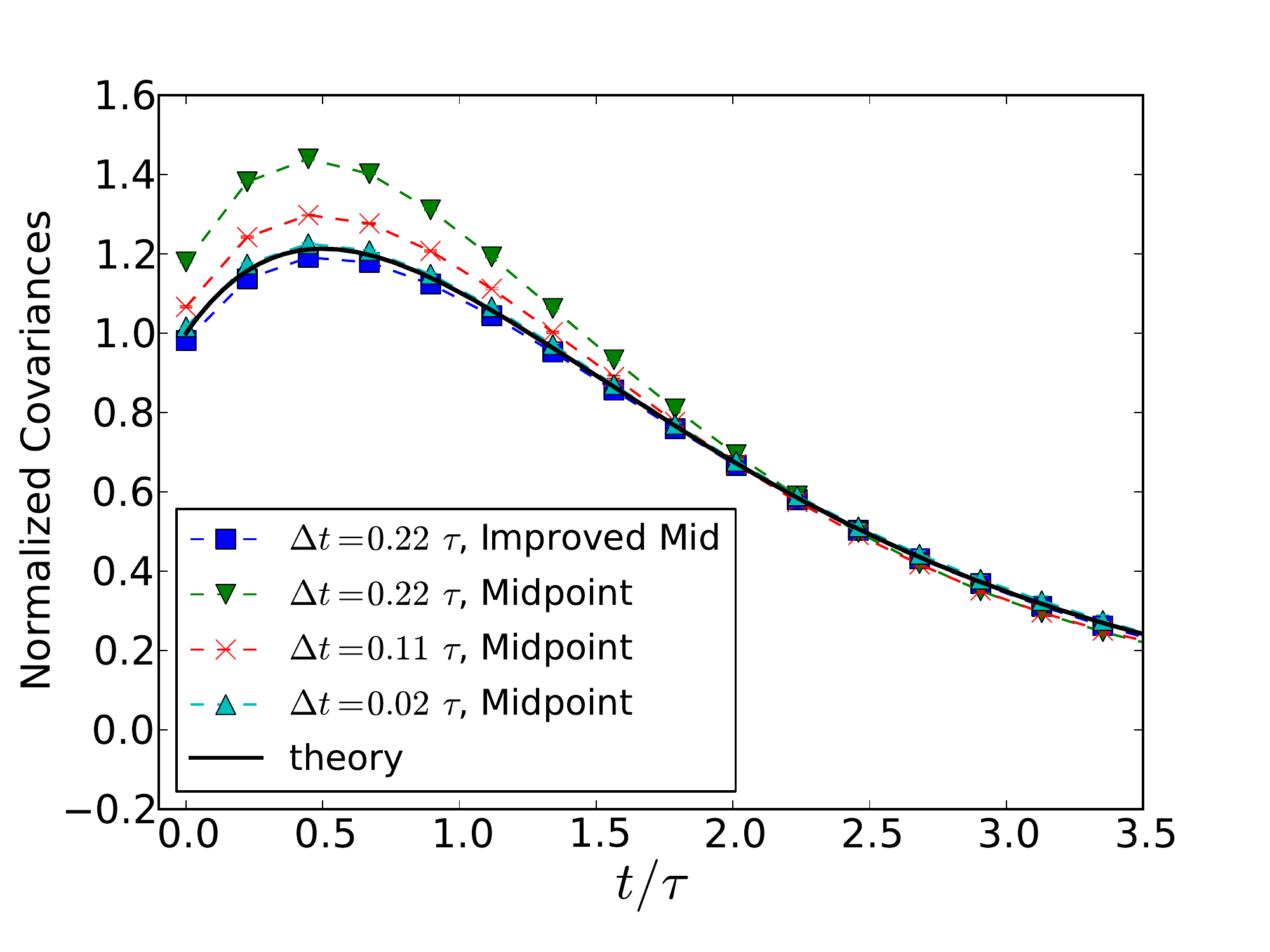}\caption{\emph{\label{fig:HarmonicShear}}Normalized time correlation functions
for a single particle in a harmonic potential in the presence of shear
flow. As the time step size is reduced, the numerical results converge
to the theoretical expressions, substantially faster for the improved
midpoint algorithm. Error bars of two standard deviations are drawn,
and are generally on the order of symbol size.\emph{ (Left panel)}
Autocorrelation of the displacement of the particle in the direction
of flow, $\avv{\tilde{x}(t)\tilde{x}(0)}$. \textit{(Right panel)}
Cross correlation of the displacements in flow and shear directions,
$\av{\tilde{x}(t)\tilde{y}(0)}$. }
\end{figure*}

\subsubsection{\label{sub:SingleParticle}A Single Particle}

In this simulation, a single particle is placed in a background shear
flow and is attached to an anchor location, $\V q_{0}$, by a harmonic
spring with potential $U=(k/2)\norm{\V q-\V q_{0}}^{2}$. The strength
of the shear flow relative to the harmonic force is measured with
the dimensionless Weissenberg number, $\text{Wi}=\dot{\gamma}\tau$,
where $\tau=\left(\mu k\right)^{-1}$ is the timescale of the particle's
relaxation to its anchor location due to the harmonic spring. Theoretical
results are given for three dimensions in Ref. \cite{holzer2010dynamics},
but the analysis also holds in two dimensions with the appropriate
diffusion coefficient. We perform the single-particle tests in two
dimensions and the two-particle tests in three dimensions.

The simulation parameters are given in Table \ref{tab:ShearParameters1}.
The strength of the spring is such that the equilibrium Gaussian distribution
for $\norm{\V q-\V q_{0}}$ has a standard deviation of $h$ (one
grid cell). We define the fluctuation $\tilde{x}=x-\av x$, where
$\av x$ is the average position, and similarly for $y$. The time
correlations $C_{xx}(t)=\av{\tilde{x}(t)\tilde{x}(0)}$, $C_{yy}(t)=\av{\tilde{y}(t)\tilde{y}(0)},$
and $C_{xy}(t)=\av{\tilde{x}(t)\tilde{y}(0)}$ are then calculated
and compared to the known theoretical results \cite{holzer2010dynamics},
\begin{eqnarray*}
\hat{C}_{xx}(t) & := & \frac{k_{B}T}{k}\left[1+\frac{Wi^{2}}{2}\left(1+\frac{|t|}{\tau}\right)\right]e^{-|t|/\tau}\\
\hat{C}_{xy}(t) & := & \frac{k_{B}T}{k}\frac{Wi}{2}\left(1+2\frac{t}{\tau}H(t)\right)e^{-t/\tau},
\end{eqnarray*}
where $H(t)$ is the Heaviside function. The time correlations in
the direction of shear, $C_{yy}(t)$, are not influenced by the background
flow and are omitted. It can be seen in Fig. \ref{fig:HarmonicShear}
that the simulation results converge to the correct time correlations
as the time step size is reduced, which confirms that the FIB method's
accurately captures the dynamics of an immersed particle subject to
external forcing and flow.

Significantly more accurate time correlation functions can be obtained
by using the improved midpoint scheme (\ref{eq:SecondOrderDeterministic}).
Because the equations of motion of a single blob in shear flow are
additive-noise equations, the improved midpoint scheme is second-order
accurate. This is confirmed in Fig. \ref{fig:HarmonicShear} where
we see that the second-order scheme is able to obtain the same accuracy
as the first-order scheme with a time step that is an order of magnitude
larger.

\subsubsection{Two Particles}

In the previous section, we tested the ability of our algorithm to
reproduce the dynamics of a single particle. We now test the ability
of our approach also correctly capture the hydrodynamic interactions
between particles. We extend the previous simulation to include two
particles, each in its own harmonic potential with minima separated
by vector of length $b$ in the direction of flow, $U(\V q)=(k/2)\norm{\V q-\V q_{\text{min}}}^{2}$,
where $\V q_{\text{min}}$ are the positions of the minima of the
two harmonic wells. The shear flow used is the same as in the previous
section. In this simulation we study the correlations between the
motion of particle 1 and particle 2 (cross-correlations), as well
as correlations of particle 1 with itself (self-correlations). The
self-correlations are different from the single-particle case due
to the disturbances in the fluid caused by the presence of the second
particle.

Theoretical results are calculated in Ref. \cite{bammert2010dynamics}
for an infinite domain by linearizing the equations around the equilibrium
location of the particles $\bar{\V q}$ (which is in general different
from $\V q_{\text{min}}$) and forming equations of motion for the
fluctuations $\tilde{\V q}=\V q-\bar{\V q}$ under the assumption
that $\tilde{\V q}$ is small. This leads to a simple Ornstein-Uhlenbeck
process \cite{bammert2010dynamics} 
\begin{equation}
\frac{d\tilde{\V q}}{dt}=\M A\tilde{\V q}-k\Mob(\bar{\V q})\tilde{\V q}+\M B\tilde{\V q}+\Mob^{\frac{1}{2}}(\bar{\V q})\M{\mathcal{W}}\left(t\right),\label{eq:BD_linearized}
\end{equation}
where $\M A$ is the shear rate tensor, $A_{12}=A_{45}=\dot{\gamma}$
and all other entries are zero, and $\M B_{ij}=\partial_{j}\M M_{ik}(\bar{\V q})\left(\V q_{\text{min}}^{(k)}-\bar{\V q}^{(k)}\right)$.
Because we have chosen to have the shear flow in the direction that
separates the wells, we have that $\bar{\V q}=\V q_{\text{min}}$
if we choose $\bar{\V q}_{1}^{(2)}=\bar{\V q}_{2}^{(2)}=0$, and therefore
$\M B=\M 0$ and no derivatives of the mobility are required. Since
we employ periodic boundary conditions for the velocity in our simulations,
we approximate $\Mob(\bar{\V q})$ using a periodic correction to
the Rotne-Prager-Yamakawa tensor calculated with an Ewald sum \cite{RotnePrager_Periodic}
and evaluated at position $\bar{\V q}$.

\begin{figure*}[!p]
\begin{centering}
\includegraphics[width=0.45\textwidth]{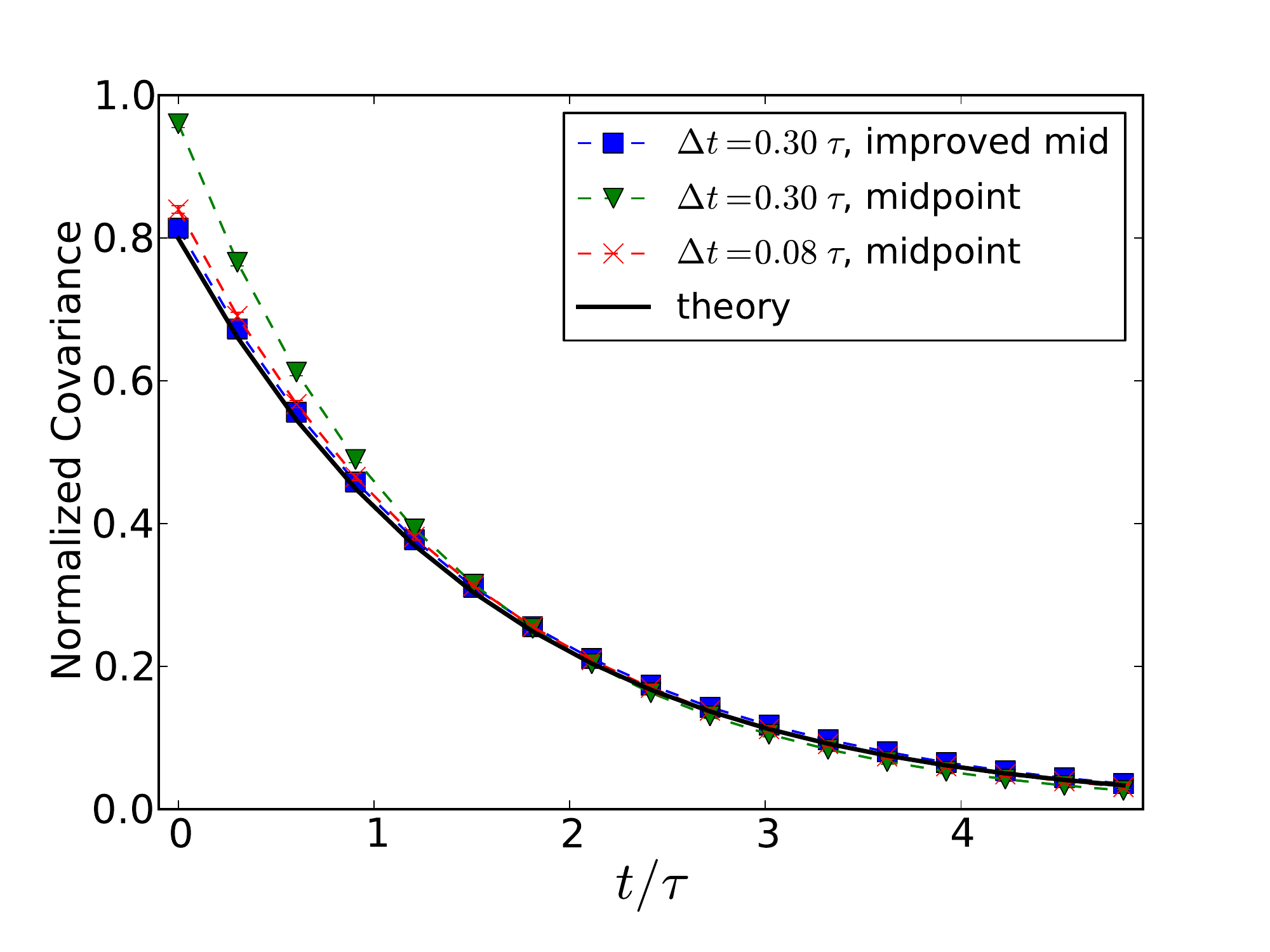}\includegraphics[width=0.45\textwidth]{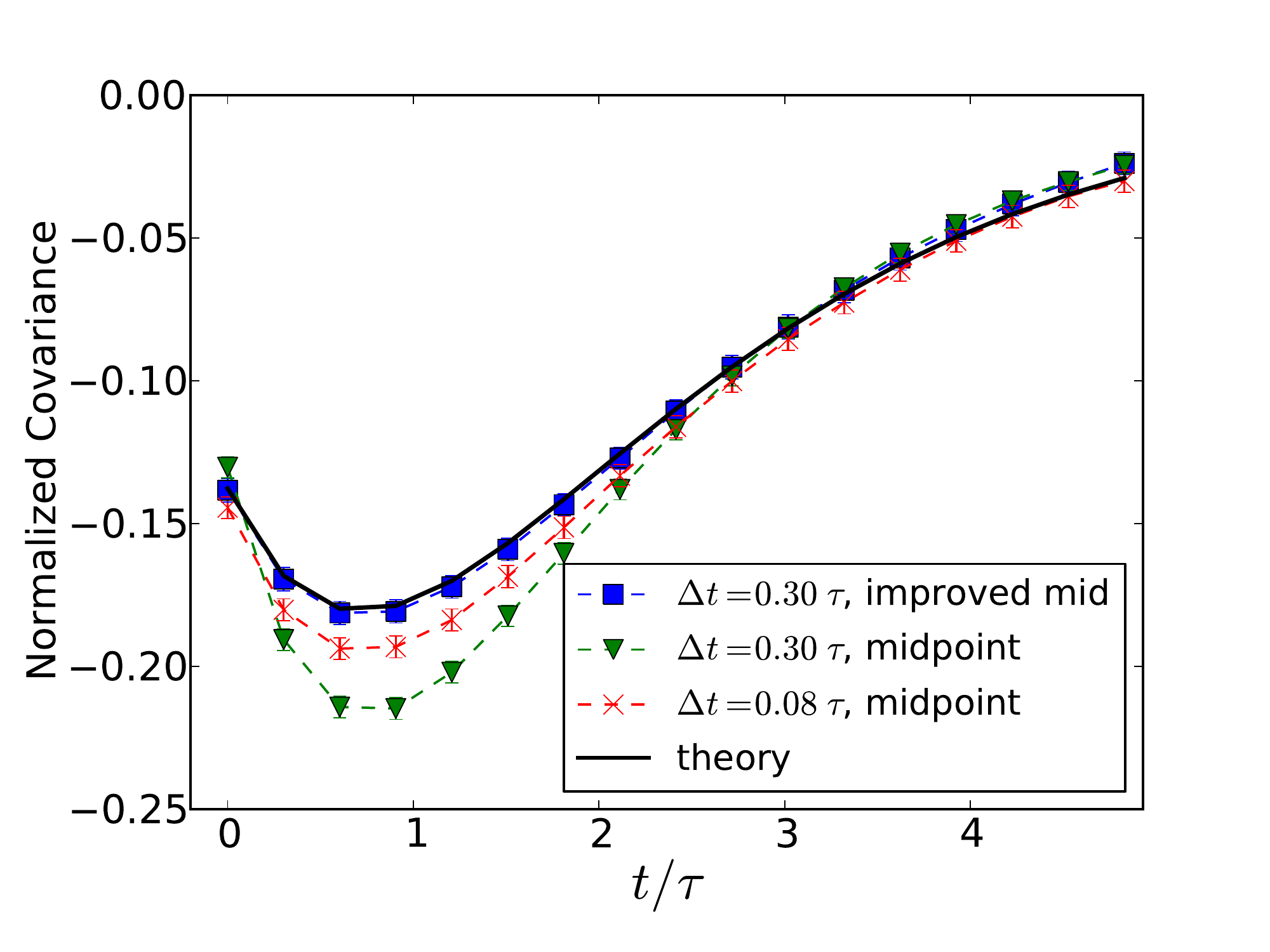}
\par\end{centering}

\caption{\label{fig:TwoShear_1}Time correlation functions for two particles
bound by harmonic potentials in the presence of shear flow. \textit{(Left
panel)} Autocorrelation of the displacement of one of the particles
in the direction of flow, $\av{\tilde{x}_{1}(t)\tilde{x}_{1}(0)}$.
\textit{(Right panel)} Correlation of the displacements of the two
particles in the direction of flow, $\av{\tilde{x}_{1}(t)\tilde{x}_{2}(0)}$. }
\end{figure*}

\begin{figure*}[!p]
\centering{}\includegraphics[width=0.45\textwidth]{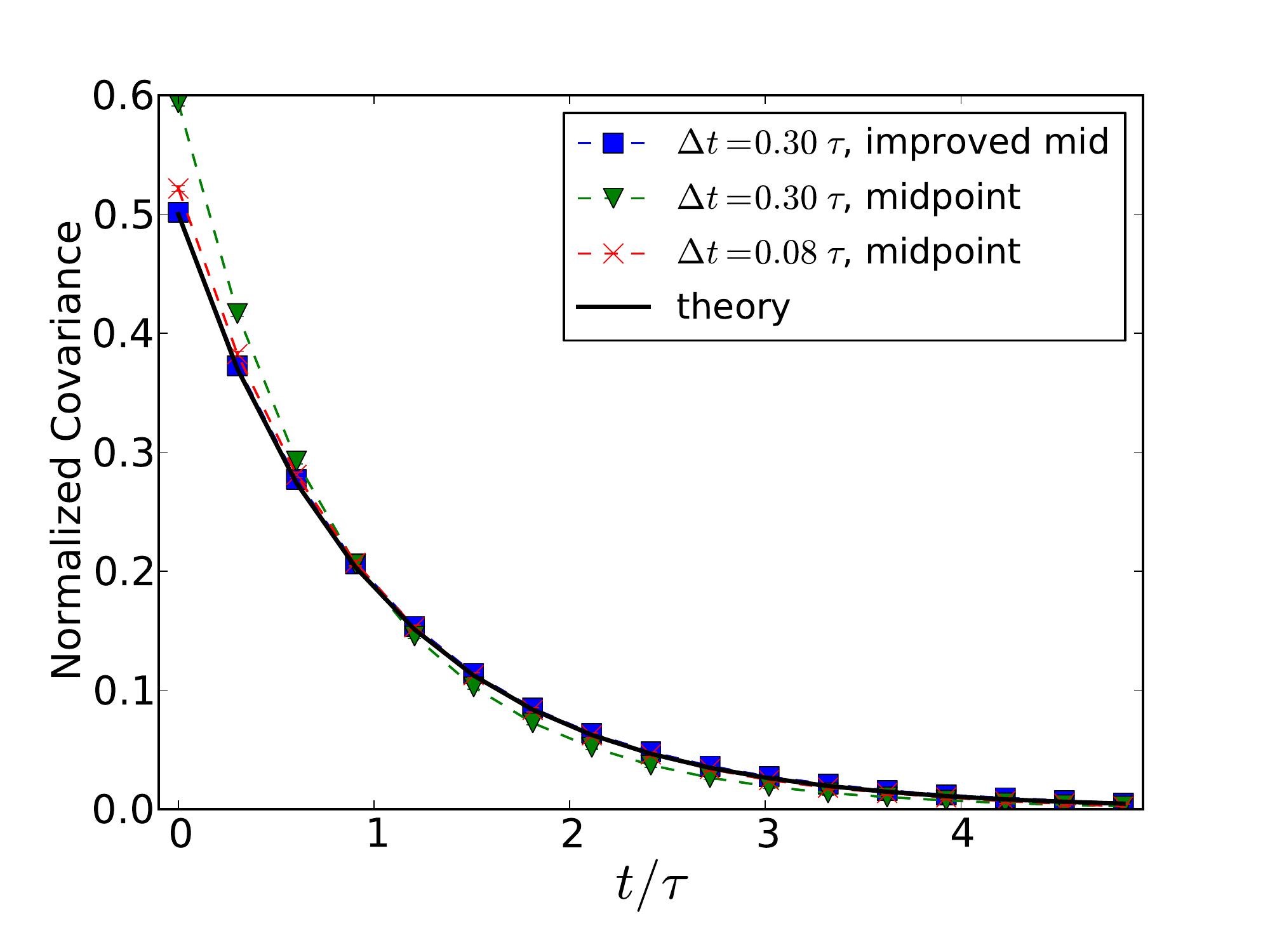}\includegraphics[width=0.45\textwidth]{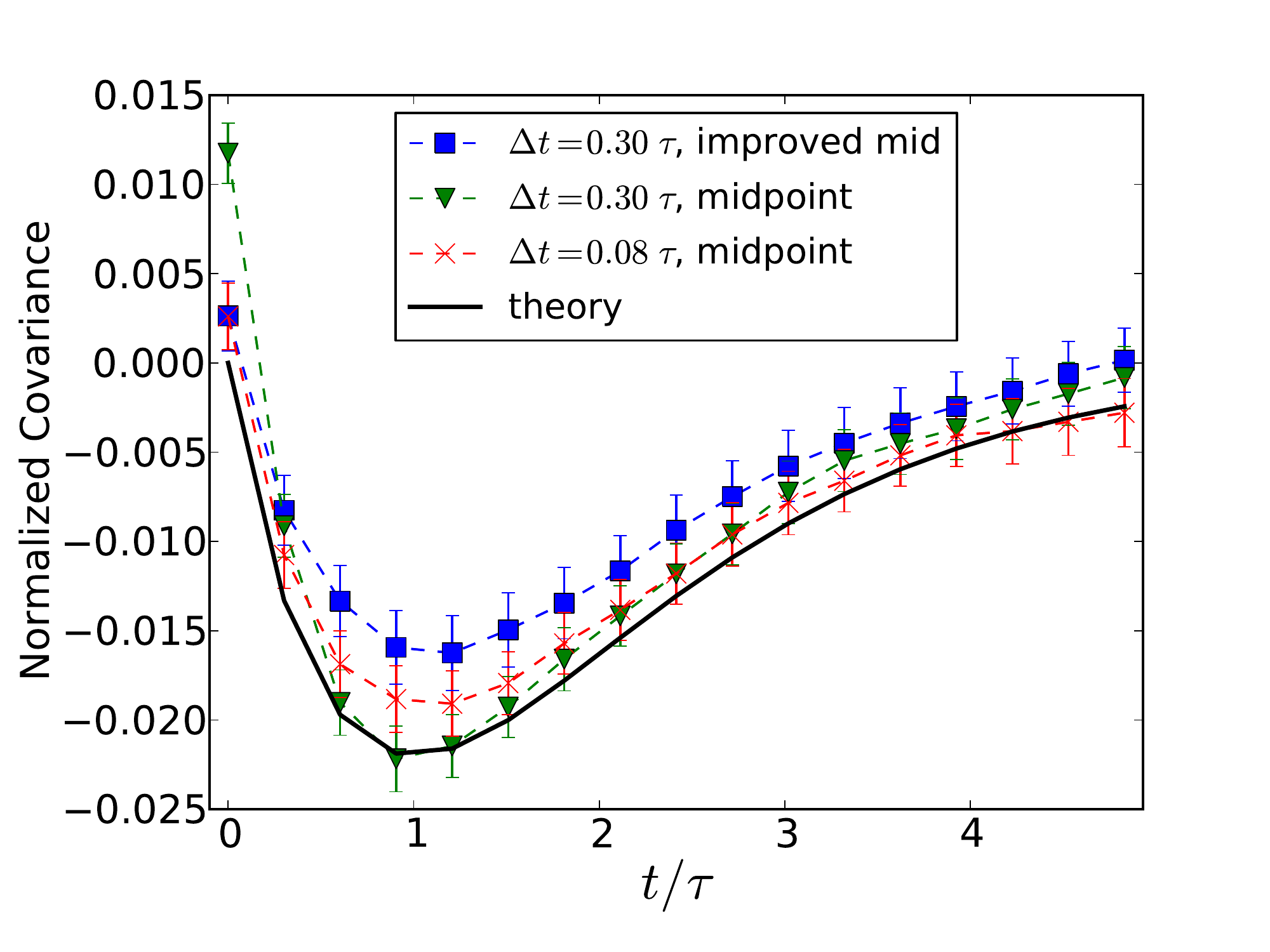}\caption{\label{fig:TwoShear_2}\textit{(Left panel)} Autocorrelation of the
displacement of one of the particles in the direction of shear $\av{\tilde{y}_{1}(t)\tilde{y}_{1}(0)}$.
\textit{(Right panel)} Correlation of the displacements of the two
particles in the direction of the shear, $\av{\tilde{y}_{1}(t)\tilde{y}_{2}(0)}$.}
\end{figure*}

\begin{figure*}[!p]
\centering{}\includegraphics[width=0.45\textwidth]{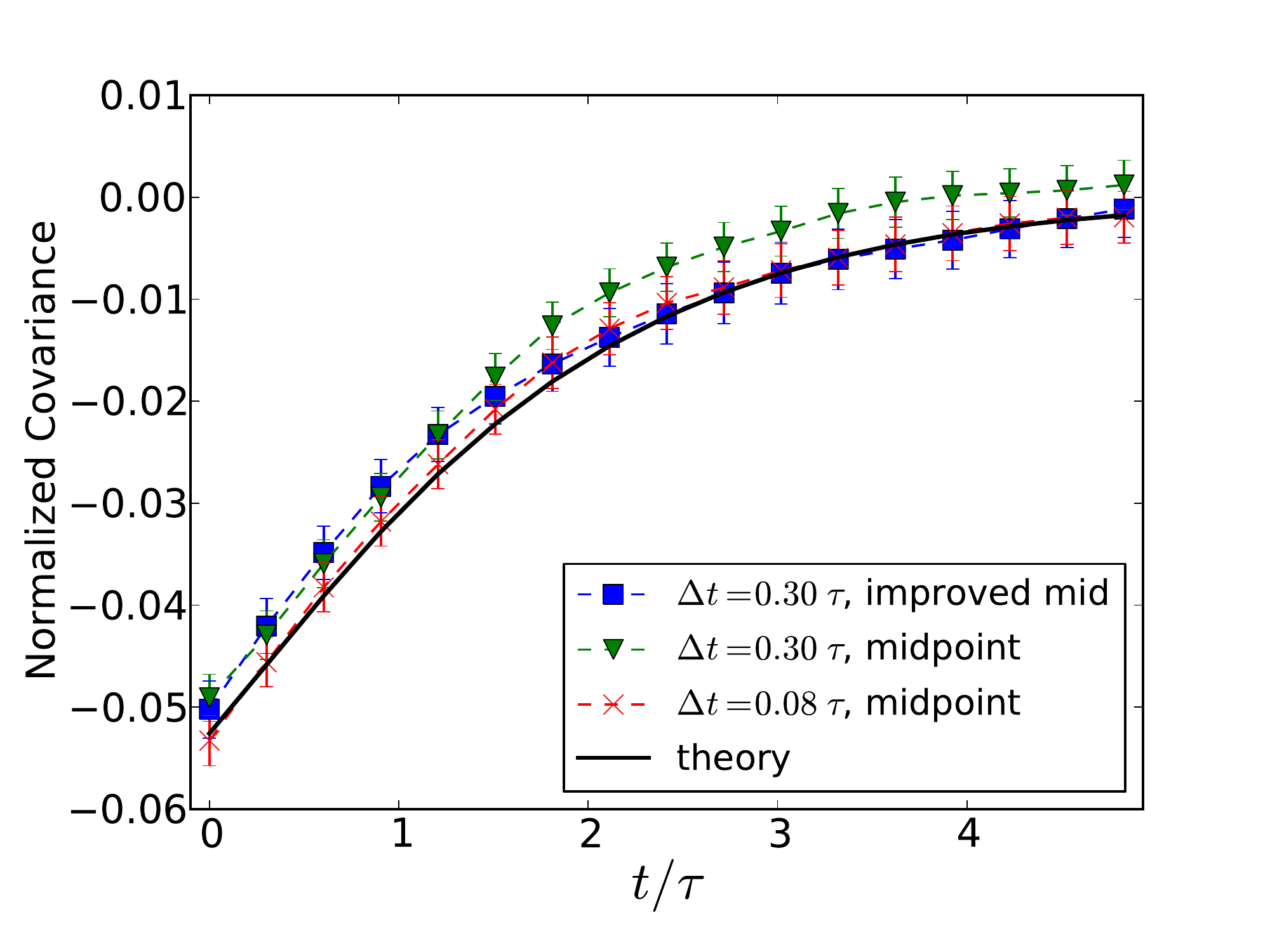}\includegraphics[width=0.45\textwidth]{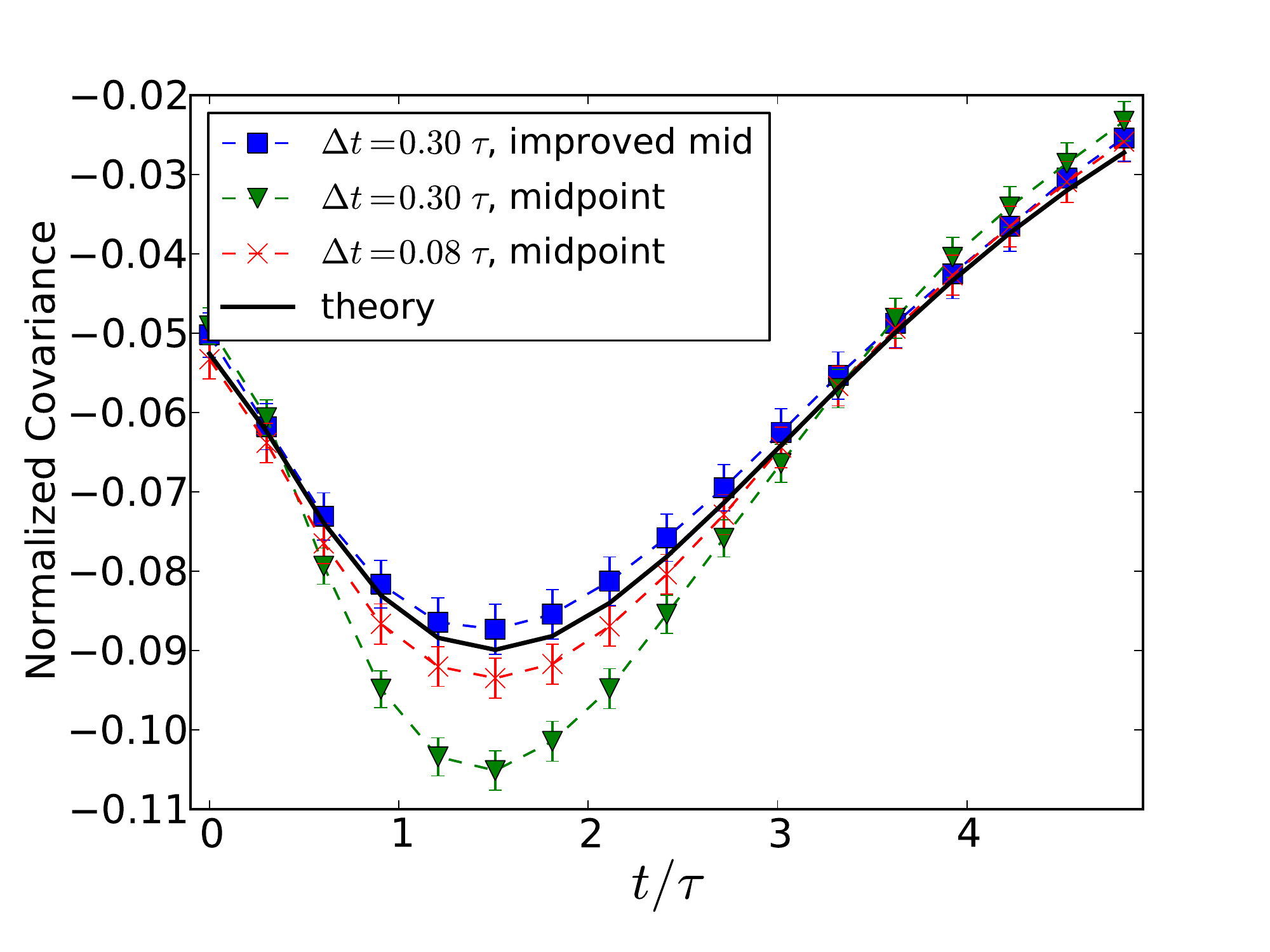}\caption{\label{fig:TwoShear_3}Cross correlation of the displacements of the
two particles in the shear and flow directions, $\av{\tilde{y}_{2}(t)\tilde{x}_{1}(0)}$
(left panel) and $\av{\tilde{x}_{1}(t)\tilde{y}_{2}(0)}=\av{\tilde{y}_{2}(-t)\tilde{x}_{1}(0)}$
(right panel).}
\end{figure*}

The simulation was run using a periodic three dimensional domain,
and the temperature was set such that the standard deviation of the
particles' displacements was $\sqrt{10}h$, keeping the particles
near the potential minima and thus giving better agreement with the
linearized theoretical calculations. The simulation parameters are
given in Table \ref{tab:ShearParameters2}. The numerical time correlation
functions shown in Figs. \ref{fig:TwoShear_1}, \ref{fig:TwoShear_2}
and \ref{fig:TwoShear_3} are in good agreement with the theoretical
results for the moderate time step size. The error is improved as
the time step size is decreased to better resolve the relevant timescale.
Note that the improved midpoint scheme (\ref{eq:SecondOrderDeterministic})
gives better agreement with theoretical results, as it is second-order
accurate for this example because the equations of motion are essentially
linear. Note that a visible mismatch with the theoretical curve is
seen for the cross-correlation $\av{\tilde{y}_{1}(t)\tilde{y}_{2}(0)}$
in the right panel of Fig. \ref{fig:TwoShear_2}; since the two midpoint
schemes are in agreement with each other this mismatch comes from
the approximations made in the theory.

\subsection{\label{sub:ColloidalGellation}Colloidal Gelation}

In this section, we confirm that the FIB method correctly reproduces
the dynamical effect of multi-particle hydrodynamic interactions for
a collection of colloidal particles interacting via excluded-volume
(non-bonded) interactions with an attractive tail. It has been demonstrated
that hydrodynamic interactions play a significant role in the process
of colloidal gelation \cite{ColloidalGelation}. Here we use the FIB
method to study a model test example of colloidal cluster dynamics,
and compare the FIB results to those of traditional Brownian Dynamics
(with hydrodynamic interactions).

As a simple test problem illustrating the effect of hydrodynamics
on gelation, a 13-particle colloidal cluster collapse example has
been constructed in Ref. \cite{ColloidalGelation}. The physical system
consists of 13 blobs initially placed at the vertices of an icosahedron
(see Fig. 4 in Ref. \cite{ColloidalGelation}), and then released
to relax toward the thermodynamically-preferred collapsed (bound)
cluster of 13 spheres. In the absence of hydrodynamic interactions
the collapse is rapid. In the presence of hydrodynamic interactions,
however, the cluster undergoes a slow rearrangement process through
multiple elongated configurations (see Fig. 4 in Ref. \cite{ColloidalGelation})
before it collapses. This results in a dramatic slowing down of the
collapse when hydrodynamics is accounted for.

\begin{figure}
\begin{centering}
\includegraphics[width=1\columnwidth]{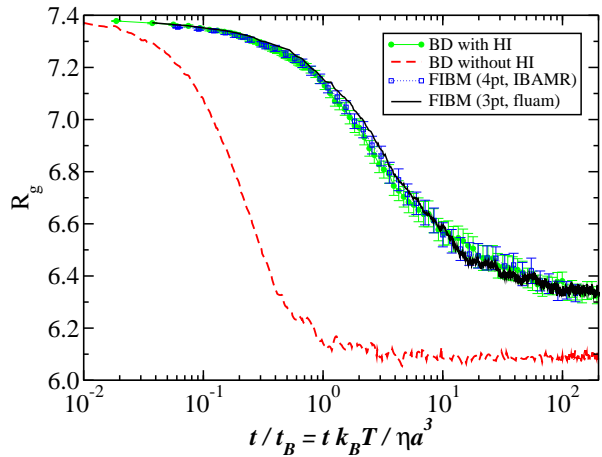}
\par\end{centering}

\caption{\label{fig:Colloidal}Relaxation of the radius of gyration of a colloidal
cluster of 13 spheres toward equilibrium, as obtained by averaging
64 independent simulations. The FIB method is compared to traditional
Brownian Dynamics (BD) with and without hydrodynamic interactions
(HI). For comparison, FIB simulations were performed both using the
\emph{fluam} code with the three-point kernel and a domain of $32^{3}$
grid cells, as well as using the IBAMR code with the four-point kernel
and $64^{3}$ grid cells.}
\end{figure}

The collapse of the cluster can be monitored via the radius of gyration
of the cluster $R_{g}(t)$. An ensemble average $\av{R_{g}(t)}$ over
64 trajectories obtained using the FIB method is shown in Fig. \ref{fig:Colloidal}.
In the first set of simulations, we employ periodic boundary conditions
with a grid of $32^{3}$ cells and use the GPU-based code \emph{fluam}
with the three-point Peskin kernel \cite{ISIBM} and the simple midpoint
integrator. The second set of simulations were performed using IBAMR
on a periodic grid of $64^{3}$ cells with the four-point kernel and
the improved midpoint integrator.\textbf{ }We use the Asakura-Oosawa
depletion force with a repulsive Lennard-Jones interaction, following
Ref. \cite{ColloidalGelation}. Important parameters of our simulations
are summarized in Table \ref{tab:ClusterCollapse}. 

\begin{table}
\centering{}%
\begin{tabular}{|c|c|}
\hline 
grid spacing $\D x$  &
$3.27$\tabularnewline
\hline 
grid size &
$32^{3}$\tabularnewline
\hline 
shear viscosity $\eta$  &
1 \tabularnewline
\hline 
time step size $\Delta t$  &
$0.05$ (simple) or $0.1$ (improved)\tabularnewline
\hline 
temperature $k_{B}T$  &
$12.3$ \tabularnewline
\hline 
LJ strength $\epsilon$  &
$10$ \tabularnewline
\hline 
LJ / hydro diameter $\sigma$  &
$6.4$\tabularnewline
\hline 
number of particles $N$ &
13\tabularnewline
\hline 
\end{tabular}\caption{\label{tab:ClusterCollapse}Parameters used in the colloidal cluster
collapse simulations shown in the right panel of Fig. \ref{fig:RDF}.
These are chosen to match those in Ref. \cite{ColloidalGelation}
as closely as possible.}
\end{table}

It is important to note that the time step size used for these simulations
is much smaller than the Brownian time scale $\tau_{B}=a^{2}/\chi\approx50$.
This is because the time step size here is severely limited by stability
considerations. The stiff hard-core repulsion between the particles
and the fact that the particles are close to each other due to the
attractive tail combine to make the simple midpoint scheme unstable
for $\D t>0.05$ (determined empirically). The improved midpoint scheme
shows slightly improved stability and we have successfully used it
for $\D t=0.1$, however, the cost per time step is approximately
doubled so this improvement is not substantial. Achieving larger time
step sizes and avoiding exploding (unstable) trajectories requires
specialized temporal integration methods such as Metropolization%
\footnote{Note, however, that Metropolization of even the simple midpoint scheme
is a rather nontrivial task because the mobility matrix is never formed
or factorized in FIB.%
} \cite{MetropolizedBD}. Note that in a small fraction of the trajectories
(we only observed two such trajectories) the cluster dissolves instead
of collapsing. This could be the signature of a rare event but it
could also be an artifact of numerical instabilities arising from
the stiff interparticle potentials; lacking better statistics we have
excluded these trajectories from the averages.

In Ref. \cite{ColloidalGelation}, the authors compare their method
to BD without hydrodynamic interactions (HI) (i.e., employ a mobility
that it a diagonal matrix), but do not compare to BD with hydrodynamic
interactions. In the right panel of Fig. \ref{fig:Colloidal} we compare
the results from the FIB method to BD with and without HI. We included
hydrodynamics using the free-space Rotne-Prager-Yamakawa (RPY) mobility
(\ref{eq:RPYTensor}), and employed a simple Euler-Maruyama integrator
with time step size $\D t=0.05$ instead of the Fixman method since
the divergence of the free-space RPY mobility vanishes identically.
The results in Fig. \ref{fig:Colloidal} demonstrate that both BD-with
HI and the FIB method reproduce the slowing down (relative to BD without
HI) in the cluster collapse and agree with each other. While inclusion
of higher-order effects such as stresslets and lubrication , may lead
to some quantitative differences, our results are already in good
agreement with those in Fig. 4 in Ref. \cite{ColloidalGelation} and
indicate that the primary effect comes from the far-field hydrodynamic
interactions.

\section{Conclusions}

We have developed a method for performing Brownian Dynamics (BD) with
hydrodynamic interactions in confined geometries such as slit or square
channels or chambers. Unlike traditional methods for BD, our FIB method
does not rely on analytical Green's functions, and only requires the
numerical solution of a single steady Stokes system per time step
to capture both the deterministic, stochastic, and thermal drift contributions
to the overdamped dynamics of the hydrodynamically-coupled particles.
The FIB method is particularly appealing when dealing with more complex
boundary conditions such as confined flows in non-trivial channel
geometries, since analytical solutions are quite involved and ensuring
a positive semi-definite mobility is nontrivial, even in the presence
of only a single no-slip planar wall \cite{StokesianDynamics_Wall,BD_LB_Ladd}.
Computing analytical solutions in cases where there are osmophoretic
flows at the boundaries, as in active suspensions of particles \cite{Hematites_Science},
is essentially impossible because the boundary condition itself comes
from the solution of another nontrivial reaction-diffusion problem.
The only alternative would be to use relatively-expensive and complex
boundary integral methods \cite{Stokes2D_Greengard,Stokes3D_Greengard,StokesSecondKind_Shelley},
none of which, to our knowledge, include the effects of thermal fluctuations.

Following the completion of this work we learned about a related recent
extension of the SELM approach to use a (P1-MINI) finite-element Stokes
solver to generate the hydrodynamic response \cite{SELM_FEM}, very
similar to the approach we independently took in this work. Note,
however, that our temporal integrators are different from the Euler-Maruyama
scheme used in Ref. \cite{SELM_FEM}, which requires calculating the
divergence of the mobility by other means (We remark that the issue
of computing the divergence of the mobility does not appear to be
addressed directly in Ref. \cite{SELM_FEM}.). In terms of spatial
discretizations, the key relation $-\L^{-1}\M L\L^{-1}=\L^{-1}$ is
used in both works to generate the correct stochastic increments by
simply solving the saddle-point steady Stokes problem. A key difference,
however, is that on the structured MAC grid used in this work the
generation of a stochastic stress tensor with covariance $\sim-\M L$
is straightforward \cite{LLNS_Staggered}, where as accomplishing
the same for unstructured FEM grids appears to require an iterative
stochastic multigrid method \cite{SELM_FEM}. Furthermore, the P1-MINI
discretization is only first-order spatially accurate and requires
more degrees of freedom (DOF) per cell, where as the structured staggered
(MAC) grid (which can be thought of as a particular FEM discretization)
achieves second-order spatial accuracy with only a single velocity
DOF per grid face and a single pressure DOF per cell center. This
makes the methods developed here particularly attractive, due to their
simplicity and efficiency, in simple confined geometries such as channels
or chambers. At the same time, unstructured FEM discretizations have
a notable advantage for complex geometries. Additionally, achieving
variable spatial resolution is natural on unstructured grids \cite{SELM_FEM}
but requires (block structured) adaptive mesh refinement (AMR) techniques
\cite{IBAMR} on structured grids. Adaptive resolution is very important
at low densities of suspended particles to avoid using a fine spatial
grid to resolve long-ranged hydrodynamics; this is in fact a key advantage
of using Green's functions instead of numerical solvers. In future
work we will consider solving the fluctuating Stokes equations on
block-structured refined staggered grids.

The FIB method presented here and related methods \cite{ForceCoupling_Fluctuations,SELM_FEM}
are only a first step toward the ultimate goal of performing Brownian
(i.e., overdamped) dynamics for a collection of rigid and flexible
bodies in flow in the presence of complex boundaries. Achieving that
goal may ultimately require a combination of techniques, such as multipole
series, immersed boundary \cite{IBM_Viscoelasticity} and immersed
finite-element \cite{ImmersedFEM_Patankar}, or boundary integral
representations for the suspended structures, together with cut cell
(embedded boundary) or finite-element methods \cite{SELM_FEM} for
representing the complex geometry. What our work makes evident is
that thermal fluctuations are most easily and consistently included
by using fluctuating hydrodynamics combined with appropriate multiscale
temporal integrators. This illustrates the power of a bottom-up approach
in which one starts with the fundamental formulation of the fluid
dynamics of suspensions \cite{VACF_FluctHydro,Faxen_FluctuatingHydro,VACF_Langevin}
and then coarse-grains in space and time to reach larger length scales
and longer time scales, instead of starting at the top from a formulation
of the equations of motion that contains difficult-to-calculate objects
such as multi-body mobility or resistance tensors that hide all of
the coarse-grained information inside them.
\begin{acknowledgments}
We thank Eric Vanden-Eijnden and Anthony Ladd for informative discussions,
and Ranojoy Adhikari and Eric Keaveny for helpful comments on the
manuscript. S. Delong was supported by the DOE office of Advanced
Scientific Computing Research under grant DE-FG02-88ER25053. A. Donev
was supported in part by the Air Force Office of Scientific Research
under grant number FA9550-12-1-0356. B. Griffith acknowledges research
support from the National Science Foundation under awards OCI 1047734
and DMS 1016554. R. Delgado-Buscalioni and F. Balboa acknowledge funding
from the Spanish government FIS2010-22047-C05 and from the Comunidad
de Madrid MODELICO-CM (S2009/ESP-1691). Collaboration between A. Donev
and R. Delgado-Buscalioni was fostered at the Kavli Institute for
Theoretical Physics in Santa Barbara, California, and supported in
part by the National Science Foundation under Grant No. NSF PHY05-51164.
\end{acknowledgments}
\appendix

\section*{Appendix}

\section{\label{sec:WeakTemporalAccuracy}Weak Temporal Accuracy}

In this Appendix we show that the algorithms outlined in Section \ref{sec:TemporalIntegrator}
are first order weakly accurate temporal integrators for the system
(\ref{eq:OverdampedEq}). It suffices to show that the first three
moments of the numerical one-step increment in time match to first
order the moments of the exact increment \cite{MilsteinSDEBook}.
Without loss of generality, we consider the case $\eta=1$ for this
analysis. The RFD term (\ref{eq:RFDTerm}) introduces an error proportional
to $\delta^{2}$. This is a spatial truncation error and will be ignored
in the context of temporal accuracy. Note that in practice, $\delta$
will be a very small fixed value that introduces a negligible truncation
error to the approximation of the thermal drift. 

For the continuous equation, we have, to first order,
\begin{eqnarray}
\Delta_{\alpha}^{q} & \equiv & q_{\alpha}((n+1)\Delta t)-q_{\alpha}(n\Delta t)\label{eq:dq_th}\\
 & = & \Delta t(\,\mathcal{J}_{\alpha\mu}^{n}\mathcal{L}_{\mu\nu}^{-1}\mathcal{S}_{\nu\beta}^{n})F_{\beta}^{n}\nonumber \\
 & + & \sqrt{2k_{B}T}\int_{t'}^{t}\mathcal{J}_{\alpha\mu}^{n}\mathcal{L}_{\mu\nu}^{-1}\widetilde{D}_{\nu\beta}d\mathcal{W}_{\beta}\nonumber \\
 & + & \Delta t\: k_{B}T\partial_{\gamma}\cdot(\mathcal{J}_{\alpha\mu}^{n}\mathcal{L}_{\mu\nu}^{-1}\mathcal{S}_{\nu\gamma}^{n})+O\left(\D t^{\frac{3}{2}}\right),\nonumber 
\end{eqnarray}
where $\J$, $\S$ and $\V F$ are evaluated at the beginning of the
time step. The first moment of the true increment is
\begin{eqnarray*}
E[\Delta_{\alpha}^{q}] & \equiv & \Delta t\,\mathcal{J}_{\alpha\mu}^{n}\mathcal{L}_{\mu\nu}^{-1}\mathcal{S}_{\nu\beta}^{n}F_{\beta}^{n}+O\left(\D t^{2}\right)\\
 & + & \D t\: k_{B}T\,\Biggl[\:\mathcal{J}_{\alpha\mu}^{n}\mathcal{L}_{\mu\nu}^{-1}\partial_{\gamma}(\mathcal{S}_{\nu\gamma}^{n})+\partial_{\gamma}(\mathcal{J}_{\alpha\mu}^{n})\mathcal{L}_{\mu\nu}^{-1}\mathcal{S}_{\nu\gamma}^{n}\Biggr].
\end{eqnarray*}

\subsection{First Order Midpoint Scheme}

Looking at the discrete increment to first order, we get,\begin{widetext}
\begin{eqnarray*}
\bar{\Delta}_{\alpha}^{q} & \equiv & q_{\alpha}^{n+1}-q_{\alpha}^{n}=\Delta t\,\mathcal{J}_{\alpha\mu}^{n}\mathcal{L}_{\mu\nu}^{-1}\mathcal{S}_{\nu\beta}^{n}F_{\beta}^{n}\\
 & + & \Delta t\,\mathcal{J}_{\alpha\mu}^{n}\mathcal{L}_{\mu\nu}^{-1}\frac{k_{B}T}{\delta}[\mathcal{S}_{\nu\beta}(\V q^{n}+\frac{\delta}{2}\widetilde{\V W}^{n})\widetilde{W}_{\beta}^{n}-\mathcal{S}_{\nu\beta}(\V q^{n}-\frac{\delta}{2}\widetilde{\V W}^{n})\widetilde{W}_{\beta}^{n}]\\
 & + & (\Delta t)^{\frac{1}{2}}\left(\frac{k_{B}T}{2\D V}\right)^{\frac{1}{2}}\partial_{\gamma}(\mathcal{J}_{\alpha\mu}^{n})(q_{\gamma}^{n+\frac{1}{2}}-q_{\gamma}^{n})\mathcal{L}_{\mu\nu}^{-1}\widetilde{D}_{\nu\beta}W_{\beta}^{n}\\
 & + & \sqrt{\frac{2k_{B}T\Delta t}{\D V}}\mathcal{J}_{\alpha\mu}\mathcal{L}_{\mu\nu}^{-1}\widetilde{D}_{\nu\beta}W_{\beta}^{n}+O\left(\D t^{\frac{3}{2}}\right).
\end{eqnarray*}
\end{widetext}Inserting the expression for the predictor increment
$\V q^{n+\frac{1}{2}}-\V q^{n}$, simplifying and ignoring terms of
order $\Delta t^{2}$, along with terms of order $\Delta t^{\frac{3}{2}}$
with zero expectation, and terms of order $\delta^{2}$, we obtain
\begin{eqnarray*}
\bar{\Delta}_{\alpha}^{q} & = & \Delta t\,\mathcal{J}_{\alpha\mu}^{n}\mathcal{L}_{\mu\nu}^{-1}\mathcal{S}_{\nu\beta}^{n}F_{\beta}^{n}\\
 & + & \Delta t\, k_{B}T\:\mathcal{J}_{\alpha\mu}^{n}\mathcal{L}_{\mu\nu}^{-1}\partial_{\gamma}(\mathcal{S}_{\nu\beta}^{n})\widetilde{W}_{\gamma}^{n}\widetilde{W}_{\beta}^{n}\\
 & + & \frac{\Delta t\, k_{B}T}{\D V}\partial_{\gamma}(\mathcal{J}_{\alpha\mu}^{n})\left(\mathcal{J}_{\gamma\epsilon}\mathcal{L}_{\epsilon\zeta}^{-1}\widetilde{D}_{\zeta\eta}W_{\eta}^{n}\right)\mathcal{L}_{\mu\nu}^{-1}\widetilde{D}_{\nu\beta}W_{\beta}^{n}\\
 & + & \sqrt{\frac{2k_{B}T\Delta t}{\D V}}\mathcal{J}_{\alpha\mu}\mathcal{L}_{\mu\nu}^{-1}\widetilde{D}_{\nu\beta}W_{\beta}^{n}+O\left(\D t^{\frac{3}{2}}\right).
\end{eqnarray*}
The first moment of this increment is obtained by using the adjoint
relation $\mathcal{J}_{\gamma\epsilon}=\mathcal{S}_{\epsilon\gamma}\D V$,
as well as $\mathcal{L}_{\epsilon\zeta}^{-1}\widetilde{D}_{\zeta\eta}\mathcal{L}_{\mu\nu}^{-1}\widetilde{D}_{\nu\beta}\av{W_{\eta}^{n}W_{\beta}^{n}}=\mathcal{L}_{\epsilon\zeta}^{-1}\widetilde{D}_{\zeta\eta}\widetilde{D}_{\nu\eta}\mathcal{L}_{\mu\nu}^{-1}=\mathcal{L}_{\mu\epsilon}^{-1}$
by virtue of (\ref{eq:LLL}), 
\begin{eqnarray*}
E[\bar{\Delta}_{\alpha}^{q}] & = & \Delta t\,\mathcal{J}_{\alpha\mu}^{n}\mathcal{L}_{\mu\nu}^{-1}\mathcal{S}_{\nu\beta}^{n}F_{\beta}^{n}+O\left(\Delta t^{2}\right)\\
 & + & \Delta t\: k_{B}T\,\Biggl[\:\mathcal{J}_{\alpha\mu}^{n}\mathcal{L}_{\mu\nu}^{-1}\partial_{\gamma}(\mathcal{S}_{\nu\gamma}^{n})+\partial_{\gamma}(\mathcal{J}_{\alpha\mu}^{n})\mathcal{L}_{\mu\epsilon}^{-1}\mathcal{S}_{\epsilon\gamma}^{n}\Biggr],
\end{eqnarray*}
which matches the $O\left(\Delta t\right)$ terms in the continuous
increment (\ref{eq:dq_th}).

The second moment of the discrete increment,
\begin{eqnarray*}
E[\bar{\Delta}_{\alpha}^{q}\bar{\Delta}_{\eta}^{q}] & = & 2\Delta t\, k_{B}T\:\mathcal{J}_{\alpha\mu}\mathcal{L}_{\mu\nu}^{-1}\mathcal{\mathcal{S}}_{\nu\eta}+O(\Delta t^{2}),
\end{eqnarray*}
also matches the continuous second moment to second order. Finally
the third moments are both $O(\Delta t^{2})$, because the order $\Delta t^{\frac{3}{2}}$
terms are mean zero.

\subsection{Improved Midpoint Scheme}

We show here that the scheme given by Eq. (\ref{eq:SecondOrderDeterministic})
is also first order weakly accurate. The discrete increment for this
scheme to first order is\begin{widetext}
\begin{eqnarray*}
\bar{\Delta}_{\alpha}^{q} & \equiv & q_{\alpha}^{n+1}-q_{\alpha}^{n}=\D t\,\mathcal{J}_{\alpha\mu}^{n}\mathcal{L}_{\mu\nu}^{-1}\left[\mathcal{S}_{\nu\beta}^{n}F_{\beta}^{n}+\sqrt{\frac{k_{B}T}{\D V\Delta t}}\widetilde{D}_{\nu\beta}\left(W_{\beta}^{n,1}+W_{\beta}^{n,2}\right)\right]\\
 & + & \sqrt{\frac{\D t\, k_{B}T}{\D V}}\partial_{\eta}\left(\mathcal{J}_{\alpha\mu}^{n}\right)\left(q_{\eta}^{n+\frac{1}{2}}-q_{\eta}^{n}\right)\mathcal{L}_{\mu\nu}^{-1}\widetilde{D}_{\nu\beta}\left(W_{\beta}^{n,1}+W_{\beta}^{n,2}\right)\\
 & + & \frac{\D t\, k_{B}T}{\delta}\:\mathcal{J}_{\alpha\mu}^{n}\mathcal{L}_{\mu\nu}^{-1}\left[\mathcal{\mathcal{S}}_{\nu\beta}(\V q^{n}+\frac{\delta}{2}\widetilde{\V W}^{n})\widetilde{W}_{\beta}^{n}-\mathcal{S}_{\nu\beta}(\V q^{n}-\frac{\delta}{2}\widetilde{\V W}^{n})\widetilde{W}_{\beta}^{n}\right]+O\left(\Delta t^{\frac{3}{2}}\right).
\end{eqnarray*}
\end{widetext}Inserting the expression for the predictor increment
and removing terms that are either $O(\D t^{2})$, $O(\D t^{\frac{3}{2}})$
with mean zero, or $O(\delta^{2})$, we get, \begin{widetext} 
\begin{eqnarray*}
\bar{\Delta}_{\alpha}^{q} & = & \Delta t\mathcal{J}_{\alpha\mu}^{n}\mathcal{L}_{\mu\nu}^{-1}\mathcal{S}_{\nu\beta}^{n}F_{\beta}^{n}\\
 & + & \frac{\D{t\,}k_{B}T}{\D V}\partial_{\eta}\left(\mathcal{J}_{\alpha\mu}^{n}\right)\left(\mathcal{J}_{\eta\kappa}^{n}\mathcal{L}_{\kappa\rho}^{-1}\widetilde{D}_{\rho\zeta}W_{\zeta}^{n,1}\right)\mathcal{L}_{\mu\nu}^{-1}\widetilde{D}_{\nu\beta}\left(W_{\beta}^{n,1}+W_{\beta}^{n,2}\right)\\
 & + & \Delta t\, k_{B}T\:\mathcal{J}_{\alpha\mu}^{n}\mathcal{L}_{\mu\nu}^{-1}\partial_{\beta}(\mathcal{S}_{\nu\beta}^{n})\\
 & + & \sqrt{\frac{\Delta t\, k_{B}t}{\D V}}\mathcal{J}_{\alpha\mu}^{n}\mathcal{L}_{\mu\nu}^{-1}\widetilde{D}_{\nu\beta}\left(W_{\beta}^{n,1}+W_{\beta}^{n,2}\right)+O\left(\D t^{\frac{3}{2}}\right).
\end{eqnarray*}
\end{widetext}The second and third terms on the right hand side of
this expression give us the thermal drift. The first moment is identical
to that of the simple scheme to $O\left(\Delta t\right)$, as is the
second moment. The third moments of both the discrete and continuous
terms are already each $O(\Delta t^{2})$. Note that for the special
case of additive noise, such as for example the linearized equation
(\ref{eq:BD_linearized}), the improved midpoint scheme can be shown
to match the first five moments of the true increment to $O\left(\Delta t^{2}\right)$,
and is thus weakly second-order accurate (see Appendix A1 in Ref.
\cite{DFDB}).

\section{\label{sec:Stresslets}Including Stresslet Terms}

As summarized in Section \ref{sub:FiniteSize} and discussed at length
in the work of Maxey and collaborators \cite{ForceCoupling_Monopole,ForceCoupling_Stokes,ForceCoupling_Channel},
the simple coupling between the fluid and the particles used here
correctly reproduces the hydrodynamic interactions between particles
only up to the Rotne-Prager level (including in the presence of boundaries).
This is because only the \emph{monopole} term (\emph{Stokeslet}) is
included in the fluid-particle force, along with the Faxen correction
for the resulting particle velocity. As a consequence the present
approach can only accurately resolve the fluid flow at distances larger
than the typical size of the particles. For traditional applications
of Brownian Dynamics such as polymeric fluids \cite{BrownianDynamics_DNA,BrownianDynamics_DNA2,BD_IBM_Graham,Tethered_Polymer}
this is probably sufficient, since polymer chains are themselves described
at a coarse-grained level and in reality they are not made of a collection
of rigid spheres linked with spheres or rods. For colloidal suspensions,
however, at higher packing densities one must include higher-order
multipole terms in order to more accurately capture the hydrodynamics,
as done in the method of Stokesian Dynamics \cite{BrownianDynamics_OrderNlogN}
and the improved Force Coupling Method (FCM) \cite{ForceCoupling_Stokes}.
This amounts to including the anti-symmetric component of the dipole
(\emph{rotlet}) and the symmetric components of the dipole (\emph{stresslet})
force terms. Note that here we do not discuss lubrication forces.

It is not difficult to extend our approach to also include the rotlet
contributions, as has been done by Keaveny \cite{ForceCoupling_Fluctuations}
in the fluctuating FCM method. Firstly, particle rotational degrees
of freedom would need to be added to the blob description, along with
an angular velocity $\V{\omega}_{i}$ for each blob. We would need
to impose an additional \emph{rotational no-slip} constraint, requiring
that the particle rotate with the locally-averaged angular velocity
of the fluid, $\V{\omega}=\J\left(\grad\times\V v\right)/2,$ and
distribute (spread) the torque $\V{\tau}$ applied on the particles
as a torque density $\M f_{\V{\tau}}=\grad\times\left(\S\V{\tau}\right)/2$
in the fluid momentum equation. This type of approach has already
been employed in deterministic immersed-boundary methods to model
suspensions of neutrally-buoyant semi-rigid rods \cite{IBM_Generalized,IBM_TwistBend}.
The FIB method and the discrete fluctuation-dissipation results we
presented continue to apply since inclusion of rotation simply amounts
to augmenting the local averaging and spreading operators. The main
difficulties are in developing a translationally-invariant and accurate
discretization for the $\text{curl}$ operator and its adjoint the$\text{rot}$
operator, including in the presence of boundaries.

Inclusion of the stresslet contributions, on the other hand, is not
trivial as it requires including an additional \emph{rigidity constraint}
on the locally-averaged deformation tensor, as proposed by Maxey and
collaborators in the context of the deterministic FCM \cite{ForceCoupling_Stokes}
and extended to account for thermal fluctuations by Keaveny \cite{ForceCoupling_Fluctuations}.
In this appendix we present the continuum formulation including stresslets,
and demonstrate that the random increments can easily be generated
by including the random stress in the Stokes equations. This has already
been observed and proven by Keaveny in the Appendices of Ref. \cite{ForceCoupling_Fluctuations};
here we present a simple proof using compact operator notation \cite{SELM}.
Importantly, we do not rely on periodic boundary conditions and Fourier
transform techniques, thus demonstrating that the power of the fluctuating
hydrodynamics approach to modeling Brownian motion in confined suspensions.

We will omit rotlet contributions here by focusing on the case when
there are no torques applied on the particles and assuming that the
particles are spherical, so that their orientation does not affect
the external or inter-particle forces. We can account for the rigidity
of the particle by including a constraint that the locally the rate
of strain of the fluid velocity vanish inside the particle, approximated
here in a spirit similar to the no-slip constraint (\ref{eq:ParticleVelocity})
\cite{ForceCoupling_Stokes},
\begin{equation}
\int\delta_{b}\left(\V q_{i}-\V r\right)\left[\grad\V v\left(\V r,t\right)+\grad^{T}\V v\left(\V r,t\right)\right]d\V r\equiv\left(\M{\mathcal{K}}\V v\right)_{i}=\V 0.\label{eq:stresslet}
\end{equation}
Here $\M{\mathcal{K}}\left(\V q\right)$ is linear integro-differential
operator that locally-averages the strain rate. In principle a different
kernel $\delta_{b}\neq\delta_{a}$ can be used here in order to better
approximate the behavior of a rigid sphere \cite{ForceCoupling_Stokes}.
In the discrete setting one would have to construct a discrete operator
(matrix) $\M{\mathcal{K}}$ that gives good translational invariance;
this is a nontrivial task that amounts to constructing an immersed
boundary representation of force dipoles similar to that for monopoles
constructed by Peskin \cite{IBM_PeskinReview}. Enforcing the constraint
(\ref{eq:stresslet}) requires including Lagrange multipliers (stresslets)
$\M{\Lambda}_{i}$ in the velocity equation (\ref{eq:ContFluid})
\cite{ForceCoupling_Stokes}, $\grad\cdot\V v=0$ and 
\begin{eqnarray}
\rho\partial_{t}\V v+\grad\pi & = & \eta\grad^{2}\V v+\sqrt{2\eta k_{B}T}\,\grad\cdot\mathcal{\M{\mathcal{Z}}}+\V f_{\text{th}}\label{eq:v_eq}\\
 & + & \sum_{i}\V F_{i}\delta_{a}\left(\V q_{i}-\V r\right)+\sum_{i}\M{\Lambda}_{i}\grad\delta_{b}\left(\V q_{i}-\V r\right),\nonumber 
\end{eqnarray}
where the unknown stresslets $\M{\Lambda}_{i}$ are symmetric traceless
$d\times d$ tensors that need to be solved for.

Putting the pieces together and using compact composite operator notation
we can write the equations of motion including stresslet terms in
a form that applies either to the continuum or the spatially-discretized
equations,
\begin{eqnarray}
\rho\partial_{t}\V v & = & \eta\M L\bv+\M D^{\star}\pi+\M{\mathcal{K}}^{\star}\M{\Lambda}+\V f\label{eq:stresslet_system}\\
\M D\V v & = & 0\nonumber \\
\M{\mathcal{K}}\V v & = & \V 0\nonumber 
\end{eqnarray}
where $\V{\Lambda}=\left\{ \V{\Lambda}_{1},\dots,\V{\Lambda}_{N}\right\} $
are unknown stresslets and 
\begin{equation}
\V f=\J^{\star}\V F+\sqrt{\eta k_{B}T}\,\widetilde{\M D}\mathcal{\M{\mathcal{W}}}+\V f_{\text{th}}.\label{eq:forcing_f}
\end{equation}
The (translational) dynamics of the particles continues to be described
by the no-slip condition $d\V q/dt=\J\V v$. It is evident from the
form of (\ref{eq:stresslet_system}) that the rigidity constraint
$\M{\mathcal{K}}\V v=\V 0$ is in principle no different from the
divergence-free constraint, except for the fact that $\M{\mathcal{K}}\left(\V q\right)$
depends on the configuration of the particles just like $\J\left(\V q\right)$
does. In fact, as observed by Keaveny \cite{ForceCoupling_Fluctuations}
and also used in Stokesian Dynamics, the inclusion of stresslets simply
amounts to redefining the mobility matrix, and the overdamped limiting
dynamics for the positions of the particles is still given by (\ref{eq:BDKinetic}).
Equation (\ref{eq:BDKinetic}) in fact applies much more generally
and is not specific to multipole expansions.

The form of the modified mobility matrix can be obtained by deleting
the inertial term $\rho\partial_{t}\V v$ and solving the augmented
steady Stokes system (\ref{eq:stresslet_system}) using a Schur complement
approach, to obtain
\[
\V v=\left[\L^{-1}-\L^{-1}\M{\mathcal{K}}^{\star}\left(\M{\mathcal{K}}\L^{-1}\M{\mathcal{K}}^{\star}\right)^{-1}\M{\mathcal{K}}\L^{-1}\right]\V f=\M{\mathcal{N}}\V f,
\]
where $\L^{-1}$ is the (discrete) inverse Stokes operator given by
(\ref{eq:ExpliclitStokesSolve}) (note the identical structure of
$\L^{-1}$ and $\M{\mathcal{N}}$). This gives the mobility matrix
\[
\Mob=\J\M{\mathcal{N}}\J^{\star}=\J\M{\mathcal{N}}\S.
\]
Just as without stresslets, by solving the steady Stokes equation
with the random forcing (\ref{eq:forcing_f}) one can compute both
$\Mob\V F$ and the stochastic increments. That, is the square root
of the mobility matrix can be taken to be $\Mob^{\frac{1}{2}}=\J\M{\mathcal{N}}\widetilde{\M D}$,
as evident from the identity
\[
\J\M{\mathcal{N}}\left(\widetilde{\M D}\widetilde{\M D}^{\star}\right)\M{\mathcal{N}}^{\star}\J^{\star}=\J\M{\mathcal{N}}\S=\Mob,
\]
which is the generalization of (\ref{eq:DFDB_monopoles}) to account
for stresslets.

This shows that, in principle, it is relatively straightforward to
incorporate both boundary conditions and stresslets into the FIB algorithm
and thus do Stokesian dynamics without Green's functions. In practice,
there are significant challenges to surmount to accomplish this goal.
The main difficulty is that solving the system (\ref{eq:stresslet_system})
efficiently in the presence of nontrivial boundary conditions is hard
and requires the development of novel preconditioners. For periodic
domains one can use Fourier transform techniques to diagonalize the
Stokes operator, as used by Keaveny \cite{ForceCoupling_Fluctuations},
but in general one cannot easily decouple the computation of the stresslets
from solving the Stokes system. Furthermore, a nontrivial generalization
of the temporal algorithms developed here is required to obtain the
correct thermal drift term which comes from the fact that $\M{\mathcal{N}}\left(\V q\right)$
depends on the configuration because $\M{\mathcal{K}}\left(\V q\right)$
does.


\begin{thebibliography}{10}

\bibitem{ParticleMesoscaleHydrodynamics}
H.~Noguchi, N.~Kikuchi, and G.~Gompper.
\newblock {Particle-based mesoscale hydrodynamic techniques}.
\newblock {\em Europhysics Letters}, 78:10005, 2007.

\bibitem{SHSD_PRL}
A.~Donev, A.~L. Garcia, and B.~J. Alder.
\newblock {Stochastic Hard-Sphere Dynamics for Hydrodynamics of Non-Ideal
  Fluids}.
\newblock {\em Phys. Rev. Lett}, 101:075902, 2008.

\bibitem{DiffusionRenormalization_PRL}
A.~Donev, A.~L. Garcia, Anton de~la Fuente, and J.~B. Bell.
\newblock {Diffusive Transport by Thermal Velocity Fluctuations}.
\newblock {\em Phys. Rev. Lett.}, 106(20):204501, 2011.

\bibitem{Landau:Fluid}
L.D. Landau and E.M. Lifshitz.
\newblock {\em Fluid Mechanics}, volume~6 of {\em Course of Theoretical
  Physics}.
\newblock Pergamon Press, Oxford, England, 1959.

\bibitem{FluctHydroNonEq_Book}
J.~M.~O.~De Zarate and J.~V. Sengers.
\newblock {\em {Hydrodynamic fluctuations in fluids and fluid mixtures}}.
\newblock Elsevier Science Ltd, 2006.

\bibitem{OttingerBook}
H.~C. {\"O}ttinger.
\newblock {\em Beyond equilibrium thermodynamics}.
\newblock Wiley Online Library, 2005.

\bibitem{DiffusionRenormalization}
A.~Donev, A.~L. Garcia, Anton de~la Fuente, and J.~B. Bell.
\newblock {Enhancement of Diffusive Transport by Nonequilibrium Thermal
  Fluctuations}.
\newblock {\em J. of Statistical Mechanics: Theory and Experiment},
  2011:P06014, 2011.

\bibitem{StagerredFluct_Inhomogeneous}
B.~Z. Shang, N.~K. Voulgarakis, and J.-W. Chu.
\newblock {Fluctuating hydrodynamics for multiscale simulation of inhomogeneous
  fluids: Mapping all-atom molecular dynamics to capillary waves}.
\newblock {\em J. Chem. Phys.}, 135:044111, 2011.

\bibitem{StagerredFluctHydro}
N.~K. Voulgarakis and J.-W. Chu.
\newblock {Bridging fluctuating hydrodynamics and molecular dynamics
  simulations of fluids}.
\newblock {\em J. Chem. Phys.}, 130(13):134111, 2009.

\bibitem{StaggeredFluct_Energy}
B.Z. Shang, N.K. Voulgarakis, and J.W. Chu.
\newblock Fluctuating hydrodynamics for multiscale modeling and simulation:
  Energy and heat transfer in molecular fluids.
\newblock {\em J. Chem. Phys.}, 137(4):044117--044117, 2012.

\bibitem{LowMachExplicit}
A.~Donev, A.~J. Nonaka, Y.~Sun, T.~G. Fai, A.~L. Garcia, and J.~B. Bell.
\newblock {Low Mach Number Fluctuating Hydrodynamics of Diffusively Mixing
  Fluids}.
\newblock {\em Communications in Applied Mathematics and Computational
  Science}, 9(1):47--105, 2014.

\bibitem{FluctuatingHydro_FluidOnly}
N.~Sharma and N.~A. Patankar.
\newblock {Direct numerical simulation of the Brownian motion of particles by
  using fluctuating hydrodynamic equations}.
\newblock {\em J. Comput. Phys.}, 201:466--486, 2004.

\bibitem{LatticeBoltzmann_Polymers}
O.~B. Usta, A.~J.~C. Ladd, and J.~E. Butler.
\newblock {Lattice-Boltzmann simulations of the dynamics of polymer solutions
  in periodic and confined geometries}.
\newblock {\em J. Chem. Phys.}, 122(9):094902, 2005.

\bibitem{BD_LB_Comparison}
Anthony~JC Ladd, Rahul Kekre, and Jason~E Butler.
\newblock {Comparison of the static and dynamic properties of a semiflexible
  polymer using lattice Boltzmann and Brownian-dynamics simulations}.
\newblock {\em Physical Review E}, 80(3):036704, 2009.

\bibitem{BD_LB_Ladd}
Rahul Kekre, Jason~E. Butler, and Anthony J.~C. Ladd.
\newblock {Comparison of lattice-Boltzmann and Brownian-dynamics simulations of
  polymer migration in confined flows}.
\newblock {\em Phys. Rev. E}, 82:011802, 2010.

\bibitem{LB_SoftMatter_Review}
B.~D{\"u}nweg and A.J.C. Ladd.
\newblock {Lattice Boltzmann simulations of soft matter systems}.
\newblock {\em Adv. Comp. Sim. for Soft Matter Sciences III}, pages 89--166,
  2009.

\bibitem{SELM}
P.~J. Atzberger.
\newblock {Stochastic Eulerian-Lagrangian Methods for Fluid-Structure
  Interactions with Thermal Fluctuations}.
\newblock {\em J. Comp. Phys.}, 230:2821--2837, 2011.

\bibitem{ImmersedFEM_Patankar}
Adrian~M Kopacz, Neelesh~A Patankar, and Wing~K Liu.
\newblock The immersed molecular finite element method.
\newblock {\em Computer Methods in Applied Mechanics and Engineering},
  233:28--39, 2012.

\bibitem{ISIBM}
F.~Balboa Usabiaga, R.~Delgado-Buscalioni, B.~E. Griffith, and A.~Donev.
\newblock {Inertial Coupling Method for particles in an incompressible
  fluctuating fluid}.
\newblock {\em Comput. Methods Appl. Mech. Engrg.}, 269:139--172, 2014.
\newblock Code available at \url{https://code.google.com/p/fluam}.

\bibitem{ForceCoupling_Fluctuations}
Eric~E. Keaveny.
\newblock Fluctuating force-coupling method for simulations of colloidal
  suspensions.
\newblock {\em J. Comp. Phys.}, 269(0):61 -- 79, 2014.

\bibitem{CompressibleBlobs}
F.~Balboa~Usabiaga and R.~Delgado-Buscalioni.
\newblock {A minimal model for acoustic forces on Brownian particles}.
\newblock {\em Phys. Rev. E}, 88:063304, 2013.

\bibitem{BrownianParticle_IncompressibleSmoothed}
T.~Iwashita, Y.~Nakayama, and R.~Yamamoto.
\newblock {A Numerical Model for Brownian Particles Fluctuating in
  Incompressible Fluids}.
\newblock {\em Journal of the Physical Society of Japan}, 77(7):074007, 2008.

\bibitem{FIMAT_Patankar}
Y.~Chen, N.~Sharma, and N.~Patankar.
\newblock {Fluctuating Immersed Material (FIMAT) dynamics for the direct
  simulation of the Brownian motion of particles}.
\newblock In {\em IUTAM Symposium on Computational Approaches to Multiphase
  Flow}, pages 119--129. Springer, 2006.

\bibitem{BD_Fixman}
M.~Fixman.
\newblock {Simulation of polymer dynamics. I. General theory}.
\newblock {\em J. Chem. Phys.}, 69:1527, 1978.

\bibitem{BD_Hinch}
P.S. Grassia, E.J. Hinch, and L.C. Nitsche.
\newblock Computer simulations of brownian motion of complex systems.
\newblock {\em Journal of Fluid Mechanics}, 282:373--403, 1995.

\bibitem{BrownianDynamics_DNA}
R.~M. Jendrejack, J.~J. de~Pablo, and M.~D. Graham.
\newblock {Stochastic simulations of DNA in flow: Dynamics and the effects of
  hydrodynamic interactions}.
\newblock {\em J. Chem. Phys.}, 116(17):7752--7759, 2002.

\bibitem{BrownianDynamics_DNA2}
Richard~M Jendrejack, David~C Schwartz, Michael~D Graham, and Juan~J de~Pablo.
\newblock {Effect of confinement on DNA dynamics in microfluidic devices}.
\newblock {\em J. Chem. Phys.}, 119:1165, 2003.

\bibitem{BrownianDynamics_OrderN}
J.~P. Hernandez-Ortiz, J.~J. de~Pablo, and M.~D. Graham.
\newblock {Fast Computation of Many-Particle Hydrodynamic and Electrostatic
  Interactions in a Confined Geometry}.
\newblock {\em Phys. Rev. Lett.}, 98(14):140602, 2007.

\bibitem{BD_IBM_Graham}
Yu~Zhang, Juan~J de~Pablo, and Michael~D Graham.
\newblock An immersed boundary method for brownian dynamics simulation of
  polymers in complex geometries: Application to dna flowing through a nanoslit
  with embedded nanopits.
\newblock {\em The Journal of Chemical Physics}, 136:014901, 2012.

\bibitem{BrownianDynamics_FMM}
Shidong Jiang, Zhi Liang, and Jingfang Huang.
\newblock A fast algorithm for brownian dynamics simulation with hydrodynamic
  interactions.
\newblock {\em Mathematics of Computation}, 82(283):1631--1645, 2013.

\bibitem{BrownianDynamics_OrderNlogN}
A.~Sierou and J.~F. Brady.
\newblock {Accelerated Stokesian Dynamics simulations}.
\newblock {\em J. Fluid Mech.}, 448:115--146, 2001.

\bibitem{StokesianDynamics_Wall}
James~W. Swan and John~F. Brady.
\newblock {Simulation of hydrodynamically interacting particles near a no-slip
  boundary}.
\newblock {\em Physics of Fluids}, 19(11):113306, 2007.

\bibitem{HYDROLIB}
K~Hinsen.
\newblock {HYDROLIB: a library for the evaluation of hydrodynamic interactions
  in colloidal suspensions}.
\newblock {\em Computer physics communications}, 88(2):327--340, 1995.

\bibitem{SphereConglomerate}
B~Cichocki and K~Hinsen.
\newblock Stokes drag on conglomerates of spheres.
\newblock {\em Physics of Fluids}, 7:285, 1995.

\bibitem{HYDROPRO}
A~Ortega, D~Amor{\'o}s, and J~Garc{\'\i}a~de La~Torre.
\newblock Prediction of hydrodynamic and other solution properties of rigid
  proteins from atomic-and residue-level models.
\newblock {\em Biophysical journal}, 101(4):892--898, 2011.
\newblock Code available at
  \url{http://leonardo.inf.um.es/macromol/programs/hydropro/hydropro.htm}.

\bibitem{HYDROPRO_Globular}
Jos{\'e} Garc{\'\i}a de~la Torre, Mar{\'\i}a~L Huertas, and Beatriz Carrasco.
\newblock Calculation of hydrodynamic properties of globular proteins from
  their atomic-level structure.
\newblock {\em Biophysical Journal}, 78(2):719--730, 2000.

\bibitem{StokesianDynamics_Slit}
James~W Swan and John~F Brady.
\newblock Particle motion between parallel walls: Hydrodynamics and simulation.
\newblock {\em Physics of Fluids}, 22:103301, 2010.

\bibitem{StokesianDynamics_Confined}
James~W Swan and John~F Brady.
\newblock The hydrodynamics of confined dispersions.
\newblock {\em Journal of Fluid Mechanics}, 687:254, 2011.

\bibitem{SD_TwoWalls}
Rapha{\"e}l Pesch{\'e} and Gerhard N{\"a}gele.
\newblock Stokesian dynamics study of quasi-two-dimensional suspensions
  confined between two parallel walls.
\newblock {\em Physical Review E}, 62(4):5432, 2000.

\bibitem{StokesianDynamics_Brownian}
Adolfo~J Banchio and John~F Brady.
\newblock Accelerated stokesian dynamics: Brownian motion.
\newblock {\em The Journal of chemical physics}, 118:10323, 2003.

\bibitem{VACF_FluctHydro}
E.~H. Hauge and A.~Martin-Lof.
\newblock {Fluctuating hydrodynamics and Brownian motion}.
\newblock {\em J. Stat. Phys.}, 7(3):259--281, 1973.

\bibitem{Faxen_FluctuatingHydro}
D.~Bedeaux and P.~Mazur.
\newblock {Brownian motion and fluctuating hydrodynamics}.
\newblock {\em Physica}, 76(2):247--258, 1974.

\bibitem{VACF_Langevin}
E.~J. Hinch.
\newblock {Application of the Langevin equation to fluid suspensions}.
\newblock {\em J. Fluid Mech.}, 72(03):499--511, 1975.

\bibitem{LangevinDynamics_Theory}
J.~N. Roux.
\newblock {Brownian particles at different times scales: a new derivation of
  the Smoluchowski equation}.
\newblock {\em Phys. A}, 188:526--552, 1992.

\bibitem{StokesEinstein}
F.~Balboa Usabiaga, X.~Xie, R.~Delgado-Buscalioni, and A.~Donev.
\newblock {The Stokes-Einstein Relation at Moderate Schmidt Number}.
\newblock {\em J. Chem. Phys.}, 139(21):214113, 2013.

\bibitem{LBM_vs_BD_Burkhard}
Tri~T Pham, Ulf~D Schiller, J~Ravi Prakash, and Burkhard D{\"u}nweg.
\newblock Implicit and explicit solvent models for the simulation of a single
  polymer chain in solution: Lattice boltzmann versus brownian dynamics.
\newblock {\em J. Chem. Phys.}, 131:164114, 2009.

\bibitem{Hematites_PRL}
I.~Theurkauff, C.~Cottin-Bizonne, J.~Palacci, C.~Ybert, and L.~Bocquet.
\newblock Dynamic clustering in active colloidal suspensions with chemical
  signaling.
\newblock {\em Phys. Rev. Lett.}, 108:268303, 2012.

\bibitem{Hematites_Science}
Jeremie Palacci, Stefano Sacanna, Asher~Preska Steinberg, David~J Pine, and
  Paul~M Chaikin.
\newblock Living crystals of light-activated colloidal surfers.
\newblock {\em Science}, 339(6122):936--940, 2013.

\bibitem{NonProjection_Griffith}
B.E. Griffith.
\newblock {An accurate and efficient method for the incompressible
  Navier-Stokes equations using the projection method as a preconditioner}.
\newblock {\em J. Comp. Phys.}, 228(20):7565--7595, 2009.

\bibitem{StochasticImmersedBoundary}
P.~J. Atzberger, P.~R. Kramer, and C.~S. Peskin.
\newblock {A stochastic immersed boundary method for fluid-structure dynamics
  at microscopic length scales}.
\newblock {\em J. Comp. Phys.}, 224:1255--1292, 2007.

\bibitem{ForceCoupling_Stokes}
S.~Lomholt and M.R. Maxey.
\newblock {Force-coupling method for particulate two-phase flow: Stokes flow}.
\newblock {\em J. Comp. Phys.}, 184(2):381--405, 2003.

\bibitem{ForceCoupling_Monopole}
M.~R. Maxey and B.~K. Patel.
\newblock {Localized force representations for particles sedimenting in Stokes
  flow}.
\newblock {\em International journal of multiphase flow}, 27(9):1603--1626,
  2001.

\bibitem{ForceCoupling_Ellisoids}
D~Liu, EE~Keaveny, Martin~R Maxey, and George~E Karniadakis.
\newblock Force-coupling method for flows with ellipsoidal particles.
\newblock {\em Journal of Computational Physics}, 228(10):3559--3581, 2009.

\bibitem{IBM_PeskinReview}
C.S. Peskin.
\newblock {The immersed boundary method}.
\newblock {\em Acta Numerica}, 11:479--517, 2002.

\bibitem{SELM_FEM}
Pat Plunkett, Jonathan Hu, Christopher Siefert, and Paul~J Atzberger.
\newblock Spatially adaptive stochastic methods for fluid--structure
  interactions subject to thermal fluctuations in domains with complex
  geometries.
\newblock {\em Journal of Computational Physics}, 277:121--137, 2014.

\bibitem{KineticStochasticIntegral_Ottinger}
M.~H{\"u}tter and H.C. {\"O}ttinger.
\newblock Fluctuation-dissipation theorem, kinetic stochastic integral and
  efficient simulations.
\newblock {\em J. Chem. Soc., Faraday Trans.}, 94(10):1403--1405, 1998.

\bibitem{BD_Fixman_sqrtM}
Marshall Fixman.
\newblock Construction of langevin forces in the simulation of hydrodynamic
  interaction.
\newblock {\em Macromolecules}, 19(4):1204--1207, 1986.

\bibitem{RotnePrager}
Jens Rotne and Stephen Prager.
\newblock Variational treatment of hydrodynamic interaction in polymers.
\newblock {\em The Journal of Chemical Physics}, 50:4831, 1969.

\bibitem{DiffusionJSTAT}
A.~Donev, T.~G. Fai, and E.~Vanden-Eijnden.
\newblock {A reversible mesoscopic model of diffusion in liquids: from giant
  fluctuations to Fick's law}.
\newblock {\em Journal of Statistical Mechanics: Theory and Experiment},
  2014(4):P04004, 2014.

\bibitem{blake1971note}
JR~Blake.
\newblock A note on the image system for a stokeslet in a no-slip boundary.
\newblock In {\em Proc. Camb. Phil. Soc}, volume~70, pages 303--310. Cambridge
  Univ Press, 1971.

\bibitem{SIBM_Brownian}
P.~J. Atzberger.
\newblock {A note on the correspondence of an immersed boundary method
  incorporating thermal fluctuations with Stokesian-Brownian dynamics}.
\newblock {\em Physica D: Nonlinear Phenomena}, 226(2):144--150, 2007.

\bibitem{ForceCoupling_Channel}
Kyongmin Yeo and Martin~R Maxey.
\newblock Dynamics of concentrated suspensions of non-colloidal particles in
  couette flow.
\newblock {\em Journal of Fluid Mechanics}, 649(1):205--231, 2010.

\bibitem{LB_IB_Points}
R.~W. Nash, R.~Adhikari, and M.~E. Cates.
\newblock Singular forces and pointlike colloids in lattice boltzmann
  hydrodynamics.
\newblock {\em Physical Review E}, 77(2):026709, 2008.

\bibitem{DirectForcing_Balboa}
F.~Balboa Usabiaga, I.~Pagonabarraga, and R.~Delgado-Buscalioni.
\newblock {Inertial coupling for point particle fluctuating hydrodynamics}.
\newblock {\em J. Comp. Phys.}, 235:701--722, 2013.

\bibitem{ReactiveBlobs}
A.~Pal~Singh Bhalla, B.~E. Griffith, N.~A. Patankar, and A.~Donev.
\newblock {A Minimally-Resolved Immersed Boundary Model for Reaction-Diffusion
  Problems}.
\newblock {\em J. Chem. Phys.}, 139(21):214112, 2013.

\bibitem{IBMDelta_Boundary}
B.E. Griffith, X.~Luo, D.M. McQueen, and C.S. Peskin.
\newblock Simulating the fluid dynamics of natural and prosthetic heart valves
  using the immersed boundary method.
\newblock {\em International Journal of Applied Mechanics}, 1(01):137--177,
  2009.

\bibitem{SELM_Reduction}
G.~Tabak and P.J. Atzberger.
\newblock Systematic stochastic reduction of inertial fluid-structure
  interactions subject to thermal fluctuations.
\newblock {\em arXiv preprint arXiv:1211.3798}, 2013.

\bibitem{DFDB}
S.~Delong, B.~E. Griffith, E.~Vanden-Eijnden, and A.~Donev.
\newblock {Temporal Integrators for Fluctuating Hydrodynamics}.
\newblock {\em Phys. Rev. E}, 87(3):033302, 2013.

\bibitem{LLNS_Staggered}
F.~Balboa Usabiaga, J.~B. Bell, R.~Delgado-Buscalioni, A.~Donev, T.~G. Fai,
  B.~E. Griffith, and C.~S. Peskin.
\newblock {Staggered Schemes for Fluctuating Hydrodynamics}.
\newblock {\em SIAM J. Multiscale Modeling and Simulation}, 10(4):1369--1408,
  2012.

\bibitem{AdiabaticElimination_1}
C.~W. Gardiner and M.~L. Steyn-Ross.
\newblock {Adiabatic elimination in stochastic systems. I-III}.
\newblock {\em Phys. Rev. A}, 29:2814--2844, 1984.

\bibitem{AveragingHomogenization}
Grigorios~A Pavliotis and Andrew~M Stuart.
\newblock {\em Multiscale methods: averaging and homogenization}, volume~53.
\newblock Springer, 2008.

\bibitem{StokesLaw}
A.~Donev, T.~G. Fai, and E.~Vanden-Eijnden.
\newblock {Reversible Diffusion by Thermal Fluctuations}.
\newblock Arxiv preprint 1306.3158, 2013.

\bibitem{Mobility2D_Hasimoto}
H~Hasimoto.
\newblock On the periodic fundamental solutions of the stokes equations and
  their application to viscous flow past a cubic array of spheres.
\newblock {\em J. Fluid Mech}, 5(02):317--328, 1959.

\bibitem{StaggeredIBM}
Alexandre~M Roma, Charles~S Peskin, and Marsha~J Berger.
\newblock An adaptive version of the immersed boundary method.
\newblock {\em J. Comput. Phys.}, 153(2):509--534, 1999.

\bibitem{EulerHeun}
Thomas Schaffter.
\newblock Numerical integration of sdes: a short tutorial.
\newblock {\em Swiss Federal Institute of Technology in Lausanne (EPFL),
  Switzerland, Unpublished manuscript}, 2010.

\bibitem{WeakSecondOrder_RK}
A~Abdulle, G~Vilmart, and K~Zygalakis.
\newblock Weak second order explicit stabilized methods for stiff stochastic
  differential equations.
\newblock {\em SIAM J. Sci. Comput.}, 35(4):A1792--A1814, 2013.

\bibitem{IBAMR}
B.E. Griffith, R.D. Hornung, D.M. McQueen, and C.S. Peskin.
\newblock {An adaptive, formally second order accurate version of the immersed
  boundary method}.
\newblock {\em J. Comput. Phys.}, 223(1):10--49, 2007.
\newblock Software available at \url{http://ibamr.googlecode.com}.

\bibitem{NUFFT}
L.~Greengard and J.~Lee.
\newblock Accelerating the nonuniform fast fourier transform.
\newblock {\em SIAM Review}, 46(3):443--454, 2004.

\bibitem{MobilitySlitChannel}
Thorben Benesch, Sotira Yiacoumi, and Costas Tsouris.
\newblock Brownian motion in confinement.
\newblock {\em Phys. Rev. E}, 68:021401, 2003.

\bibitem{SingleWallPerpMobility}
Peter Huang and Kenneth~S Breuer.
\newblock Direct measurement of anisotropic near-wall hindered diffusion using
  total internal reflection velocimetry.
\newblock {\em Physical review E}, 76(4):046307, 2007.

\bibitem{NearWallSphereMobility}
AJ~Goldman, Raymond~G Cox, and Howard Brenner.
\newblock Slow viscous motion of a sphere parallel to a plane wall - i motion
  through a quiescent fluid.
\newblock {\em Chemical engineering science}, 22(4):637--651, 1967.

\bibitem{ConfinedSphere_Sedimented}
Luc~P. Faucheux and Albert~J. Libchaber.
\newblock Confined brownian motion.
\newblock {\em Phys. Rev. E}, 49:5158--5163, 1994.

\bibitem{RPY_Shear_Wall}
Eligiusz Wajnryb, Krzysztof~A Mizerski, Pawel~J Zuk, and Piotr Szymczak.
\newblock Generalization of the rotne--prager--yamakawa mobility and shear
  disturbance tensors.
\newblock {\em Journal of Fluid Mechanics}, 731:R3, 2013.

\bibitem{Hydro_Tweezers}
John~C Crocker.
\newblock Measurement of the hydrodynamic corrections to the brownian motion of
  two colloidal spheres.
\newblock {\em J. Chem. Phys.}, 106:2837, 1997.

\bibitem{holzer2010dynamics}
Lukas Holzer, Jochen Bammert, Roland Rzehak, and Walter Zimmermann.
\newblock Dynamics of a trapped brownian particle in shear flows.
\newblock {\em Physical Review E}, 81(4):041124, 2010.

\bibitem{bammert2010dynamics}
Jochen Bammert, Lukas Holzer, and Walter Zimmermann.
\newblock Dynamics of two trapped brownian particles: Shear-induced
  cross-correlations.
\newblock {\em The European Physical Journal E}, 33(4):313--325, 2010.

\bibitem{Atzberger_Shear}
Paul~J. Atzberger.
\newblock Incorporating shear into stochastic eulerian-lagrangian methods for
  rheological studies of complex fluids and soft materials.
\newblock {\em Physica D: Nonlinear Phenomena}, 265(0):57 -- 70, 2013.

\bibitem{RotnePrager_Periodic}
C.~W.~J. Beenakker.
\newblock {Ewald sum of the Rotne-Prager tensor}.
\newblock {\em J. Chem. Phys.}, 85:1581, 1986.

\bibitem{ColloidalGelation}
A.~Furukawa and H.~Tanaka.
\newblock Key role of hydrodynamic interactions in colloidal gelation.
\newblock {\em Phys. Rev. Lett.}, 104(24):245702, 2010.

\bibitem{MetropolizedBD}
N.~Bou-Rabee, A.~Donev, and E.~Vanden-Eijnden.
\newblock {Metropolis Integration Schemes for Self-Adjoint Diffusions}.
\newblock {\em SIAM J. Multiscale Modeling and Simulation}, 12(2):781--831,
  2014.

\bibitem{Stokes2D_Greengard}
Leslie Greengard and Mary~Catherine Kropinski.
\newblock An integral equation approach to the incompressible navier--stokes
  equations in two dimensions.
\newblock {\em SIAM Journal on Scientific Computing}, 20(1):318--336, 1998.

\bibitem{Stokes3D_Greengard}
Anna-Karin Tornberg and Leslie Greengard.
\newblock A fast multipole method for the three-dimensional stokes equations.
\newblock {\em Journal of Computational Physics}, 227(3):1613--1619, 2008.

\bibitem{StokesSecondKind_Shelley}
Eric~E Keaveny and Michael~J Shelley.
\newblock Applying a second-kind boundary integral equation for surface
  tractions in stokes flow.
\newblock {\em Journal of Computational Physics}, 230(5):2141--2159, 2011.

\bibitem{IBM_Viscoelasticity}
D.~Devendran and C.~S. Peskin.
\newblock An immersed boundary energy-based method for incompressible
  viscoelasticity.
\newblock {\em J. Comp. Phys.}, 231(14):4613--4642, 2012.

\bibitem{MilsteinSDEBook}
G.N. Milstein and M.V. Tretyakov.
\newblock {\em Stochastic numerics for mathematical physics}.
\newblock Springer, 2004.

\bibitem{Tethered_Polymer}
Y.~Zhang, A.~Donev, T.~Weisgraber, B.~J. Alder, M.~D. Graham, and J.~J.
  de~Pablo.
\newblock {Tethered DNA Dynamics in Shear Flow}.
\newblock {\em J. Chem. Phys}, 130(23):234902, 2009.

\bibitem{IBM_Generalized}
B.E. Griffith and S.~Lim.
\newblock {Simulating an elastic ring with bend and twist by an adaptive
  generalized immersed boundary method}.
\newblock {\em Commun. Comput. Phys.}, 12:433--461, 2012.

\bibitem{IBM_TwistBend}
S.~Lim, A.~Ferent, X.~S. Wang, and C.~S. Peskin.
\newblock Dynamics of a closed rod with twist and bend in fluid.
\newblock {\em SIAM~J~Sci~Comput}, 31(1):273--302, 2008.

\end{thebibliography}

\end{document}